\documentclass[useAMS,a4paper,usenatbib]{mnras}

\usepackage{newtxtext}

\usepackage[T1]{fontenc}
\usepackage{ae,aecompl}

\usepackage{graphicx}	
\usepackage{float}
\usepackage{caption}

\usepackage{amsmath}	
\usepackage{amssymb}	
\usepackage{listings}
\usepackage{color}
\usepackage{booktabs}   
\usepackage{natbib}
\usepackage{hyperref}    
\hypersetup{ 			
    colorlinks=true,
    linkcolor=blue,
    filecolor=magenta,      
    urlcolor=cyan,
}

\title[Disk Winds, Turbulence, \& Planet Populations]{Combined Effects of Disk Winds \& Turbulence-Driven Accretion on Planet Populations}

\author[M. Alessi \& R. E. Pudritz]{
Matthew Alessi$^{1}$\thanks{E-mail:
matthew.alessi3@gmail.com (MA); pudritz@mcmaster.ca (REP)} and Ralph E. Pudritz$^{1,2}$\footnotemark[1]\\
$^{1}$Department of Physics and Astronomy, McMaster University, Hamilton, ON L8S 4M1, Canada\\
$^{2}$Origins Institute, McMaster University, Hamilton, ON L8S 4M1, Canada\\
}

\date{Accepted XXX. Received YYY; in original form ZZZ}

\pubyear{2020}

\begin{document}
\label{firstpage}
\pagerange{\pageref{firstpage}--\pageref{lastpage}}
\maketitle

\begin{abstract}
Recent surveys  show that protoplanetary disks have lower levels of turbulence than expected based on their observed accretion rates.  A viable solution to this is that magnetized disk winds dominate angular momentum transport.  This has several important implications for planet formation processes. We compute the physical and chemical evolution of disks and the formation and migration of planets under the combined effects of angular momentum transport by turbulent viscosity and disk winds.  We take into account the critical role of planet traps to limit Type I migration in all of these models and compute thousands of planet evolution tracks for single planets drawn from a distribution of initial disk properties and turbulence strengths.  We do not consider multi-planet models nor include N-body planet-planet interactions.  Within this physical framework we find that populations with a constant value disk turbulence and winds strength produce mass-semimajor axis distributions in the M-a diagram with insufficient scatter to compare reasonably with observations. However,  populations produced as a consequence of  sampling disks with a distribution of the relative strengths of disk turbulence and winds fit much better.  Such models give rise to a substantial super Earth population at orbital radii 0.03-2 AU, as well as a clear separation between the produced hot Jupiter and warm Jupiter populations.   Additionally, this model results in a good comparison with the exoplanetary mass-radius distribution in the M-R diagram after post-disk atmospheric photoevaporation is accounted for.  
\end{abstract}




\section{Introduction} \label{Results4_Intro}

How are the observed properties of exoplanet populations connected to physical processes and planet formation in disks?    Signatures of planet formation on extant populations could be encoded in several ways such as the distributions  of their orbital characteristics, planetary radii, or in the compositions of their atmospheres.     As an example, a classic paper argues that the C/O ratios of materials in planetary atmospheres could constrain where a planet may form in a disk in relation to water, $ CO $, or $CO_2$ ice lines \citep{Oberg2011}.  A difficulty in making a quantitative  connection is that planets likely migrate while they form so that atmospheric composition also reflects the collection of materials that they acquire from different parts of the disk \citep{ Cridland2016, Madhusudhan2018, Cridland2019, Cridland2019b}.   

Surveys have clearly shown that exoplanetary populations span a large range of properties as is apparent in the mass-semimajor axis (hereafter M-a) and mass-radius (hereafter M-R) diagrams (\citealt{Batalha2013, Rowe2014, Morton2016}; see also review by \citealt{Winn2015}).   The statistical properties of populations have been significantly improved over the last 5 years:  the most complete catalogue of transients in the Kepler data contains 4034 planet candidates \citep{Thompson2018}; RV methods are in development to reach 10 cm per second in order to detect Earth mass planets \citep{Fischer2016}; and new surveys are measuring planet masses around other kinds of stars such as  M dwarfs \citep{Reiners2018}.   

Planets form in protoplanetary disks and the surge of spatially-resolved  disk (sub)millimeter observations has revealed a striking amount of substructure  in disks' solid distributions (i.e. \citealt{ALMA2015, VanBoekel2017, Andrews2018b}). These structures may result as a consequence of gap opening in the dust due to the presence of forming planets (e.g. \citealt{Dong2015b, Fung2016}).  Young planets forming in such disks  have  now been directly observed in the PDS70 system \citep{Keppler2018,  Muller2018, Haffert2019}.    Recent disk surveys show that these  disks have an inherent scatter in their global quantities, such as masses and radii \citep{Ansdell2016, Pascucci2016, Tazzari2017, Ansdell2018, Andrews2020, vanderMarel2021, Long2022}.  

The scatter in the exoplanet M-a and M-R diagrams embodies a wide range of processes.  These  include the range of initial conditions for newly formed disks (disk masses, radii, lifetimes, metallicities) that play a huge role in planet formation;  the dominant processes that dictate disk evolution and hence planet formation within them \citep{Armitage2018};  the dynamical evolution of planetary systems due to N-body interactions between planets during both the formation phase in the disk \citep{Emsenhuber2021} as well as after disk dispersal  \citep{Chatterjee2008, Ford2008} ;  and the billions of years of planetary evolution that shaped planetary atmospheres.   
  
The central physical process that dictates the evolution of accretion disks and planet formation is how disk angular momentum is transported.  Planet formation theory has for decades assumed that turbulent stress is the sole agent of disk angular momentum transport in the  core accretion picture.   The strength of the disks' turbulent intensity is parameterized by the $\alpha_{\rm{turb}}$ parameter.  In the standard theory of viscous accretion   \citep{SS1973}: the viscosity arising from such turbulence is $\nu = \alpha_{\rm{turb}} c_s H$, where $c_s$ is the sound speed and $H$ is the pressure scale height.  The ratio of the rms turbulent to thermal pressure in the disk, measured by the value of $\alpha_{\rm{turb}}$, has a significant effect on disks' evolution timescales in viscous disk theory.  Its value determines the disk accretion rate as well as the  outward spread of material in the outer regions where angular momentum is ultimately deposited.  Its magnitude also plays a central role in planet formation controlling planet migration through co-rotation torques as well as gap opening processes ( for a recent review, see \citet{Nelson2018}).   The physical mechanism that generates turbulence was thought to be the magneto-rotational instability (MRI)  \cite{ BalbusHawley1991} although weaker turbulence can arise  from a number of hydrodynamic and thermal instabilities can arise in vertically stratified disks \citep{Klahr2018}.  

In order to test whether turbulent viscosity by itself is sufficient to explain the data,  it is necessary to measure  $\alpha$.   This, until recently, has proven to be frustratingly difficult  to do.  Previous indirect attempts  include matching the related evolution timescale to observationally inferred disk lifetimes \citep{Hernandez2007}, or matching the related accretion rates to those observationally inferred through H$\alpha$ emission \citep{Hartmann1998b}. These approaches result in estimates for $\alpha_{\rm{turb}} \sim 10^{-4}-10^{-2}$. It is important to note that these estimations assume that disk evolution is \emph{solely} driven by turbulence.  
The first direct measurements of disk turbulence have now been acquired by ALMA observations.  The results show that disks exhibit a lower levels than expected from the earlier inferences. Observed levels of disk line-broadening \citep{Flaherty2018} and studies of dust properties within pressure traps \citep{Dullemond2018, Risotti2020} have constrained the strength of turbulence in protoplanetary disks to an upper limit of $\alpha_{\rm{turb}} \lesssim 0.007 $.   The observations of disk accretion rates \citep{Manara2019} have orders of magnitude greater variation than can be accounted for by  a fixed fiducial value of $\alpha = 2 \times 10^{-3}$ \citep{Mordasini2012b} used in population synthesis modeling.     \citet{Mulders2017} argued that a dispersion in the value of $\alpha $ by 2 dex is needed to explain the variations in the observations in comparison to the use of a single value of this parameter.   Limited amounts of outward disk spreading observed for disks in the Lupus star forming region suggest low disk viscosity \citep{Trapman2019}.   Detailed study of the HL Tau disk has shown that whereas the accretion rate onto its central star by viscous torques would require $\alpha_{\rm{turb}} \simeq 10^{-2} $, the actual dust scale height is a factor of 5 smaller than the gas scale height which puts an observational upper limit of  $\alpha_{\rm{turb}} \simeq 10^{-4}  $ on the disk turbulence \citep{Pinte2016}.  Theoretical studies show that when the disk turbulence is high, dust trapping in pressure maxima is far less efficient than in the case of low turbulent amplitudes \citep{Pinilla2020}. The most recent compilation of a wide range of data and methods finds that typically  $\alpha_{\rm{turb}} \sim 10^{-3} - 10^{-4} $ \citep{Pinte2022}.   This extensive set of observations suggests that a range of turbulent viscosities is intrinsic to protoplanetary disk populations and we incorporate this into the evolution of our planetary populations. 

Disks with low values of disk viscosity $\alpha_{\rm{turb}} \sim 10^{-4}$ have particularly interesting implications for planetary migration.  Low viscosity allows outward directed co-rotation torques to push low mass forming planets out to 10s of AU via Type I migration before they eventually turn inward as they get caught up in the heat transition dynamical trap \citep{Speedie2022}.  Low viscosities are expected in more massive disks which are better shielded from external sources of radiation which ionize disks and promote MRI turbulence.  These theoretical results may provide an explanation as to why only a minority of disks are large ring/gap systems \citep{vanderMarel2021}. 

Hydromagnetic disk winds have long been known to be effective in transporting disk angular momentum.  Crucially, they provide a robust explanation for the ubiquitous appearance of  jets and outflows  during star formation.  They are connected with protostellar disks associated with the full range of stellar masses (see recent reviews \citep{Pudritz2019, Pascucci2022}.  Early theoretical work predicted  that centrifugally driven  MHD disk-winds could efficiently transport disk angular momentum  through an outflow of disk material along  magnetic field lines threading the disk \citep{Blandford1982, Pudritz1986, Pelletier1992, Ferreira1995, Ouyed1997, Spruit2010, Bai2016, Tabone2020}.  These can also be very efficient in that a small wind mass loss rate can drive much higher accretion rates.  Recent ALMA observations have resolved and measured the rotation of outflows and that they originate from extended disk scales \citep{Chen2016, Bjerkeli2016, Louvet2018, Zhang2018, deValon2020}.  These measurements show that up to 60$\%$ or more of the disk's angular momentum is being carried out in some of the observed rotating outflows.

Numerical MHD simulations are now sophisticated enough to follow the coupling and evolution of magnetic fields in dense poorly ionized protoplanetary disks.  This involves careful simulations of non-ideal MHD processes; Ohmic dissipation, ambipolar diffusion, and the Hall effect.    Such works have explored how disk evolution is tied to the detailed ionization structure that depends on non-equilibrium chemistry (e.g. \citet{BaiStone2013, Lesur2014, Gressel2015, Bai2016b, Gressel2020, Rodenkirch2020}) and that disk winds carry off the bulk of the angular momentum.  As an example, MRI is suppressed by a combinateion of  ambipolar diffusion and Hall effect down to $\alpha_{\rm{turb}} \sim 10^{-4}$ at the disk midplane  in the outer regions (5-60 AU) of the disk \citep{Bai2015}.    Several papers have investigated disk wind effects on disk structure including  \citet{Suzuki2016, Hasegawa2016, Bai2016}.   The lack of magnetically driven turbulence in wide reaches of the disk implies that a variety of hydrodynamic instabilities such as   the vertical shear, convective overstability, and Zombie Vortex instabilities can provide weak to modest levels of turbulence in disks (see reviews \citet{Klahr2018, Lesur2022}).  These could be important to help drive planetesimal formation.

Turbulence and disk winds act quite differently in controlling disk evolution.  In the former angular momentum is transported radially outward resulting in the gradual expansion of the disk while the latter moves it  vertically and away from the disk leading to disk contraction, known as advective disks \citep{Nelson2018}.  One of the most significant consequences of the advective nature of wind driven disks is that they can in principle have higher surface densities in their inner regions than turbulence models.  This may be significant in promoting planet formation there \citep{Suzuki2016, Chambers2019}.  A caveat to this is that the central surface densities will also depend on the mass loss rate carried by the disk wind, a point discussed in the next section.  Migration in purely wind-driven disks has been analyzed  by \citet{McNally2017, McNally2018, Kimmig2020}.  Winds that are strong enough can drive higher disk inflow speeds which in turn power co-rotation torques that drive outward migration \citep{Kimmig2020, Speedie2022}.  For low mass planetary migration, the absence of turbulent diffusion in wind driven disks means that local vortensity gradients, important for the corotation torques on the planets, are not dissipated so that the direction and magnitude of the wind driven torque depends upon the migration history of the planet \citep{McNally2017, McNally2018, Nelson2018, Paardekooper2022}.

Our earlier papers focused on  MRI driven turbulence, and incorporated the results into a planet population synthesis method. We assumed that angular momentum transport from the inner dead zone of such disks would be carried off by a disk wind \citep{Alessi2017} which was not treated in any great physical detail.  We found correspondence with many features of the planetary M-a and M-R relations by considering observationally-constrained ranges of disk properties as inputs to the core accretion planet formation model. In \citet{Alessi2018}, we inferred the value of forming planets' envelope opacities of $\sim$0.001 cm$^2$ g$^{-1}$ by comparing our models' gas giant orbital radius distributions with that of observations. Next, in \citet*{Alessi2020}, we incorporated a full treatment of dust evolution and radial drift into our approach, and discovered having large super Earth populations requires initial disk radii of the order $\sim$50 AU.  This is in accord with recent ALMA observations of disk radii, the majority of which are rather compact in structure \citep{vanderMarel2021}. Finally, in \citet*{Alessi2020b}, we incorporated a full solid disk chemistry treatment to track planets' compositions and also considered post-disk phase atmospheric mass loss via photoevaporation to determine the fraction of accreted atmospheres that escape. These both shaped our populations' M-R distributions, with photoevaporaton reducing the masses of our close-in Hot Jupiters to give better agreement with the M-a diagram.  

In this paper, we extend our previous investigations to address planet formation and migration in chemically evolving disks  whose dynamical evolution is driven by \emph{both} turbulence and magnetohydrodynamic (hereafter MHD) disk winds in much greater physical detail.  The remainder of this paper is structured as follows: In section \ref{Results4_Model}; we will summarize the \citet{Chambers2019} self-similar analytical disk model that handles both viscous and disk wind driven evolution.  This is a new addition to our treatment  and we detail the additions and constraints we have made to it, summarize our planet formation model, and define individual models' parameter settings that will be used in this paper. In section \ref{Results4_1}, we first examine the effects of different relative disk winds and turbulence strengths on individual planet formation tracks using fiducial disk parameters. This analysis is extended to disks with low turbulence levels in Appendix \ref{Results4_2}, where the disk outflow strength is examined. We then extend our investigation to full planet population synthesis models in section \ref{Results3}, where synthetic M-a and M-R distributions are shown that incorporate a distribution of turbulent $\alpha$ settings as well as post-disk atmospheric mass loss through photoevaporation. Lastly, in section \ref{Results4_Conclusion}, we summarize our main conclusions and discuss extensions of this treatment that will be considered in future work.

\section{Protoplanetary Disk Model: Combined Evolution Through Turbulence and Disk Winds} \label{Results4_Model}

It is currently not feasible to compute the complete evolution of planet formation and migration for the lifetime (3 - 10 Myr) of protoplanetary disks  by using the full 3D MHD treatments discussed in the Introduction.   One useful approach that captures the essential physics is to build on the simplicity of the $\alpha$ formalism and to adopt two types of $\alpha$ for the disk - one for the turbulence and another for a disk wind.   These two-$\alpha$ models have proven to be successful in creating better matches to observed populations in the M-a diagram \citep{Alessi2017, Ida2018, Bitsch2019, Matsumura2021}.  The addition of disk winds appears to produce a pile up of planets at around 1AU without invoking photoevaporation of the disk  \citep{Matsumura2021} .

Here we focus entirely on the disk phase of planetary evolution.  Planet population synthesis models utilize the best and most basic physical models of  the various aspects of planet formation; the evolution of disk structure by both turbulence and disk winds; disk astrochemistry; and the details of Type I and II planetary migration.  By computing thousands of planetary evolution tracks which sample the initial distributions of mass, radius, and metallicity in evolving disks, we  build synthetic populations using a statistical approach \citep{IdaLin2004, IdaLin2008, Alibert2011, Benz2014, Mordasini2015, Alessi2017, Alessi2020, Alessi2020b, Pudritz2018} which are compared with the observations.   As far as possible, we use the distributions of disk initial conditions (mass, radius, metallicity)  inferred via observations. 

Computing entire populations of thousands of planets forming in such evolving disks is a daunting task when population synthesis is coupled  with astrochemical evolution of the disks.   This is where the power of accurate semi-analytical treatments for disk evolution comes into its own as they are an effective and computationally inexpensive means to deduce the basic effects of winds on disk evolution and  planetary populations.  

\subsection{General Model Description} 

We adopt the formalism of  \citet{Chambers2019} to calculate the evolution of an evolving disk under the combined action of turbulence and disk winds.   This analytic model is an extension of disk evolution via disk turbulence \citep{Chambers2009} and provides an elegant analytical treatment of the numerical approach to disk winds explored by  \citet{Suzuki2016}.  Our own approach builds upon our previous successful integration of   \citet{Chambers2009} into our existing planet population synthesis framework  \citep{Alessi2018, Alessi2020, Alessi2020b}.  Another advantage of our approach is that it allows for the strengths of MRI and MHD-winds to be individually specified, so that their relative effects can be ascertained. 

There are three basic components that are linked together in our model.  The first is to include the physics of turbulent and disk wind torques on planets to compute their migration and accretion of materials as a function of time and disk radius.  This analysis makes critical use of our disk astrochemistry treatment.  The second component is to perform population synthesis computations involving thousands of models to confront our results of planets formation under these combined torques with observed exoplanet populations in the M-a diagram.   Our final step is to to compute the related M-R diagrams by solving planet structure and atmosphere equations using the  cumulative accreted 
materials and gases as well as photoevaporation effects on atmospheres of close in planets .  This is an ambitious program and its success depends critically upon striking an optimal balance between the use of  evolving disk models that provide an accurate treatment of the physics of planet formation and migration and the ability to compute thousands of model planet formation histories over 10 Myr of disk evolution and planet formation.   

Mass loss rates in disk winds are an important input into these models.   It has been pointed out that if disk winds are strong enough, they can carve out a significant portion of the interior of the disk so that planet migration can be slowed and a population of close in SuperEarths established \citep{ Ogihara2018}.    This requires that  wind mass loss rates are comparable to or exceed the disk accretion rate; $ \dot{M}_{\rm{wind}} /  \dot{M}_{\rm{acc}}  \simeq 1.0  $.   This limit of heavy disk wind mass loss conflicts with observations of  87 sources ranging from Class 0 to Class II protostars for which a large number of which have a  ratio of wind mass loss to accretion rates 
$ \dot{M}_{\rm{wind}} /  \dot{M}_{\rm{acc}}  \simeq 0.1  $  \citep{Watson2016, Pascucci2022}.  Observations  jet outflows with detectable [OI] line emission show that  $ 30 \% $ of T Tauri stars with ages between 1-3 Myr  have high velocity component ( $ v > 40 km s^{-1} $)  emission wherein the average value of $ \dot{M}_{\rm{jet}} /  \dot{M}_{\rm{acc}} $ is 0.07,  with a spread of 0.01 - 0.5   \citep{Nisini2018}.  While there is a class of disk wind solutions known as tower flows that feature slow accretion flows and heavy wind mass loss rates \citep{Lesur2022} it is not clear that these are representative of typical situations in protoplanetary disks.  Thus, in this work we focus on   examining the limit of light fast winds that extract angular momentum very efficiently with low values of jet mass loss to disk accretion rates \citep{Pelletier1992, ZhuStone2018}.

We now summarize in a non-technical way, the basic components of our models that we used in the previous papers.  The reader may refer to \citep{Alessi2020} for all of the mathematical details.  For disks undergoing purely visccous stress we have always employed  the \citet{ Chambers2009} self-similar disk model for turbulent disk evolution.   Most importantly, it enables  a semi-analytic treatment of disk and planet evolution that allows a numerical approach capable of tracking planetary populations over disk life times (up to 10 Myrs).   At any moment in time, the solution prescribes the spatial variation of the disk surface density and temperature ranging from the inner, viscously heated region that gradually transitions to the outer radiatively heated disk region.  The time evolution of the disk is dictated by the self similar evolution and accretion due to viscous stress.  We showed that these models also correspond very well with observationally tested disk models by \cite{DAlessio1999}.   We also incorporated equilibrium chemistry in the evolving disks using an effective Gibbs free energy solver that allows the composition of materials accreted onto the forming planets \cite{Alessi2018, Alessi2020b} as well as the makeup of their atmospheres.  We used this technology in \citet{Alessi2020b} to compute the composition of planetary populations as well as their atmospheres and employ it in this paper.  

 Throughout this and our earlier papers we have adopted a core accretion picture of planet formation.   Recent advances show that planetesimals are most likely constructed by the  rapid accretion of pebbles as a consequence of streaming instability \citep{ Youdin2005, Johansen2007, Bitsch2015}.    In earlier papers we have employed dust models that have both a constant dust to gas ratio  \citep{Alessi2018} as well as including the radial drift of dust \citep{Alessi2020}.   Here we use  a constant dust to gas density which effectively leads to less compact planetary populations in the M-a diagram.  Given that planetesimal formation by the streaming instability is so rapid (within $10^5$ yr) in comparison with the migration and planetary accretion time scales, we assume the local surface density of planetesimals follows that of the dust.  This allows us to better focus on the interplay of turbulent and disk wind torques.   
 
 Planetary accretion from the surrounding disk is treated by a series of approximations adopted from standard planetesimal accretion scenarios.  The details of this are given in \citet{Alessi2018} which we briefly summarize. We start each computation of a planet formation track in a given disk model from a so-called oligarch or planetary embryo whose mass is  a hundredth of an Earth mass, $10^{-2} M_E$.  The first stage is the growth of oligarchs that takes place via planetesimal accretion, which we compute following  the standard model of  \citet{Kokuboida2002}. The transition from the oligarchic growth phase to gas accretion phases occurs when the planetesimal accretion decreases to the point that the envelope pressure is insufficient to support of the surrounding gas, which then accretes on to the planet, following \citet{ Ikoma2000, IdaLin2008, HP12}. 

Accretion onto forming planets with masses exceeding this critical mass takes place on the Kelvin-Helmholtz time-scale.  The ability of  a growing core to accrete gas strongly depends on how well that gas can cool, so the opacity of the atmosphere plays a major role.  As already indicated in the  Introduction, we choose an envelop opacity for accreting planets that gives the best results for  matching the position of gas giant planets in the M-a diagram, $ \kappa_{o,env}  \simeq 0.001$ cm$^2$ g$^{-1}$.  For this, we use the fits. provided by \citet{Mordasini2014} to numerical models of gas accretion to Kelvin-Helmholtz scaling parameterization.  
 
 We always apply one set of K-H parameters per disk model regardless of where the planets were accreting from.  \citet{Coleman2017}  computed the accretion of gas onto planets of various masses, opacities, and locations in the disk and found runaway gas accretion could be limited by the inability of gas to cool sufficiently quickly in the hot inner disk regions.  This work used interstellar opacities from \cite{Bell1994} which were artificially reduced by up to 100  but still higher than our best fit.  We surmise as we had in our earlier work that an excellent physical analysis of atmospheric opacity, and in particular the behaviour and contributions of dust grains within them, is critical.   Given these uncertainties, we expect that the masses of Hot Jupiters formed in our dead zone may be upper limits with the consequence that one might have more SuperEarths and fewer Hot Jupiters  being produced than our models would suggest (see Results).   

It is well established that planetary migration is an inescapable consequence of planet- disk interaction.   At sufficiently low planetary mass, forming planets do not open gaps and undergo well known Type I migration.   Many works have shown that Type I migration rates depend on  the interplay of co-rotation and Lindblad torques very near the planet and hence depend critically upon local conditions there  \citep{Lyra2010, HellaryNelson2012, Dittkrist2014, Baillie2016, Coleman2016}.   Growing planets rapidly migrate through disks until caught in regions of zero net torque where outward co-rotation torques balance inward directed Linblad torques.  In a recently published paper that is in a sense connected with the present work, we provide detailed simulations of planet evolution in evolving disks in \citet{Speedie2022}  that are based on detailed disk astrochemistry calculations and using the detailed torques computed in  \citep{Paardekooper2011}.  We address this further in the migration discussion below.   

\subsection{Inclusion of a disk wind into disk structure and planetary migration theory}  

Disk winds have a number of effects on disk evolution.  The magnetic torque exerted by the rotating magnetized wind removes disk angular momentum driving accretion through the disk \citep{Blandford1982,  Pudritz1986}.  Disk winds also carry away mass from the disk although  this is generally a small fraction of the accreted mass.     The extension of the self-similar treatment of viscous disk evolution to include the possibility of a disk wind was carried out by  \citet{Chambers2019}, whose basic approach is an application of detailed analysis presented in \cite{Suzuki2016} and \cite{Bai2016}.   

Wind driven advection and disk mass loss contribute two new terms to the standard
 equation that describes the evolution of the disk surface density  $\Sigma(r,t)$ : 
\begin{equation} \frac{\partial \Sigma}{\partial t} = \frac{3}{r}\frac{\partial}{\partial r}\left[r^{1/2}\frac{\partial}{\partial r} \left( r^{1/2} \nu \Sigma\right)\right] + \frac{1}{r}\frac{\partial}{\partial r} \left(rv_w\Sigma\right) -\dot{\Sigma}_w \;,\label{Results4_DiskEvolution} \end{equation}
Here, the first term is familiar in the standard treatment of disk surface density evolution driven via MRI-turbulent viscosity $\nu$.  Its self-similar solution is used in the earlier purely viscous models (i.e. the \citealt{Chambers2009} disk model that we have used previously in \citealt{Alessi2018, Alessi2020}, and \citealt{Alessi2020b}). The second and third terms correspond to the additional effects of the disk wind that contribute both a stress driving disk accretion (second term), and disk mass loss by outflow (third term).  The former is characterized by $v_w$ - the inward radial velocity caused by the disk wind, and the latter is characterized by $\dot{\Sigma}_w$ - the rate of surface density loss due to the wind outflow.

\citet{Chambers2019} introduces three parameters that correspond with each of these three physical effects:
\begin{itemize}
\item $v_0$; the inward velocity of material at $r_0$ = 1 AU, and $T_0$=150 K which is the disk temperature at $r_0$ caused \emph{solely} from radiation (see equation \ref{Results4_Temperature} and following description). This parameter in large part sets the initial disk accretion rate $\dot{M}_0$  which would be measured as the accretion rate onto the star.
\item $f_w$; the fraction of $v_0$ that is caused by disk winds. Setting $f_w$=1 corresponds to a pure disk winds scenario, while setting $f_w$=0 corresponds to a pure viscous evolution.
\item $K$; which characterizes the strength of the winds-driven outflow.
\end{itemize}

Solving the protoplanetary disk structure in this framework now introduces two additional parameters beyond the one $\alpha$ viscosity parameter \citep{SS1973} that is needed in the pure viscous scenario (i.e. \citealt{Chambers2009}), 

\begin{equation} \nu = \alpha c_s H\;, \label{SS_viscosity} \end{equation} 
where $c_s$ is the sound speed and $H$ is the disk scale height. In pure viscous models, $\alpha$ in equation \ref{SS_viscosity} sets the strength of turbulent viscosity. However, despite its definitions in the pure viscous scenario, we have regarded this $\alpha$ parameter more generally as a description of all contributions to disk evolution (an ``effective'' $\alpha$, see \citealt{Alessi2018}).

The three \citet{Chambers2019} model parameters are related to the three individual disk evolution mechanisms in equation \ref{Results4_DiskEvolution} as follows. The turbulent viscosity $\nu$ is related to $f_w$ and $v_0$ as,
\begin{equation} \nu = \frac{2}{3} (1 - f_w) r_0 v_0 \left(\frac{r}{r_0}\right)^{3/2} \left(\frac{T}{T_0}\right) \;,\end{equation}
where $T$ is the midplane temperature at radius $r$. The winds-driven velocity through the disk is,
\begin{equation} v_w = f_w v_0 \left(\frac{T}{T_0}\right)^{1/2} \;.\end{equation}
Lastly, the surface density outflow rate depends on the parameter $K$ as,
\begin{equation} \dot{\Sigma}_w = \frac{K f_w v_0 \Sigma}{r_0} \left(\frac{r}{r_0}\right)^{-3/2} \;. \label{Outflow_SD} \end{equation}
This equation can be integrated to determine the total mass outflow rate,
\begin{equation} \dot{M}_{\rm{wind}} = \int_{R_{in}}^{R_{out}} 2 \pi r \dot{\Sigma}_w \rm{d}r \;, \label{Outflow_Mdot} \end{equation}
where $R_{in}$ and $R_{out}$ are the inner and outer disk radii, respectively. 

It is important to note that the wind parameter is not freely variable, since it is the wind that carries off a portion of the disk's angular momentum.  It is safe to assume that over times much shorter than the disk's viscous evolution time scale, that the steady state version of the disk angular momentum equation can be used.  These involve the magnetic torque term of the wind upon the disk, whose solution shows that there is a link between the mass loss rate of the wind and the disk accretion rate \citep{Pudritz1986, Pelletier1992}.  We return to this point below, but for now, we merely note that within the present formalism, the angular momentum equation written in these variables yields \citep{Chambers2019}, 
\begin{equation} K =  {1 \over 2} \left(\frac{r}{r_0} \right)^{-1/4} \left( \frac{\Omega r} {v_{esc} - \Omega r} \right)
\end{equation}
\noindent where $v_{esc}$ is the  tangential velocity of the escaping disk wind.

A similar form of the disk evolution equation \ref{Results4_DiskEvolution} was numerically solved in \citet{Suzuki2016}. While three parameters are still used in their formalism (setting the strength of each of the three evolution mechanisms), the individual strengths of turbulence and disk winds are instead set using the standard $\alpha$ parameters; $\alpha_{\rm{turb}}$ and $\alpha_{\rm{wind}}$. The following equations can be used to convert between these parameters and the \citet{Chambers2019} $f_w$ and $v_0$ parameters,
\begin{equation} \alpha_{\rm{turb}} = \frac{(1-f_w) r_0 v_0 \Omega_0}{c_{s0}^2} \;, \end{equation}
and,
\begin{equation} \alpha_{\rm{wind}} = \frac{f_w v_0}{c_{s0}} \;. \end{equation}
Here, $\Omega_0$ is the Keplerian angular frequency at the reference radius $r_0$ = 1 AU, and $c_{s0}$ is the sound speed at reference temperature $T_0$ = 150 K. In viscous models using $\alpha_{\rm{turb}} = 10^{-3}$ outside the dead zone, we obtain an $\alpha_{\rm{wind}} \simeq 2\times 10^{-4}$ throughout the disk. However, even with the difference in their magnitudes, roughly 80\% of the disk angular momentum is carried by the wind in such a model.

The midplane temperature $T$ is solved using,
\begin{equation} T^4= T_0^4 \left(\frac{r}{r_0}\right)^{-2} + \left(\frac{3 G M_* F}{8 \pi \sigma_{\rm{sb}} r^3}\right)\left(\frac{3 \kappa \Sigma}{8}\right) \;,\label{Results4_Temperature} \end{equation}
where $G$ is the gravitational constant, $M_*$ is the host-star mass, $F\simeq3\pi \nu \Sigma$ is the mass flux due to turbulent viscosity-driven accretion, $\sigma_{\rm{sb}}$ is the Stefan-Boltzmann constant, and $\kappa$ is the disk opacity. Here, the first term corresponds to heating from host-star radiation, and the second to heating via viscous dissipation \citep{Ruden1991}.  In the absence of viscous heating, a pure radiative equilibrium $T_{\rm{req}}$ profile is obtained for a disk with a constant aspect ratio $H/r$ (i.e. \citet{ChiangGoldreich}), 
\begin{equation} T_{\rm{req}} = T_0 \left(\frac{r}{r_0}\right)^{-1/2} \;,\label{Results4_REQ} \end{equation}
where $T_0$=150 K is the temperature at reference radius $r_0$ = 1 AU.

We note that, in equation \ref{Results4_Temperature}, the viscous heating is only generated through turbulent viscosity, and not through a general dissipation of gravitational potential energy. In a pure winds scenario, then, there will be no viscous heating contribution, and the disk midplane temperature will be the radiative equilibrium profile. In this circumstance, the gravitational potential energy lost by accreting material will be carried away by the wind. This process has been recently investigated in \citet{Mori2019} using MHD simulations, who find that this general heating from gravitational dissipation is small compared to radiative or viscous heating ($\lesssim$ a 10\% change to the midplane temperature), which confirms this result of the treatment of equation \ref{Results4_Temperature}.

The disk opacity $\kappa$ has an effect on the strength of viscous heating. As was the case in the previously-considered pure viscous \citet{Chambers2009} model, the \citet{Chambers2019} model also takes the disk opacity to be constant $\kappa = \kappa_0$ throughout the majority of the disk's radial extent. The exception is in the innermost region above the evaporation temperature $T_{\rm{evap}}$ = 1500 K where dust grains sublimate, and the opacity becomes a steeply decreasing function of temperature \citep{Stepinski1998}. While \citet{Chambers2019} uses a small value of $\kappa_0 \simeq$ 0.1 cm$^2$ g$^{-1}$ that may arise following grain growth, we continue to use a disk opacity of $\kappa_0$ = 3 cm$^2$ g$^{-1}$ in accordance with our previous disk model's treatment.

The disk accretion rate $\dot{M}_{\rm{acc}}$ is determined by calculating the mass flux across an inner radius $R_{in}$, for which \citet{Chambers2019} uses 0.05 AU. The disk accretion rate is therefore,
\begin{equation} \dot{M}_{\rm{acc}} = 2 \pi R_{in} \Sigma(R_{in}) v(R_{in}) \;,\label{Results4_DiskAccretion} \end{equation}
where the velocity of disk material at the inner radius $v(R_{in})$ scales with the velocity at 1 AU $v_0$ (a model input parameter) following,
\begin{equation} v(R_{in}) = v_0 \left(\frac{T(R_{in})}{T_0}\right)^{1/2} \;, \end{equation}
where we recall that $T_0$ = 150 K is the temperature at reference radius 1 AU solely due to radiation. However, $T(R_{in})$ is the total midplane temperature at $R_{in}$ = 0.05 AU from combined viscous and radiative heating. 

In addition to the parameters listed, one also needs to specify the initial disk mass and radius for the model, which combine to set the characteristic surface density. The initial disk mass can be directly input, using for example $f_M = 0.1$ M$_\odot$, which is the average initial disk mass used in our population synthesis models. The initial disk radius is handled in the \citet{Chambers2019} formalism through an exponential cutoff radius $r_{\rm{exp}}$ which scales with the outer disk radius. \citet{Chambers2019} uses a setting of $r_{\rm{exp}}$ = 15 AU that we also adopt. As the name suggests, the value of $r_{\rm{exp}}$ indicates the radius where the surface density profile begins to decrease sharply with further increase in $r$. As we will see, however, significant disk surface densities can exist well outside $r_{\rm{exp}}$, so this parameter does not immediately indicate the outer disk radius. Our assumed value of $r_{\rm{exp}}$ = 15 AU results in an outer disk radius that is roughly consistent with the disk radius evolution found in our previous viscous models, using an initial disk radius of 50 AU\footnote{Since an outer disk radius is not directly calculated in the \citet{Chambers2019} framework, we compared surface densities between this disk model and that of the viscous \citet{Chambers2009} model at its outer radius to compare the models' radii evolution.}. Lastly, we consider all of our disk and planet formation models to take place around a Solar mass star.

We refer the reader to section 3 of \citet{Chambers2019} for a detailed listing of the analytic equations that are solved (equations 34-43). While this model provides the main framework for the disk models investigated throughout this paper, we make the following additions.

First we can reduce the number of disk model parameters by relating the strength of the outflow as parameterized by $K$ (i.e. equations \ref{Outflow_SD} and \ref{Outflow_Mdot}) to the disk accretion rate (equation \ref{Results4_DiskAccretion}) following a basic result of disk wind theory.  The mass loss rate of a disk wind torque depends upon its lever arm, which we first briefly discuss.   Consider a field line that has its foot point at a radius $r_o$ on the disk.  Follow the accelerating outflow along that field line until you reach a point on it where the outflow speed equals the Alfv\'en speed on that field line.  This is the Alfven critical point in the outflow and it marks the point where further acceleration basically ceases:  the field cannot enforce corotation with the disk because the outflow speed is greater than the speed at which an Alf\'en wave can propagate back to the disk.  The radial distance to that point on the field line is the Alfv\'en radius for that field line, $r_A (r_o) $.   The ratio $ r_A/r_o $ is the lever arm of the resulting torque that is exerted back on the disk. The set of all such points marks the Alfv\'en surface of the disk wind - one of the most fundamental aspects of any MHD wind.  In a self-similar disk wind model such as \citet{Blandford1982}, $ r_A/r_o = const $ whose Alfv\'en surface is a plane parallel to the disk midplane.  Analysis of the disk angular momentum equation quickly reveals a most useful scaling;  the ratio of the wind mass loss rate to the disk accretion rate is inversely proportional to the square of the lever arm:  $(r_A/r_o)^{-2} $  \citep{Pudritz1986,  Pelletier1992}.  For the efficient winds discussed in the Introduction, most theoretical and numerical studies gives values for this lever arm of $ r_A/r_o \simeq 3.  $ \citep{Ouyed1997, ZhuStone2018}.  Following the discussion of the observations in the Introduction, we adopt this limit as the typical situation for at least T-Tauri stage disk winds (less than 1 Myr old) and adopt following constraint on the wind outflow rate,
\begin{equation} \frac{\dot{M}_{\rm{wind}}}{\dot{M}_{\rm{acc}}} = (r_A/r_o)^{-2}  \simeq 0.1 \;. \label{Outflow_Constraint} \end{equation}
 By solving equation \ref{Outflow_Constraint} at time $t=0$ for a particular disk model's specification of $\alpha_{\rm{turb}}$ and $\alpha_{\rm{wind}}$, the constant $K$ can be solved for as opposed to being an input parameter. This method reduces our list of disk input parameters by one. We highlight that only low settings of $K \lesssim 0.1$ are needed to solve equation \ref{Outflow_Constraint}. The large values of $K=1$ that are used in example disk models in \citet{Chambers2019} and \citet{Ogihara2018} are in a different regime of what we call heavy disk winds than  provided by equation \ref{Outflow_Constraint}, since a  setting of $K=1$ results in $\dot{M}_{\rm{wind}} \gtrsim \dot{M}_{\rm{acc}}$.  We discuss this regime in the Appendix. 

We also determine the location of the dead zone throughout disk evolution, as resulting from Ohmic dissipation. Within the dead zone, the disk ionization fraction is insufficient for the MRI instability to operate. To determine its location, we have followed an approach first developed in \citet{MP2003} and applied to planet populations  in \citet{Alessi2018}.  There we showed that the best results for planetary populations arise  when disk ionization is driven by host-star X-rays.  This makes physical sense because cosmic rays will be largely scattered by shocks in the magnetized disk wind before reaching the disk \citep{Cleeves2013, Cleeves2015} .  The X-ray luminosity we use here is also the same in our previous work, namely $L_X =  10^{30} $ erg s$^{-1}$ with typical X-ray energies of $ E_X = 4$ keV. This approach results in the following criteria for an MRI-active disk, written in terms of the Ohmic Elsasser number \citep{Simon2013},
\begin{equation} \Lambda = \frac{v_A^2}{\eta_O \Omega} \lesssim 1 \;, \end{equation}
where $v_A$ is the Alv\'en speed, $\eta_O$ is the Ohmic diffusivity, and $\Omega$ is the local Keplerian orbital frequency. As we will show in later sections, this method results in the outer edge of the dead zone $\lesssim$20-30 AU and evolving inwards with time, similar to its evolution within the previously-considered \citet{Chambers2009} framework \citep{Alessi2018}.

An improvement we make in our treatment of the dead zone is that, in the \citet{Chambers2019} framework, we can set different turbulence strengths with the $\alpha_{\rm{turb}}$ parameter inside and outside the dead zone. For example, following \citet{HP10}, we reduce $\alpha_{\rm{turb}}$ by two orders of magnitude inside the Ohmic dead zone, while maintaining a constant disk accretion rate as set by the parameter $v_0$. What this means physically, is that within the Ohmic dead zone, the stress related to disk winds increases so as to maintain a radially-constant disk accretion rate (i.e. following the result of \citet{BaiStone2013}). We specify our choice of $\alpha_{\rm{turb}}$ settings for our various disk models in section \ref{Results4_DiskSettings}. This treatment is an improvement of our previous handling of the dead zone, for which the dead zone was ``passive'' in the sense that we determined its outer edge's location, but it had no physical effect on the disk structure so as to not break the assumed self-similarity of the \citet{Chambers2009} model. Here, the dead zone has a more self-consistent effect on the disk structure, as its outer edge separates two distinct regions with different strengths of $\alpha_{\rm{turb}}$ and $\alpha_{\rm{wind}}$.

We have also investigated ambipolar diffusion (hereafter AD; another non-ideal MHD effect) in its ability to affect the dead zone's structure along the disk midplane. Following \citet{BaiStone2011}, this investigation was done by solving for the AD parameter Am throughout the disk, 
\begin{equation} \rm{Am} = \frac{v_A^2}{\eta_A \Omega} \;, \end{equation}
where $\eta_A$ is the ambipolar diffusivity. The parameter Am is AD's counterpart to the Ohmic Elsasser number, in that it quantifies how effective AD and its diffusivity will be in suppressing MRI growth. This calculation resulted in Am values of 100-1000 throughout the disk's extent along the midplane for fiducial disk settings and $\alpha_{\rm{turb}} = 10^{-3}$. At this setting of $\alpha_{\rm{turb}}$, \citet{BaiStone2011} show that MRI would only be suppressed at values of the plasma $\beta \equiv P_{\rm{gas}}/P_{\rm{B}}$ (a ratio of gas to magnetic pressure) near 0.1-1. Therefore, AD will only suppress the MRI and create an ``AD dead zone'' in the most tenuous regions of the disk, such as in the disk's outer extent, or well above the disk midplane. This result is in accordance with the commonly-found conclusion that AD affects MRI turbulence only in the lowest density regions of the disk (e.g. \citealt{Armitage2011, Simon2013}). We therefore do not include AD in our disk models, and the dead zone location we determine is only due to the effect of Ohmic dissipation.

\subsection{Planet Formation \& Migration} \label{Results4_FormationModel}
\subsubsection{Detailed numerical treatment of planet migration under viscous torques}  
 
The torque experienced by the planet depends  on the planet's mass as well as important physical properties of its host disk such as its turbulence and scale height.   Our handling of migration is an extrapolation of recent detailed torque calculations carried out in  \citet{Speedie2022}.  This closely followed the method developed for computing Type I torques  by \citet{Paardekooper2011} and visualized in what we call torque maps, introduced by \cite{Coleman2014}. 

A key issue in any theoretical treatment of planet migration is how type I migration is handled.  Many models assume a smooth power-law disk model such as the well known MMSN model.  Such models cannot therefore deal with the problem of rapid Type I migration \citep{GoldreichTremaine1980} without resorting to some ad hoc assumptions.   The central point of physics here is that disks are not smoothly varying power laws in column density or other quantities.  They have inhomogeneities due to transitions in turbulence levels (at dead zone boundaries), opacities (eg ice lines where density changes occur), and in disk heating rates (transition from viscous to radiative heating).   It is well known that co-rotation torques can push planets outwards in disks, and that therefore there exist planet traps where the inward Lindblad torque balances this outward co-rotation torque.  These zero net torque solutions typically occur at the inhomogeneities previously mentioned.  \citet{Bitsch2015} demonstrated that for smooth MMSN models the lack of such traps made the formation of massive planets difficult in such models.  Subsequent work by \citet{Bitsch2019} showed that in MMSN disk models, very low viscosity ($\alpha \simeq 10^{-4}$ )  outward directed co-rotation torques saturate and are overcome by inner directed Lindblad torque.    

The fact that disks are observed to have a range of turbulence levels motivated \citet{Speedie2022} to compute how planets grow and migrate within  disk models whose detailed evolving astrochemistry and the resulting detailed Llindblad and co-rotation torques are carefully followed.   They compute two different cases:  the conventional $\alpha=10^{-3}$, and a lower $\alpha=10^{-4}$.  In all of these simulations, a heat transition trap is present.  This trap has a far more extended radial structure and is not localized to a specific radius.  This is because there is a gradual transition in the total thermal energy from the viscously heated inner region to the outer radiation dominated zone.  Nevertheless we can identify a fiducial radius for the heat transition where the heating rates are formally equal.  We refer the readers to Figures 4, 7, and 8 in that paper which clearly show the torque maps, traps, and planet evolution tracks in the M-a diagram. 
 The important and at first glance somewhat surprising result is that planet migration histories bifurcate depending on the value of $\alpha$:  outward directed co-rotation torques can carry planets outwards in low viscosity  disks($\alpha \simeq 10^{-4}$) where they are ultimately trapped in extended heat transition traps, whereas forming planets in higher viscosity disks tend to be pushed inwards.   In regions beyond the extended heat transition zone, \citet{Speedie2022} showed that the co-rotation torque does indeed saturate so that inward directed Linblad torque force planets pushes the heat transition gradually inwards as the disk reduces in mass and the inner viscously heating region shrinks.  Their work shows that all three traps are active but that in the lowest viscosity case the trapping masses is quite low, dropping to several Earth masses.   
 
 Statistical analysis of hundreds of protoplanetary disks
shows that the dust component of disks falls into two populations:
(i) radially compact  with no resolved structure or (ii)  radially
extended with ample resolved gap-ring structure \citep{vanderMarel2021} 
Given  that co-rotation torques in low viscosity disks push planets
outward to large disk radii before they return to the inner regions and that planets in high viscosity disks  on the other hand, only migrate
inwards,\citet{Speedie2022} noted that the level of disk turbulence is playing a major role.   Extended structured protoplanetary
disk systems with dust gaps and rings observed at 10s of au are likely
disks that have these low viscosities while the majority of disks
with more compact dust components have higher viscosity.   
 A plausible physical reason for this is that massive disks are better screened from their external radiation sources, which impacts the level of MRI turbulence within them.  Thus, the statistics of disk structural properties quite naturally leads to the prediction that there is an underlying distribution of turbulence levels in disks, directly linked to the distribution of disk masses  most probably deriving from their formation conditions.    These results and physical ideas  motivates our approach in the present paper to consider a population synthesis of planets whose host disks have a distribution of turbulent $\alpha_{turb} $ values, that we take to be a lognormal distribution. 

\subsubsection{ A computationally advantageous approach } 
 
 We follow the same approach of computing core accretion models and trapped type-I migration as has been used in our previous works \citep{Alessi2018, Alessi2020, Alessi2020b}.  For a complete description, the reader may consult Appendix B of \citet{ Alessi2020}.  The planet traps we include the water ice line, heat transition, and outer edge of the dead zone. In the cases of the ice line and heat transition, the method by which their locations are calculated has been altered slightly in this new disk framework.   In \citet{Speedie2022} we have demonstrated that these traps are found in both low and high viscosity disk models in detailed numerical treatments of planetary migration that includes the evolution of  disk astrochemistry.  One difference that arises in our simplified approach and these more detailed torque simulations is that it is possible for low mass planets in some models to leave their Type I planet trap before they have sufficient mass to open a gap in the disk (see also \citet{Dittkrist2014} ).   This is an important issue because such a planet could undergo significant Type I migration before reaching a mass in which a gap opens and slower Type II migration occurs.  \citet{Hasegawa2016} has analyzed the mass scale at which the co-rotation torque saturates - equivalent to the co-rotation mass scale derived in  \citet{Speedie2022} - and shown that there are conditions when growing planet achieve the saturation mass they arrive at the condition to start opening a gap and start to enter the Type II state.  We discuss this further in the Discussion section 5. 
  
We define the ice line simply as the location along the disk midplane where the temperature is 170 K. Our models previously computed the full disks' equilibrium chemical structures to determine the phase-transition point of water. We have found that in all cases the resulting ice line location (defined where the abundance of water vapour and ice are equal) has a midplane temperature of 170 K. While in the \citet{Chambers2009} model the heat transition separating viscous and radiative heating was directly calculated, in the new \citet{Chambers2019} model it is not. We define the heat transition at the point where the temperature due to viscous heating $T_{\rm{vis}} = 0.5\, T_{\rm{req}}$. Since the total midplane temperature is $T^4 = T_{\rm{vis}}^4 + T_{\rm{req}}^4$, this definition corresponds to viscous heating contributing a $\sim$ 6\% increase to the radiative equilibrium temperature. The method of computing the dead zone's outer edge remains the same as our previous works, following \citet{MP2003} for an X-ray ionized disk.

Details of our planet formation model remain unchanged and we continue to use the best fit values related to forming planets' envelope opacities found in \citep{Alessi2018}. We consider a fiducial value of the parameter $f_{\rm{max}} \equiv M_{\rm{max}}/M_{\rm{gap}}$ = 50 that determines the mass at which gas accretion onto massive planets is terminated. The fiducial setting of $f_{\rm{max}}$ applies to models where individual planet formation tracks are computed. Following our approach in previous works, a full $f_{\rm{max}}$ distribution, which we model as log-uniform between $f_{\rm{max}}$ values of 1 and 500 is stochastically sampled when full planet populations are being computed.  

We simplify the dust physics by considering only a constant dust-to-gas ratio of $f_{\rm{dtg},0}=0.01$ at Solar disk metallicity, and do not include dust evolution effects (i.e. radial drift) in this work, whose effect on planet populations was studied extensively in \citet{Alessi2020}. The setting of the constant dust-to-gas ratio scales with disk metallicity (a varied parameter in our population runs) as $f_{\rm{dtg}} = f_{\rm{dtg},0} 10^{[\rm{Fe}/\rm{H}]}$.The assumption of a constant disk dust-to-gas ratio removes the computational expense of solving the \citet{Birnstiel2012} dust evolution model. While radial drift will have an effect on planet formation models and certainly adds a layer of complexity to this problem, our goal is to focus first on understanding the basic effects of disk evolution via winds on planet formation. In this regard, we are following the approach we took for the turbulent disk model \citep{Chambers2009} where we first considered a constant dust-to-gas ratio model \citep{Alessi2018} before the more complex scenario of incorporating dust evolution \citep{Alessi2020}.


\subsection{Disk Parameter Settings} \label{Results4_DiskSettings}

We now define the various disk models and their parameter settings that we investigate in the following results sections. 

First, in section \ref{Results4_1}, we will compare disk models at two different relative strengths of $\alpha_{\rm{turb}}$ and $\alpha_{\rm{wind}}$. In the first case, which we refer to as the ``turbulent dominated model'' (or ``turbulent model'' for short), we set $\alpha_{\rm{turb}}$ = 10$^{-3}$ which is the same setting we have used in our previous works, considering a pure viscous disk model. The second, ``combined turbulence \& winds model'' (which we refer to as the combined model, for short) considers $\alpha_{\rm{turb}}$ = 10$^{-4}$, an order of magnitude lower. In both cases, the fiducial models are normalized using the setting of parameter $v_0$ such that they have the same initial accretion rate $\dot{M}_{0,\rm{fid}} \simeq 6 \times 10^{-8}$ M$_\odot$ yr$^{-1}$. This value of $\dot{M}_{0,\rm{fid}}$ compares reasonably with initial accretion rates calculated using the previous disk model \citep{Chambers2009}, and also with observationally-inferred accretion rates from \citet{Watson2016}. Our normalization of $\dot{M}$ between the two disk models results in them having the same \emph{effective} $\alpha$,
\begin{equation} \alpha_{\rm{eff}} \simeq \frac{\dot{M}_{0,\rm{fid}}}{3 \pi c_{s0} H_0 \Sigma_0} \;,\end{equation}
where $H_0$ is the disk scale height, and subscripts of 0 follow the convention of indicating temperature-dependent quantities whose values are calculated as caused solely from radiation. The disk surface density $\Sigma_0$ is set by the initial disk mass $f_M$ and exponential cutoff radius $r_{\rm{exp}}$. As we will further detail later in this section, we set these parameters such that the turbulence-dominated and combined models have the same $\Sigma_0$ values, which again results in the two disk models having the same setting of $\alpha_{\rm{eff}}$. The normalizations of $\dot{M}$ and $\Sigma$ between the two models results in the turbulent model having an $\alpha_{\rm{wind}} \simeq 2.5\times10^{-4}$, and the combined model $\alpha_{\rm{wind}} \simeq 2.7\times10^{-4}$. 

\begin{figure*}
\centering
\includegraphics[width = 0.45\textwidth]{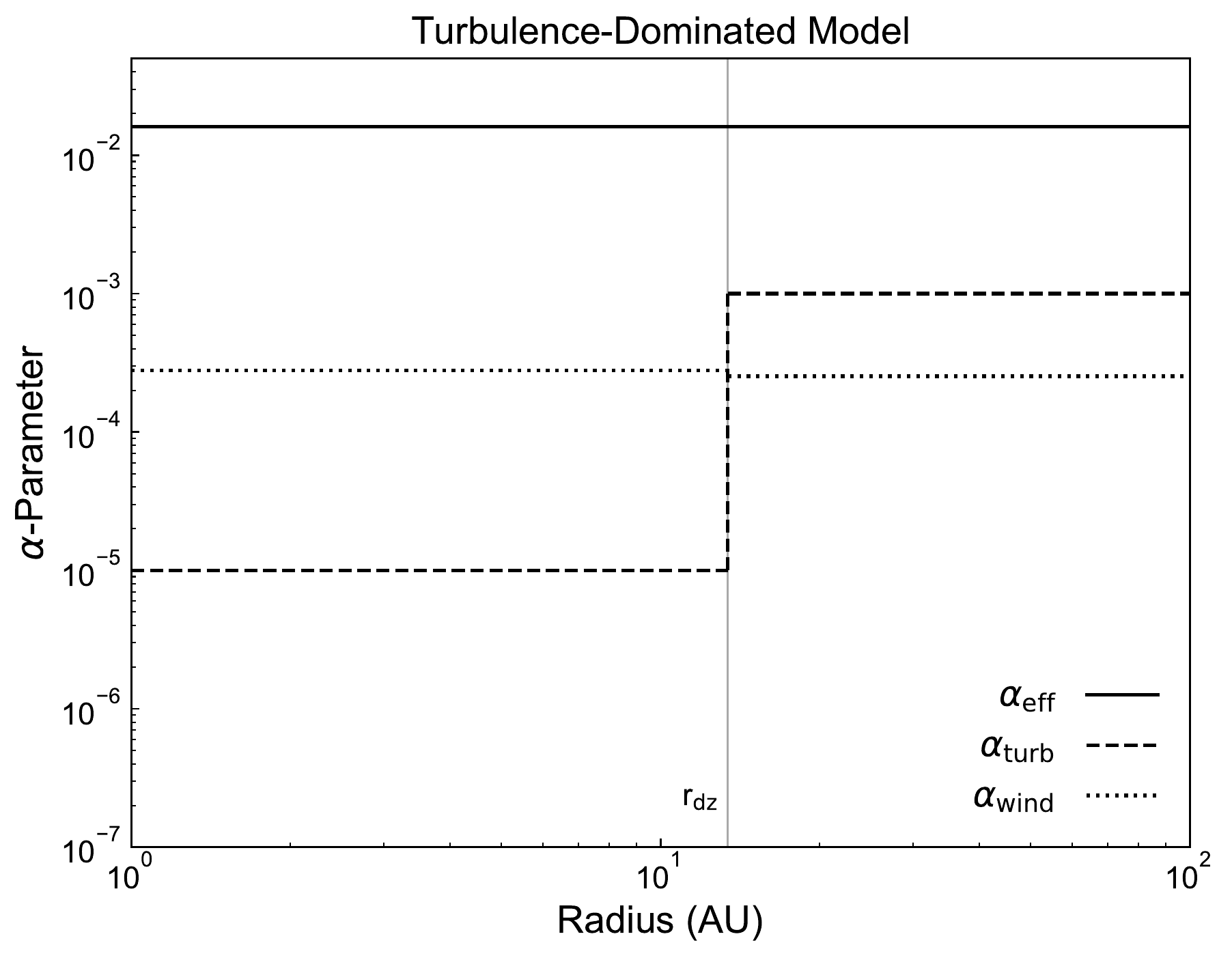} \includegraphics[width = 0.45\textwidth]{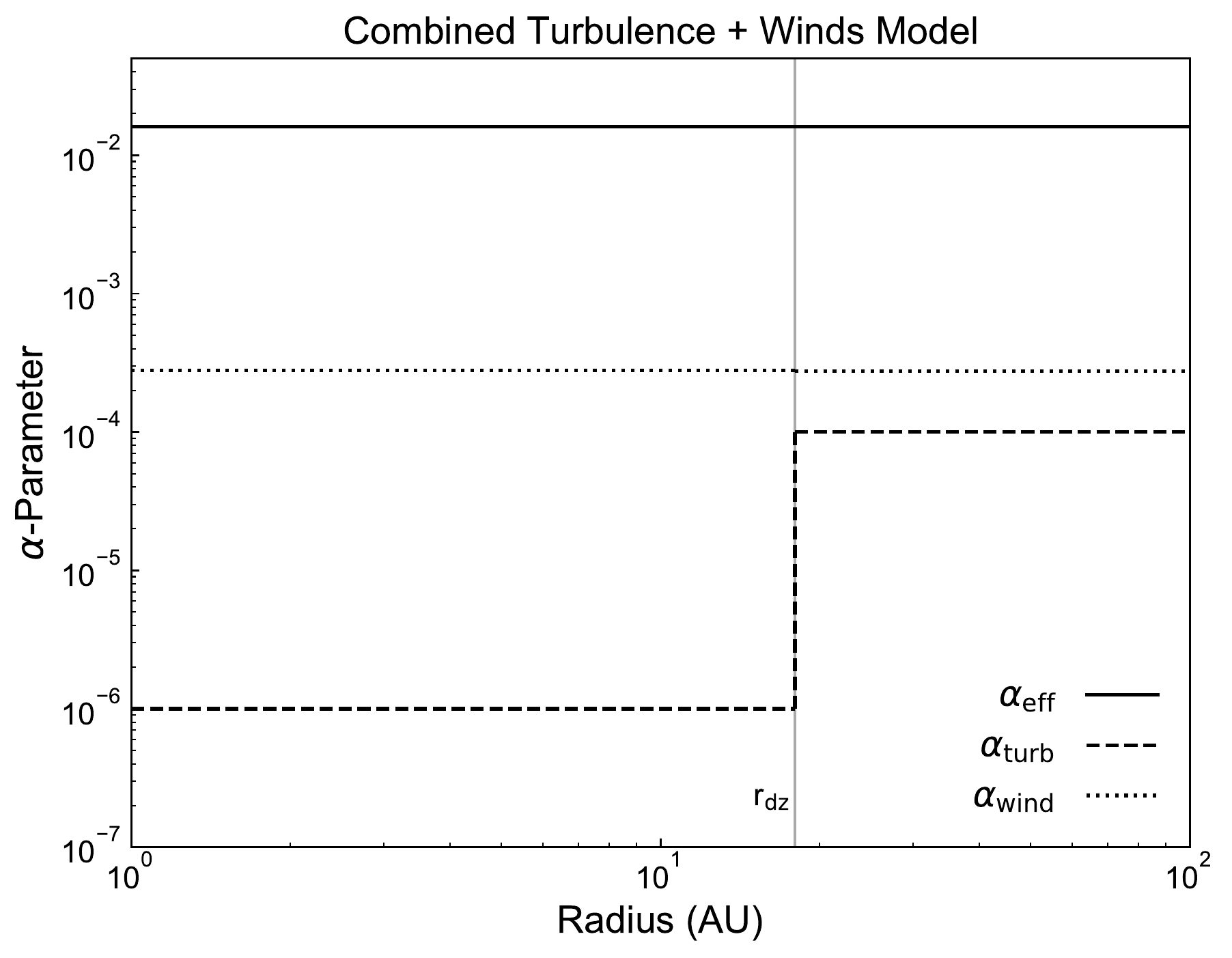}
\caption[$\alpha$ parameters of investigated models]{We show snapshots at $t$=0 of the turbulent dominated (left) and combined turbulence \& winds (right) models by plotting the disk $\alpha$ parameters across their disks' extents, where the decrease in $\alpha_{\rm{turb}}$ by two orders of magnitude at the outer edge of the dead zone ($r_{dz}$) is shown. This transition point (the dead zone trap) evolves inwards with time throughout the disk's evolution. We normalize both models by using the same initial disk accretion rate $\dot{M}_0 \simeq 6\times10^{-8}$ M$_\odot$ yr$^{-1}$, resulting in both models having the same $\alpha_{\rm{eff}}$.}
\label{Alpha_Plots}
\end{figure*}

In figure \ref{Alpha_Plots}, we summarize the turbulent-dominated and combined models by plotting snapshots of their $\alpha$ parameters across the disks' extents. In both models, within the MRI dead zone, $\alpha_{\rm{turb}}$ is decreased by two orders of magnitude. This reduction is shown clearly on the plots of figure \ref{Alpha_Plots} as a transition at the outer edge of the dead zone $r_{dz}$, whose location is shown for this example at time $t=0$ in the disk's evolution. This transition point, being the dead zone trap, evolves inwards with time as the disk evolves. We see that only a small increase in $\alpha_{\rm{wind}}$ is needed at $r_{dz}$ to maintain a radially-constant disk accretion rate despite $\alpha_{\rm{turb}}$ decreasing by two orders of magnitude at this location. As described above, these models have the same $\alpha_{\rm{eff}}$ parameters due to the normalization of $\dot{M}_{0,\rm{fid}}$ and $\Sigma_0$. 

We note that in the \citet{Chambers2019} model, quite a large disk winds fraction $f_w \simeq 0.8$ is required to produce $\alpha_{\rm{turb}} = 10^{-3}$. Using a ``pure'' turbulence setting of $f_w = 0$ leads to $\alpha_{\rm{turb}} \simeq 0.02$ at this value of $\dot{M}_{0,\rm{fid}}$, over an order of magnitude larger than the setting we considered in previous chapters. Therefore, even in the turbulent model, a substantial fraction ($\sim$ 80\%) of disk accretion is generated from disk winds and its related stress.

Other factors that affect the disks' initial accretion rates are the settings of the initial disk mass $f_M$ and radius, which scales with the exponential cutoff radius $r_{\rm{exp}}$. Following \citet{Chambers2019}, we set $r_{\rm{exp}}$ = 15 AU for all models. We then set the initial disk mass $f_M$ such that the initial surface density at reference radius 1 AU is,
\begin{equation} \Sigma_0 \equiv \Sigma(r_0 = 1 \,\rm{AU},t=0) = 1500 \,\rm{g}\,\rm{cm}^{-2} \equiv \Sigma_{0,\rm{fid}} \;. \end{equation}
This value of $\Sigma_{0,\rm{fid}}$ corresponds to initial disk masses $\simeq$ 0.05 M$_\odot$ in both the turbulent and combined models. A surface density of 1500 g cm$^{-2}$ is similar to $\Sigma_0$ values in fiducial models investigated in the previously considered \citet{Chambers2009} framework. 

In section \ref{Results4_1}, we also determine how the setting of $\Sigma_0$ affects results planet formation, as a means of foreshadowing the outcomes of a full population synthesis calculation. For both the turbulent and combined models, in addition to the fiducial setting, we consider a high setting $\Sigma_0 = 3\,\Sigma_{0,\rm{fid}}$ and a low setting $\Sigma_0 = (1/3) \, \Sigma_{0,\rm{fid}}$. While ultimately these changes are achieved by altering the initial disk mass, we label these models in terms of their $\Sigma_0$ value since the differences in surface densities can arise from a combination of changes to the initial disk mass and radius. Furthermore, the disk surface density is the physical parameter responsible for setting planet formation timescales. We note that we would arrive at similar planet formation results if we were to instead keep $f_M$ constant and alter the disk radius $r_{\rm{exp}}$, provided the same values of $\Sigma_0$ and $\dot{M}_0$ were considered. The wind outflow is constrained according to equation \ref{Outflow_Constraint} in all models presented in section \ref{Results4_1}. 


In the Appendix \ref{Results4_2}, we continue to investigate individual disk models and planet formation tracks while shifting our focus to a winds-dominated disk model to examine the effect of the outflow strength. In these models, we set $\alpha_{\rm{turb}} = 10^{-6}$, which is the turbulent strength within the dead zone of the combined model in section \ref{Results4_1}. We continue to use an initial disk mass of $f_M$ = 0.05 M$_\odot$, the fiducial setting in both the turbulent and combined models. Obtaining an initial disk accretion rate $\dot{M}_0 = 6\times10^{-8}$ M$_\odot$ yr$^{-1}$ requires $\alpha_{\rm{wind}} \simeq 2.7\times10^{-4}$. This value is similar to settings of $\alpha_{\rm{wind}}$ in the previous models. However, given the low setting of $\alpha_{\rm{turb}} = 10^{-6}$, the relative strength of disk winds is much higher.

In the first ``constrained outflow'' model, we follow our approach of constraining the wind outflow parameter $K$ according to equation \ref{Results4_1}. This constraint results in a small value of $K \simeq 0.05$. In the second model, we do not account for the constraint provided by equation \ref{Outflow_Constraint}, and instead adopt a high setting of $K=1$ as was used in \citet{Chambers2019}. We refer to this as the ``unconstrained outflow'' model. While this model leads to very high wind outflow rates that are comparable to, or higher than the disk accretion rate (in contention with disk winds theory and observations), we will see that a very efficient wind outflow has interesting effects on disk evolution, giving rise to a surface density maximum within the disk.

\begin{table}
\centering
\caption{A summary of model parameters, pertaining to individual disks where planet formation tracks are computed. The turbulent-dominated and combined turbulence \& winds models are investigated in results section \ref{Results4_1}, while the winds-dominated models are investigated in appendix \ref{Results4_2}.}
\begin{tabular}{ccc}
\hline 
\hline
Model & Outside $r_{dz}$ & Inside $r_{dz}$ \\
\hline
\hline
Turbulent-Dominated & $\alpha_{\rm{turb}} = 10^{-3}$ & $\alpha_{\rm{turb}} = 10^{-5}$ \\
&$\alpha_{\rm{wind}} = 2.5 \times 10^{-4}$& $\alpha_{\rm{wind}} = 2.8 \times 10^{-4}$\\
\hline
\hline
Fiducial $\Sigma_0$ & \multicolumn{2}{c}{$\Sigma_0 = \Sigma_{0,\rm{fid}}$ = 1500 g cm$^{-2}$} \\
High $\Sigma_0$ & \multicolumn{2}{c}{$\Sigma_0 = 3\,\Sigma_{0,\rm{fid}}$}  \\
Low $\Sigma_0$ & \multicolumn{2}{c}{$\Sigma_0 = (1/3) \, \Sigma_{0,\rm{fid}}$} \\
\hline
\hline
Combined Turbulence \& Winds & $\alpha_{\rm{turb}} = 10^{-4}$ & $\alpha_{\rm{turb}} = 10^{-6}$ \\
&$\alpha_{\rm{wind}} = 2.7 \times 10^{-4}$& $\alpha_{\rm{wind}} = 2.8 \times 10^{-4}$\\
\hline
\hline
Fiducial $\Sigma_0$ & \multicolumn{2}{c}{$\Sigma_0 = \Sigma_{0,\rm{fid}}$ = 1500 g cm$^{-2}$} \\
High $\Sigma_0$ & \multicolumn{2}{c}{$\Sigma_0 = 3\,\Sigma_{0,\rm{fid}}$}  \\
Low $\Sigma_0$ & \multicolumn{2}{c}{$\Sigma_0 = (1/3) \, \Sigma_{0,\rm{fid}}$} \\
\hline
\hline
Winds-Dominated & $\alpha_{\rm{turb}} = 10^{-6}$ & No dead zone \\
&$\alpha_{\rm{wind}} = 2.8 \times 10^{-4}$& \\

\hline
\hline
Fiducial $\Sigma_0$ & \multicolumn{2}{c}{$\Sigma_0 = \Sigma_{0,\rm{fid}}$ = 1500 g cm$^{-2}$} \\
\hline
Constrained outflow & \multicolumn{2}{c}{$K$ from equation \ref{Outflow_Constraint}} \\
Unconstrained outflow & \multicolumn{2}{c}{$K=1$}\\
\hline
\end{tabular}
\label{Results4_Table}
\end{table}

\noindent Table \ref{Results4_Table} summarizes each of the models we investigate in section \ref{Results4_1} and appendix \ref{Results4_2}, listing their parameters' settings.

\subsection{Population Synthesis Models} \label{Model_Population}

We first compute planet populations with constant values of the disk $\alpha$ parameters within each population. Our population synthesis models are constructed by computing 3000 planet formation tracks (1000 within each of the 3 traps) where we stochastically sample from distributions of 4 input parameters. Three of these input parameters pertain to the protoplanetary disk and are set by disk formation (disk lifetime, surface density, and metallicity) and 1 of which is a true model parameter ($f_{\rm{max}}$ which controls the termination of gas accretion onto massive gas giants; see section \ref{Results4_FormationModel}). We use the same approach and distributions of our previous works (i.e. \citealt{Alessi2018, Alessi2020}) with the exception of one minor modification.

Following our convention when considering individual disk models outlined in the previous section \ref{Results4_DiskSettings}, we vary the disk surface density via changes in the disk mass $f_M$ parameter. This could have also been accomplished by changing the disk radius (through varying $r_{\rm{exp}}$), or a combination of the two. However, the disk surface density is the true driver of physical changes in disk evolution timescales and resulting planet formation. To vary the disk surface density, we sample from a log-normal distribution. As previously stated, a lower mean initial disk mass of 0.05 M$_\odot$ is needed in the \citet{Chambers2019} disk model with our parameterization conventions to obtain an initial $\Sigma_0$ value that is comparable to the fiducial setting from our previous models using the \citet{Chambers2009} disk (which used a mean initial disk mass of 0.1 M$_\odot$). Aside from this change in the mean initial disk mass, all other distributions' details remain the same, and can be found in section 2 of \citet{Alessi2018}.

All other details of our population models in constant-$\alpha$ disks remain the same as previously outlined in section \ref{Results4_DiskSettings}. Specifically, our setting of the initial disk accretion rate remains as $\dot{M}_{0,\rm{fid}} \simeq 6 \times 10^{-8}$ M$_\odot$ yr$^{-1}$, with the setting of the $f_w$ parameter varied to obtain the desired $\alpha_{\rm{turb}}$ setting. Finally, in all models the outflow parameter $K$ is constrained to follow the outflow strength criterion presented in equation \ref{Outflow_Constraint}. 

We construct the distribution of $\alpha_{\rm{turb}}$ as a log-normal distribution using a mean of $10^{-3}$. Our previous works that have used this setting of $\alpha_{\rm{turb}}$ have found reasonable correspondence with the observed M-a and M-R distributions, motivating our choice of the distribution's average. We use $\alpha_{\rm{turb}}$ values of 0.007 and $10^{-4}$ as $\pm$1$\sigma$ of the normal distribution, respectively (that is, the normal distribution is defined to be asymmetric). Additionally, we restrict the range of this parameters' setting to be between these two values. The upper limit of 0.007 was chosen to correspond with the $\alpha_{\rm{turb}}$ upper limit found in the TW-Hya disk in \citet{Flaherty2018}.   A log normal form for the distribution of  $\alpha_{\rm{turb}}$ is also physically reasonable in  that more massive disks are expected to be more poorly ionized.   Therefore we expect  there to be a correspondence between the initial disk mass distribution (which is roughly lognormal) and that of the turbulence levels they contain  \citep{Speedie2022}. 

\begin{figure}
\centering
\includegraphics[width = 0.45\textwidth]{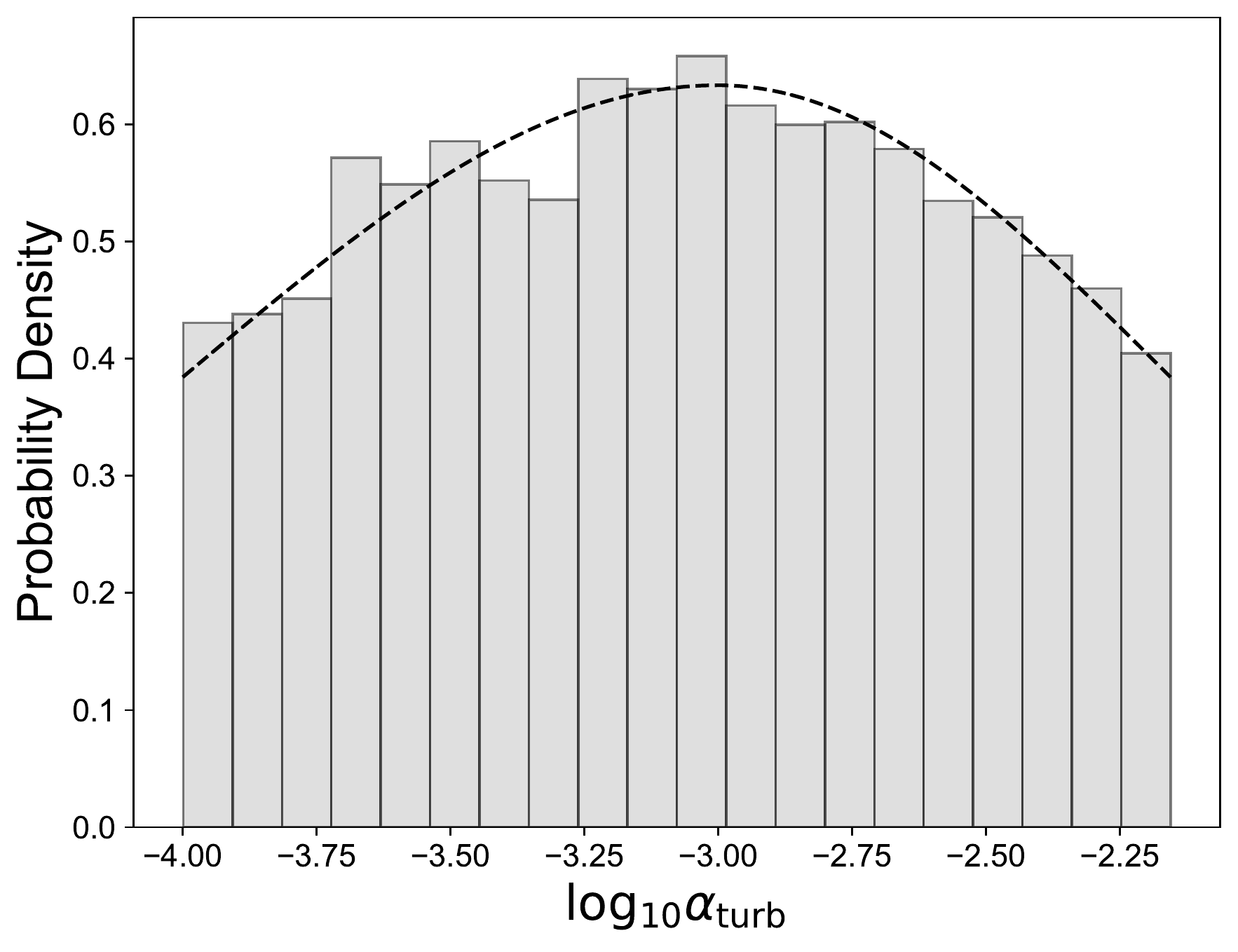}
\caption{A population's sampled $\alpha_{\rm{turb}}$ distribution is plotted as a histogram along with our constructed distribution's population density that is sampled in our population synthesis models (dashed curve).}
\label{Alpha_Dist}
\end{figure} 

We show our constructed $\alpha_{\rm{turb}}$ distribution in figure \ref{Alpha_Dist}, plotting a histogram corresponding to a population's sample of $\alpha_{\rm{turb}}$ values along with the modelled distribution's population density. While our choice to use a log-normal within the range of [10$^{-4}$, 0.007] as opposed to a log-uniform distribution may be somewhat arbitrary given the lack of observational or theoretical support, we highlight that our choice to make the upper and lower bounds of this range $\pm 1 \sigma$, respectively, results in $\alpha_{\rm{turb}}$ values near the average of 10$^{-3}$ being sampled only slightly more frequently than values near each end of the range. We therefore do not expect that changing to a log-uniform distribution across this range to greatly affect our results.  

Additionally, we note that in all populations, regardless of the setting of $\alpha_{\rm{turb}}$, we maintain a $v_0$ parameter value of $\simeq$ 20 cm s$^{-1}$, which results in an accretion rate $\dot{M} = \dot{M}_{0,\rm{fid}} \simeq 6 \times 10^{-8}$ M$_\odot$ yr$^{-1}$ at the fiducial value of surface density $\Sigma_0$. However, the accretion rate will vary at different values of disk surface density that are sampled in a population synthesis calculation. By defining both $\alpha_{\rm{turb}}$ and $v_0$ for each disk sampled in a population, $\alpha_{\rm{wind}}$ is a parameter that may be solved for but not pre-defined in a given population. This choice is somewhat arbitrary, and an alternate approach could be to set both $\alpha_{\rm{turb}}$ and $\alpha_{\rm{wind}}$. However, our chosen approach allows for a transparent means to compare with previous versions of our disk model. We note that both $\alpha$ parameters are likely to show variance among different systems, and the ratio of the turbulence to wind strength is an important quantity that affects disk evolution, and therefore planet formation results. Since we are directly investigating changes in $\alpha_{\rm{turb}}$, we will effectively be changing the relative strength of the two drivers of disk evolution, and therefore will be probing the effect of both parameters.

\subsection{Planet \& Atmospheric Modelling}

We use the same core and atmospheric structure models as \citet{Alessi2020b} to compute planet radii, and refer the reader to that work for a more detailed description. A planet's core radius will be dependent upon its solid composition (i.e. \citealt{Valencia2007}), and we performed detailed solid disk chemistry models to investigate this relationship in our previous paper \citep{Alessi2020b}. However, when using the standard approach of binning the solid refractories and minerals throughout the disk into bulk components of irons, silicates, and water ice, the biggest cause of change in these components' compositions is their location with respect to the ice line. On this basis, rather than computing a detailed chemistry model in this work, we simplify this process by calculating the time-dependent location of the water ice line (tracking the midplane temperature of 170 K), and fitting the radial profiles of the bulk abundances of irons, silicates, and ices with a reasonably high degree of accuracy to the profiles we obtained in \citet{Alessi2020b} where a full equilibrium chemistry model was performed using Solar abundances from \citet{Asplund2009}. We fit to equilibrium chemistry models corresponding to Solar C/O and Mg/Si ratios, which were investigated separately in our previous work. The disk compositions are used to track accreted materials onto planets throughout their formation, which is a necessary input to the core radius calculation.

We also explored the importance of post-disk atmospheric photoevaporation on our synthetic populations' M-R distributions in \citet{Alessi2020b}. We continue to use the same atmospheric mass-loss model, which combines approaches of \citet{murray-clay_chiang_murray_2009} and \citet{Jackson_2012}. We model atmospheric mass loss as caused by photoevaporation, driven by host-stellar X-rays and UV for up to 2 Gyr (or until convergence) following the dissipation of the protoplanetary disk at its disk lifetime $t_{\rm{lt}}$. 
We did this by combining the UV and X-ray
driven models of these two papers, respectively.  We use the power law fits to measured integrated
fluxes from \citet{ribas_guinan_gudel_audard_2005} for young solar-type stars in
the X-ray (1-20 \AA) and extreme ultra-violet (EUV) (100-360
\AA) wavelengths.
In the early evolution of the planet  X-ray driven photoevaporation dominates due to the high x-ray fluxes from
a young star \citep{ribas_guinan_gudel_audard_2005, Jin2014}. Mass loss is modelled by assuming that the energy
from incident photons is converted into work to remove gas
from the gravitational potential of the planet \citep{Jackson_2012}.  For the extreme EUV driven regime we adopt the two regimes highlighted in the model of \citet{murray-clay_chiang_murray_2009}. Finally,  between X-ray and UV-driven mass flows there is an important transition in the flows.  Above a certain UV
flux, X-rays are no longer able to penetrate the UV ionization front, resulting in a UV-dominated flow \citep{Owen_2012}. We start the  mass loss evolution of our planets immediately after the protoplanetary disk evaporates, a parameter
that is stochastically-varied throughout our population of
planets according to the observed range of disk lifetimes.
We evolve each planet forward until it is 1 Gyr old beyond which mass loss is negligible.  We again refer the reader to Appendix C in \citet{Alessi2020b} for a complete description of our approach.

 An atmospheric mass loss model is an important inclusion when considering out populations' M-R distributions, as the amount of gas accreted onto a planet greatly affects its radius since it is the lightest constituent material out of which planets form. Therefore, it is important to not only track the amount of accreted gas during the disk phase (as we do with our planet formation model), but also the fraction that is retained when exposed to high-energy radiation from the host star.

\section{Results I: Effect of relative strength of turbulence and disk winds} \label{Results4_1}

We begin by investigating individual disk models to compare the effects of different relative turbulent and disk winds strengths on disk evolution and planet formation. Specifically, we compare models at $\alpha_{\rm{turb}} = 10^{-3}$ and $\alpha_{\rm{turb}} = 10^{-4}$, using otherwise fiducial disk and model parameters that we have outlined in detail in section \ref{Results4_DiskSettings}. 

\begin{figure*}
\centering
\includegraphics[width = 0.45 \textwidth]{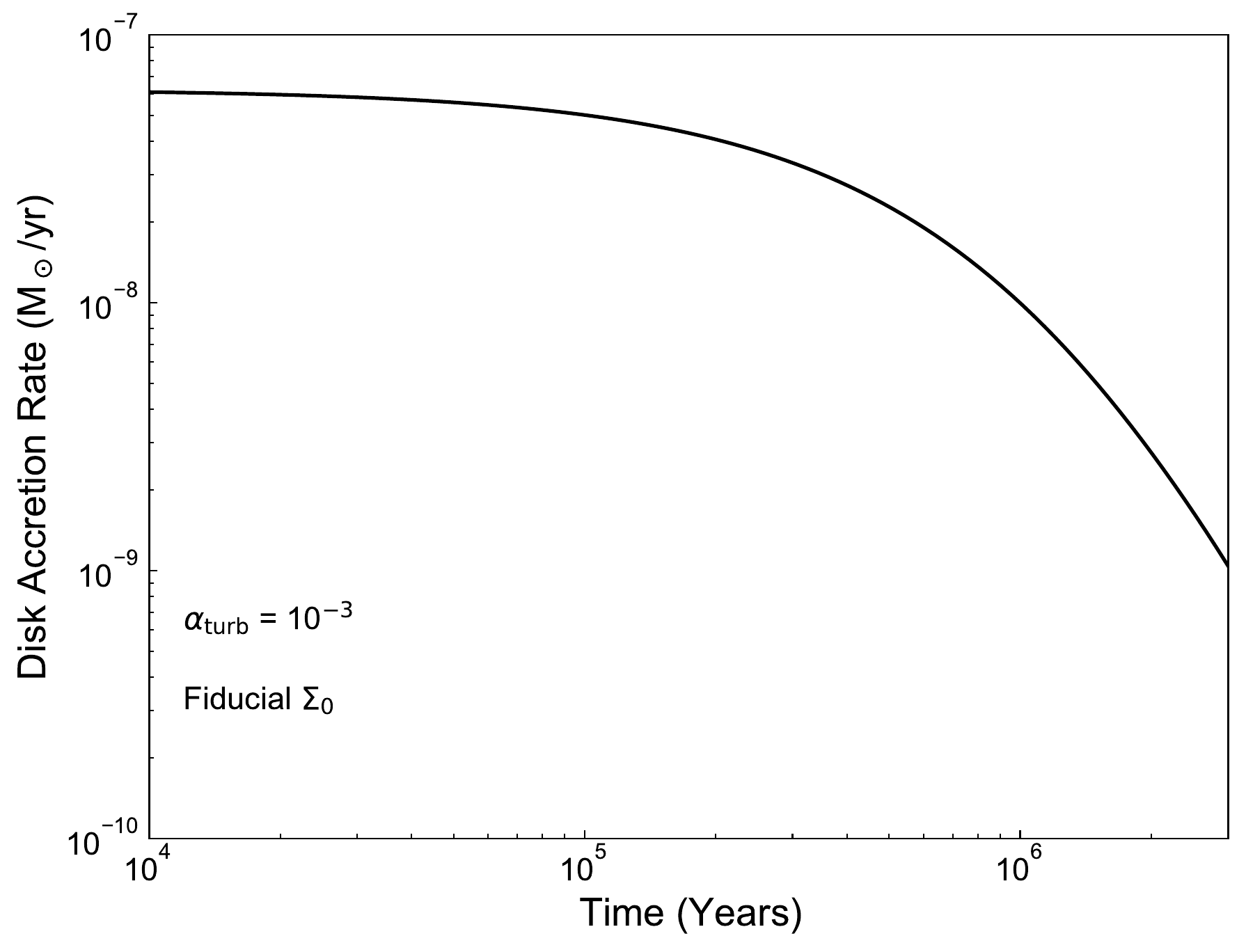} \includegraphics[width = 0.45 \textwidth]{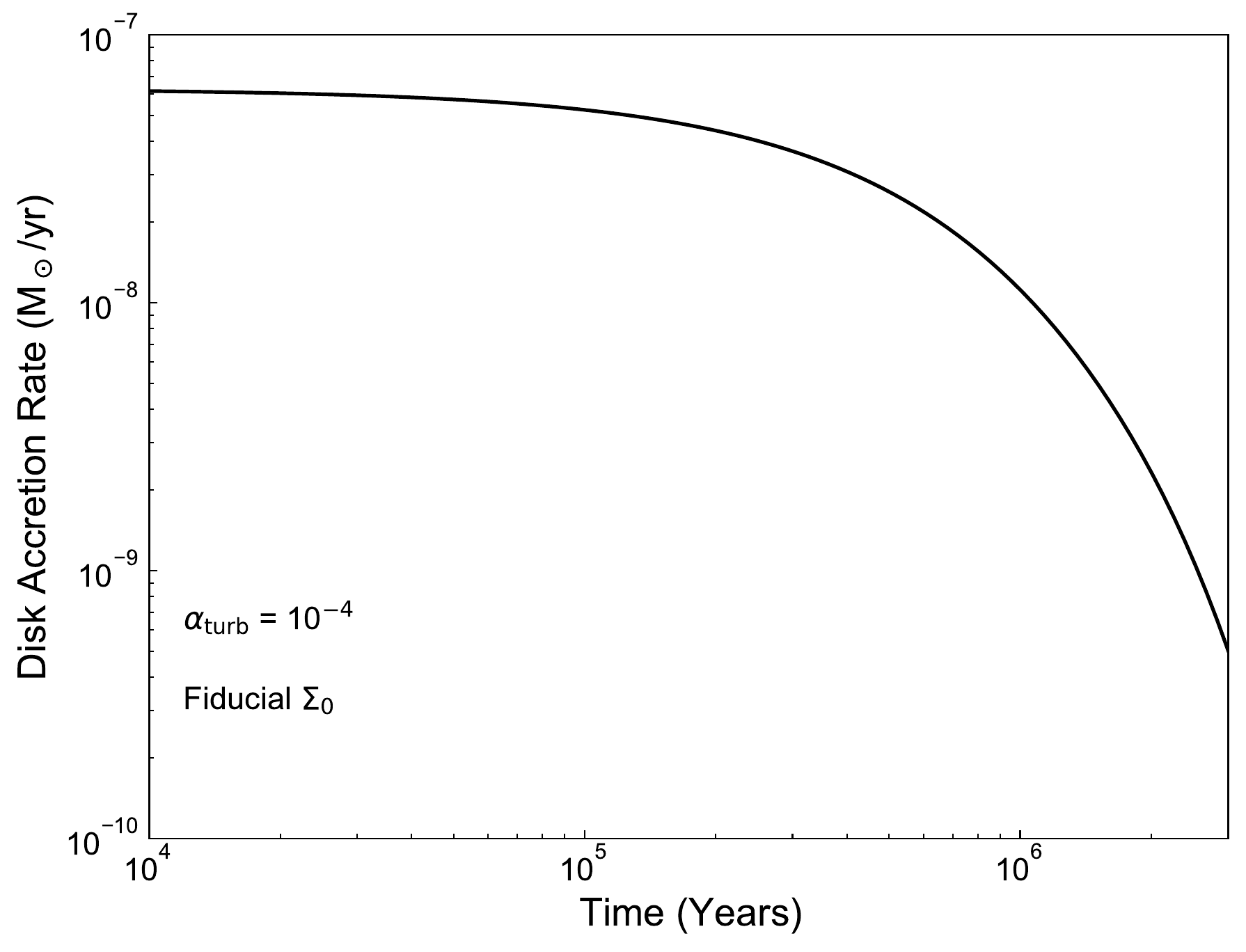} \\
\includegraphics[width = 0.45 \textwidth]{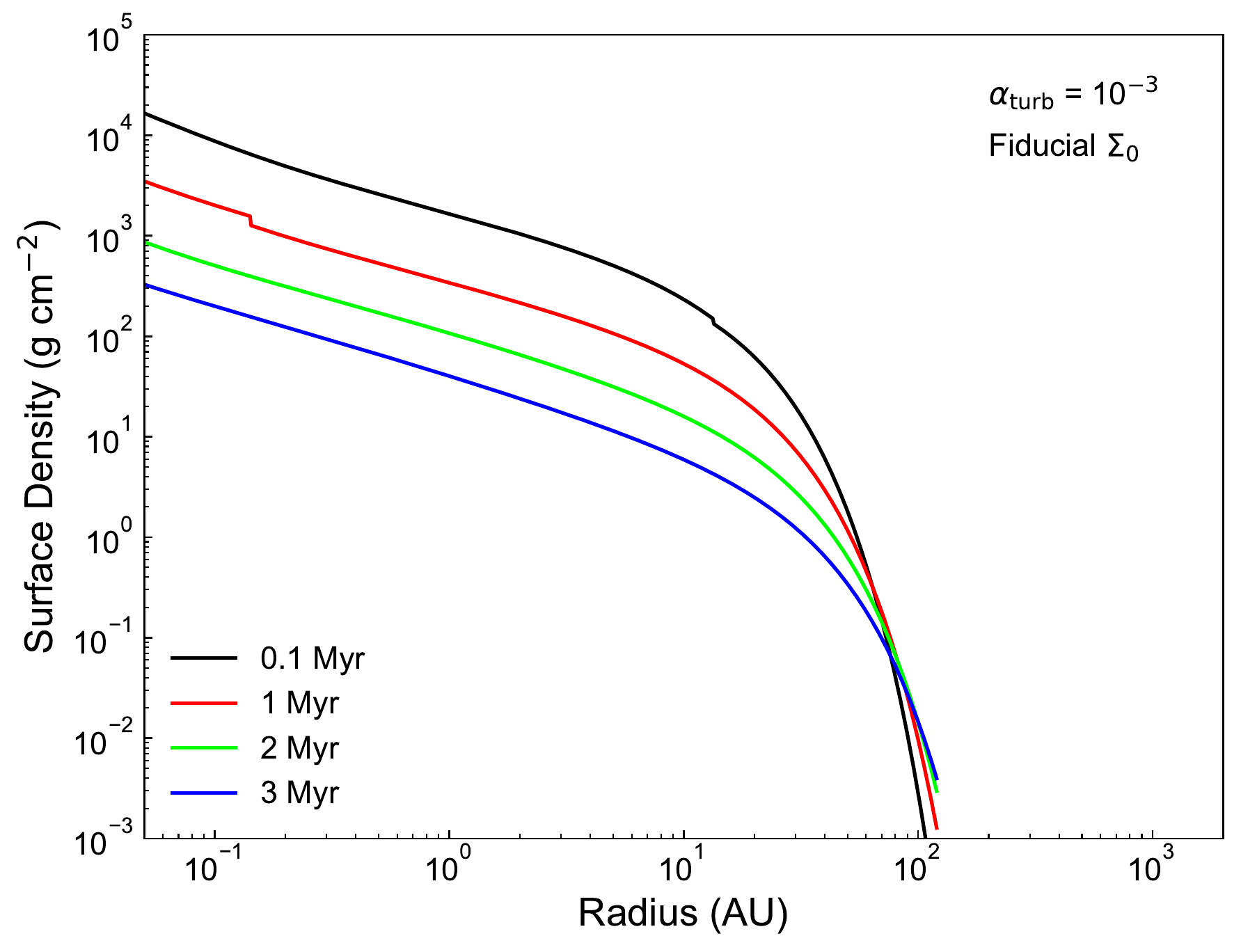} \includegraphics[width = 0.45 \textwidth]{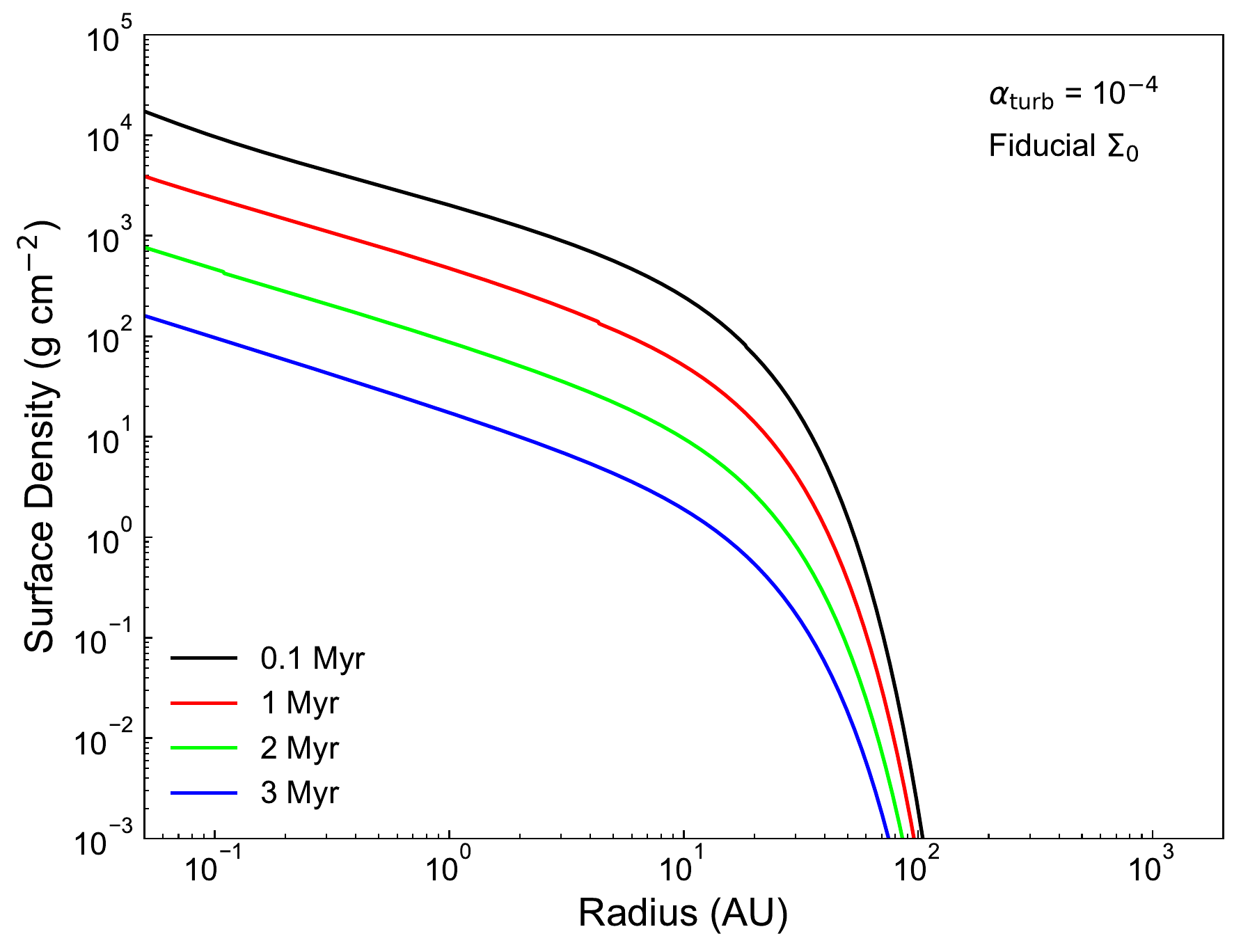} \\
\includegraphics[width = 0.45 \textwidth]{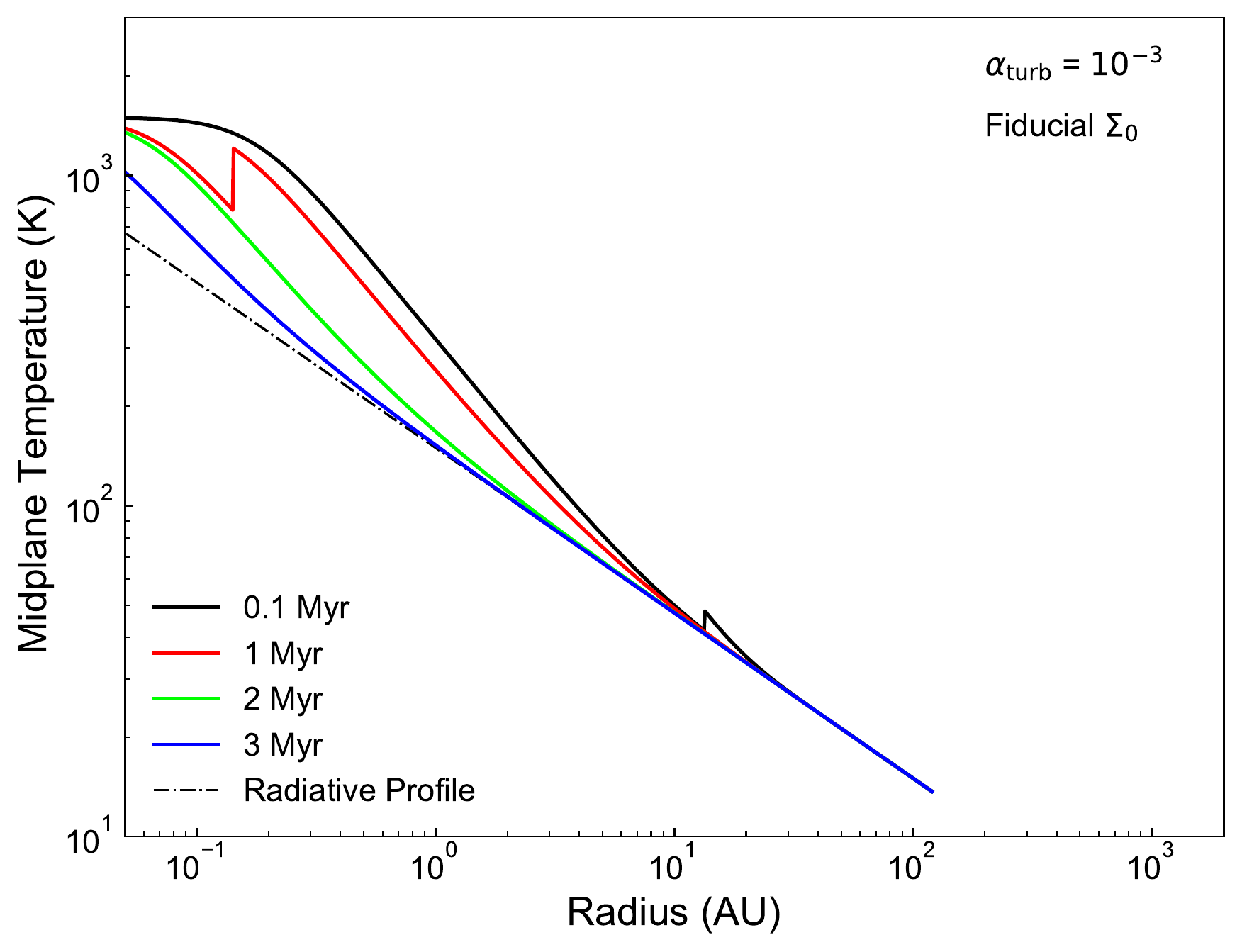} \includegraphics[width = 0.45 \textwidth]{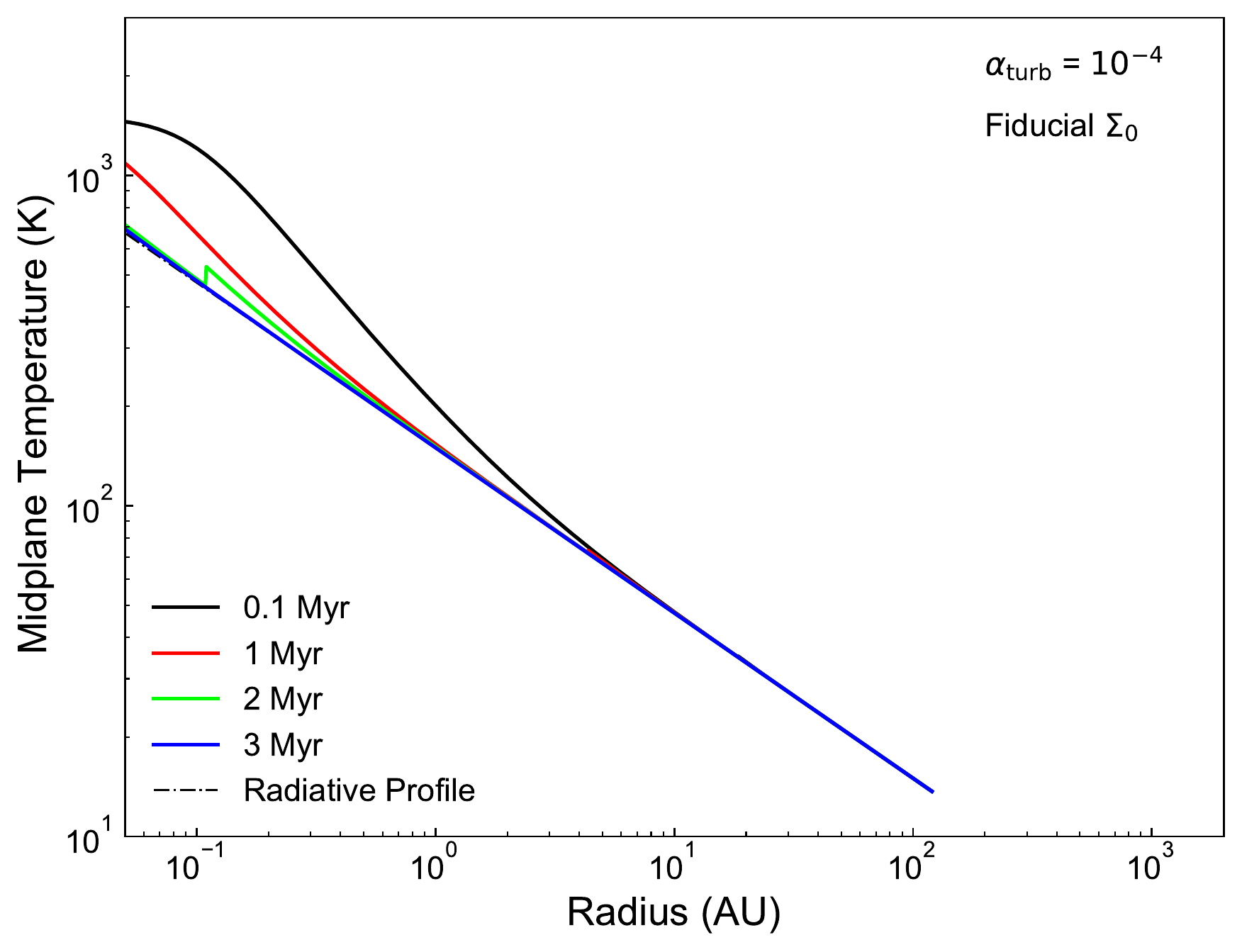}
\caption[Comparing disk evolution models at two settings of relative strength of turbulence and disk winds]{We compare protoplanetary disk evolution in the `turbulent' ($\alpha_{\rm{turb}}$=10$^{-3}$; left column) and `combined' turbulence and winds ($\alpha_{\rm{turb}}$=10$^{-4}$; right column) cases. Both disks are normalized to the fiducial setting of $\Sigma_0$. Time evolution of the disk accretion rate is plotted in the top row, and profiles of surface density and midplane temperature are plotted throughout disk evolution in the middle and bottom rows, respectively.}
\label{Results1_Disk}
\end{figure*}

In figure \ref{Results1_Disk}, we compare the evolution of the turbulent-dominated and combined disk models to discern the effect of  the relative strength of turbulence and disk winds. We plot the disks' accretion rates, as well as radial profiles of surface densities and midplane temperatures throughout 3 Myr of evolution at the fiducial $\Sigma_0$ setting. For both models the disk accretion rate decreases by roughly 1.5 orders of magnitude below the initial accretion rate of 6$\times10^{-8}$ M$_\odot$ yr$^{-1}$ after 3 Myr, decreasing slightly more in the combined model.  This range of $\dot{M}$ is quite comparable to those resulting from the previous \citet{Chambers2009} models with fiducial parameters. 

The inhomogeneities present within the radial surface density and temperature profiles correspond to the outer edge of the dead zone, where there is a local change in the strength of $\alpha_{\rm{turb}}$. Comparing the two models' surface density evolutions in the outer disk (at radii near 70-100 AU), we see a key difference between disk evolution via turbulence and winds. In the viscous evolution case, a small amount of spreading in the outer disk occurs - a necessary consequence of angular momentum conservation within the viscous evolution mechanism.  Disk spreading is absent in the combined model where winds-driven evolution is more prominent, and in this circumstance the disk, in fact, slightly contracts. We recall that the former `turbulent' model required quite a high disk winds fraction $f_w \simeq 0.8$ to set $\alpha_{\rm{turb}} = 10^{-3}$. Since a small fraction ($\simeq$ 20\%) of accretion is a consequence of turbulent viscosity, there is only a small amount of spreading in the outer disk. We note that a more extreme turbulent model of $\alpha_{\rm{turb}} = 10^{-2}$, as considered in \citet{Chambers2019}, shows more significant amount of viscous spreading.

While these differences between the two models exist in the outer disk regions owing to differences in relative disk wind strength, the inner 10 AU of the two disks are relatively comparable in terms of their surface densities. This is the region where most planet formation occurs due to low solid accretion rates at larger radii (with core accretion rate scaling with $\Sigma$).   The $\Sigma$ profiles of the two models are quite comparable in the inner 10 AU, particularly for the first 2 Myr. The more rapid evolution of the combined model becomes apparent on the surface density profile at 3 Myr, where $\Sigma$ is noticeably smaller in the combined model than in the turbulent scenario. On the basis of comparing the two models' (somewhat similar) surface densities, and noting that planetary growth via solid accretion is a relatively fast formation regime ($<$ 1 Myr), we do not expect significant differences in core accretion rates between the two models.

When plotting the disk midplane temperature profiles in figure \ref{Results1_Disk}, we include the radiative equilibrium profile (equation \ref{Results4_REQ}) to indicate the regions where viscous heating is effective in both models. The turbulent and combined models are  quite different in terms of their temperature structures. Viscous heating is more pronounced in the turbulent model due to the higher value of $\alpha_{\rm{turb}}$, resulting in the viscously heated region (where $T > T_{\rm{req}}$) extending to larger radii (10-20 AU at early disk evolution times). We also see that at the location of the dead zone, where $\alpha_{\rm{turb}}$ decreases by a factor of 100, the midplane temperatures decrease towards the radiative equilibrium profile. This decrease is due to the lower turbulence strength within the disks' dead zones, which results in less effective viscous heating. We therefore find that there are substantial differences between the turbulent and combined models in terms of their midplane temperatures. Since two of the traps in our model (the ice line and heat transition) depend on the disk midplane temperature, we expect the locations of these traps to be quite different between the two models. The difference in the traps' locations will also cause a significant difference between the models' chemical abundance profiles, which will depend most sensitively on the disks' temperature structures.

\subsection{Planet traps and trap crossings} 

\begin{figure*}
\centering
\includegraphics[width = 0.45 \textwidth]{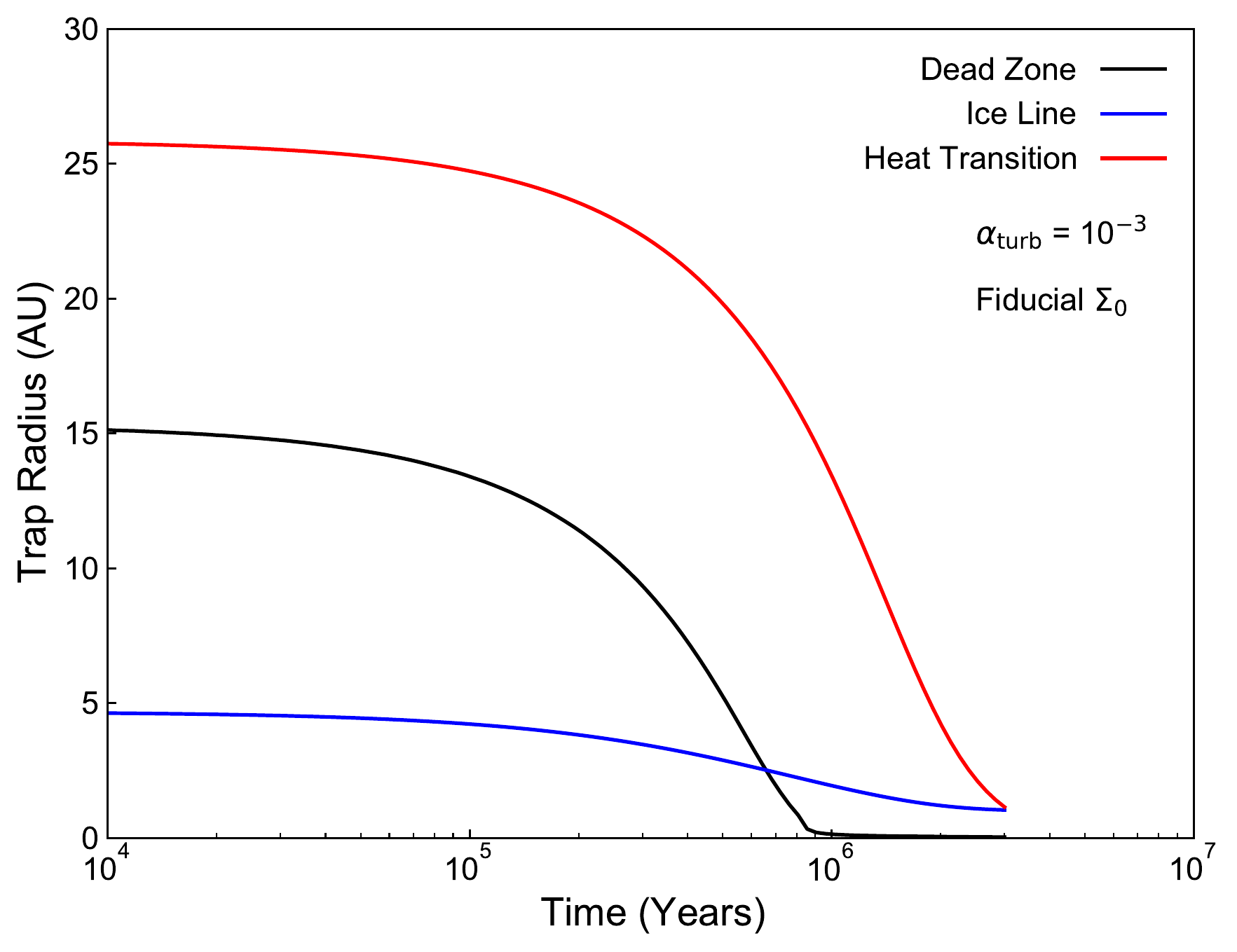} \includegraphics[width = 0.45 \textwidth]{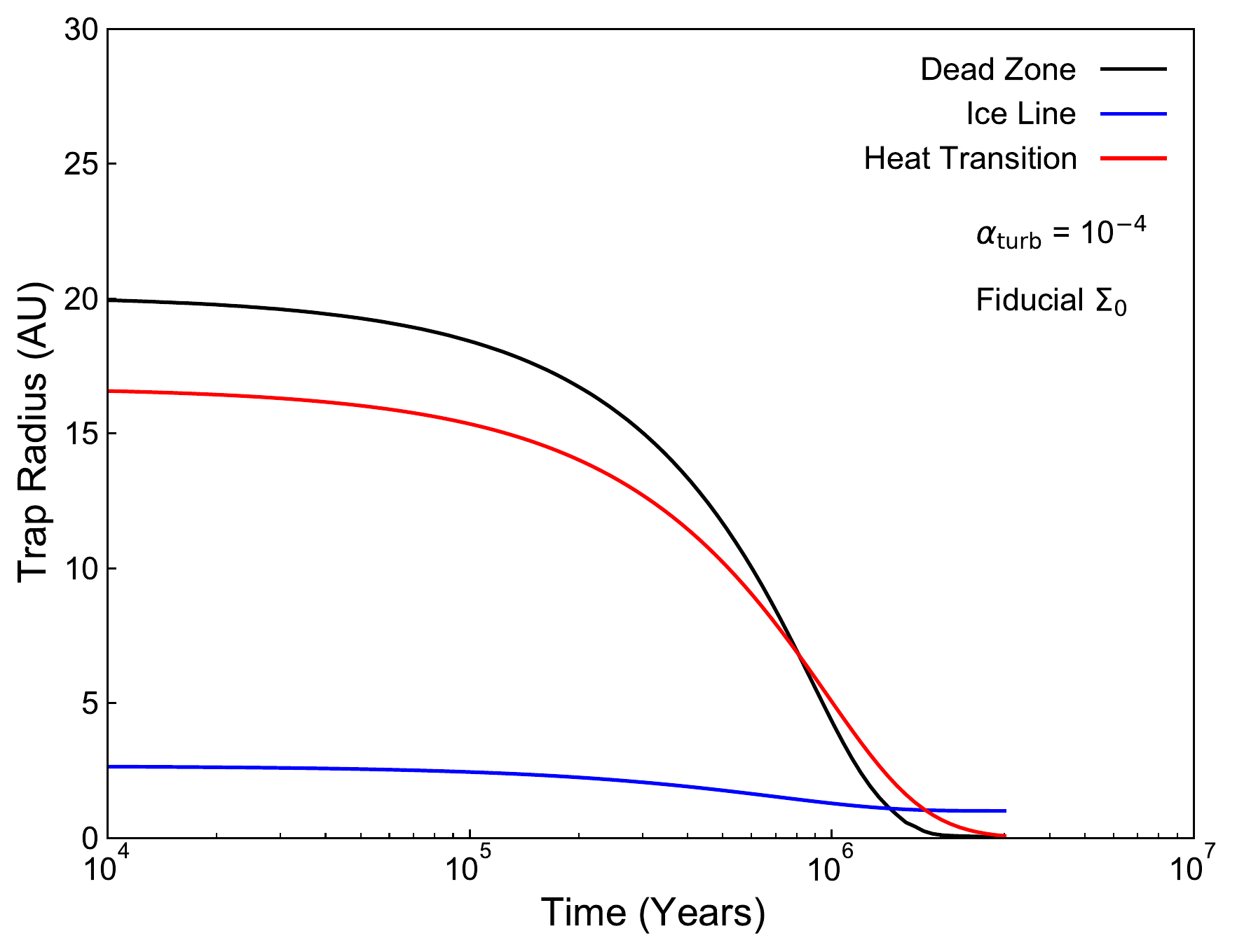}
\caption[Comparing planet traps' positions and evolution at two settings of relative strength of turbulence and disk winds]{Planet traps' evolutions are plotted for the turbulent (left) and combined (right) disk models, both considering a fiducial setting of $\Sigma_0$.} 
\label{Results1_Traps}
\end{figure*}

In figure \ref{Results1_Traps}, we explore differences in the planet traps' locations and evolutions between the two models at the fiducial setting of $\Sigma_0$. We indeed find that the temperature-dependent traps, the heat transition and ice line, exist at larger radii in the turbulent model due to the increased viscous heating. Conversely, the outer edge of the dead zone lies farther out in the combined model than in the turbulent case, despite their surface density profiles being similar. While the midplane X-ray ionization is sensitively dependent on the $\Sigma$ profile, there are several temperature-dependent factors in the dead zone model (recombination rates, Ohmic diffusivity $\eta_O$, etc). These factors, combined with the differences in the disks' temperature structures and evolution, ultimately affect the location of the outer edge of the dead zone where the Elsasser number $\Lambda = 1$. 

We see from figure \ref{Results1_Traps} that within a typical disk lifetime of 3 Myr, all the traps in our model converge to within $\lesssim$ 3 AU. We therefore predict that substantial solid accretion rates can exist on planetary cores undergoing trapped type-I migration at these traps' locations, as high surface densities persist in both disk models' inner regions even after 2-3 Myr of evolution. In the combined model, all traps exist within 10 AU after only 1 Myr of disk evolution. Conversely, in the turbulent model, this is only the case for the ice line and dead zone, and the heat transition evolves within 10 AU after $\sim$ 2 Myr of disk evolution. Planet formation will be most efficient within 10-20 AU based on the disks' surface density profiles (which set the solid accretion timescales) that decrease sharply outside of this radius range. At the fiducial $\Sigma_0$, the traps' evolution to within this region by 1 Myr in the combined model indicates that planet formation will be effective in each of the three traps. In the turbulent model, we expect this to be also true for formation in the ice line and dead zone traps. Formation within the heat transition in the turbulent model, however, will likely result in much longer formation timescales, due to its large radius at times up to 2 Myr.

The evolution of planet traps in the turbulent-dominated model using the new disk framework of \citet{Chambers2019} can be compared with trap evolution corresponding to the pure turbulence disk model \citep{Chambers2009} we previously investigated in \citet{Alessi2018}\footnote{We refer the reader to figure 5 of \citet{Alessi2018}, which shows the traps' evolution for a fiducial set of model parameters the purely viscous disk model of \citet{Chambers2009}.}.  In the case of  turbulent-dominated model treated in this paper using  the \citet{Chambers2019} formalism,  the initial ordering of traps is that the heat transition is the outermost trap at 25 AU, followed by the dead zone at 15 AU, and then  the ice line as the innermost and closest to star at around 5 AU.  On the other hand, the pure turbulence case has a different initial ordering of the outer two planet traps, with the X-ray dead zone being the farthest out in the disk, followed by the heat transition.  In this regard, we see that, despite using the same $\alpha_{\rm{turb}}=10^{-3}$ value, the different disk models do change the initial locations and relative ordering of the outer two planet traps.   The explanation is that early on in these different models, the pure turbulence disk column density at 0.1 Myr at 10 AU as an example (about 300 g cm$^{-1}$ )  is larger than that of the turbulence dominated case (about 200 g cm$^{-1}$) .  The greater column depth means that the disk will have lower disk ionization by X-rays out to larger distances leading to a more extensive region of  MRI  damping and dead zone.   We attribute this difference to the fact that in our turbulent-dominated model  the \citet{Chambers2019}  80\% of disk accretion is driven by disk winds and this has  the effect of reducing the column density below what the pure turbulence driven case can do.  The two disk models are similar, however, in that the traps rather quickly evolve into the inner disk ($\lesssim$ 10 AU). In both disk models, the dead zone does so faster ($\lesssim$ 1 Myr) than the heat transition, which shifts to within 10 AU after about 2 Myr.   

The relative evolution of the different types of traps shown in figure \ref{Results1_Traps} plays a particularly important role in the character of the populations that they produce.   There are three important points to notice.  First consider the evolution of the ice line in both models.  In the high viscosity model, the ice line trap starts at about 5 AU.  With time we see that it drifts slowly down to 1 to 2 AU where it stalls.  The reason here is that the viscosity heats the disk to temperatures higher than would result from just pure radiative heating by the host star.  As the column density drops so does the viscous heating rate and the ice line moves inward to smaller disk radii.  In the low viscosity model, the iceline starts at around 3 AU due to lower viscous heating and moves relatively less.  This marks the second characteristic of trap evolution in that the ice line become static in disk radius once viscous heating and heating by stellar radiation are comparable - which begins precisely when the heat transition reaches the ice line position.  That asymptotic limit is at around 1-2 AU for a solar mass star and this marks an important feature in the M-a diagram as we shall see.

The third, and equally important aspect  is the crossing of the dead zone and ice line trap in low and high viscosity models.  From the two panels in  figure \ref{Results1_Traps}  the high viscosity case, this takes place at about 0.7 Myr in our fiducial model, whereas for the low viscosity case it occurs later at about 2 Myr.   This has possible implications for a planet that is moving inwards while trapped on the dead zone because it could be intercepted by the ice line and remain at 1-2 AU.   This suggests that the ice line can potentially act as a filter that prevents a larger population of Hot Jupiters from forming.   We investigate this point further  in the Discussion section after the the population synthesis results are laid out.

\begin{figure*}
\centering
\includegraphics[width = 0.45 \textwidth]{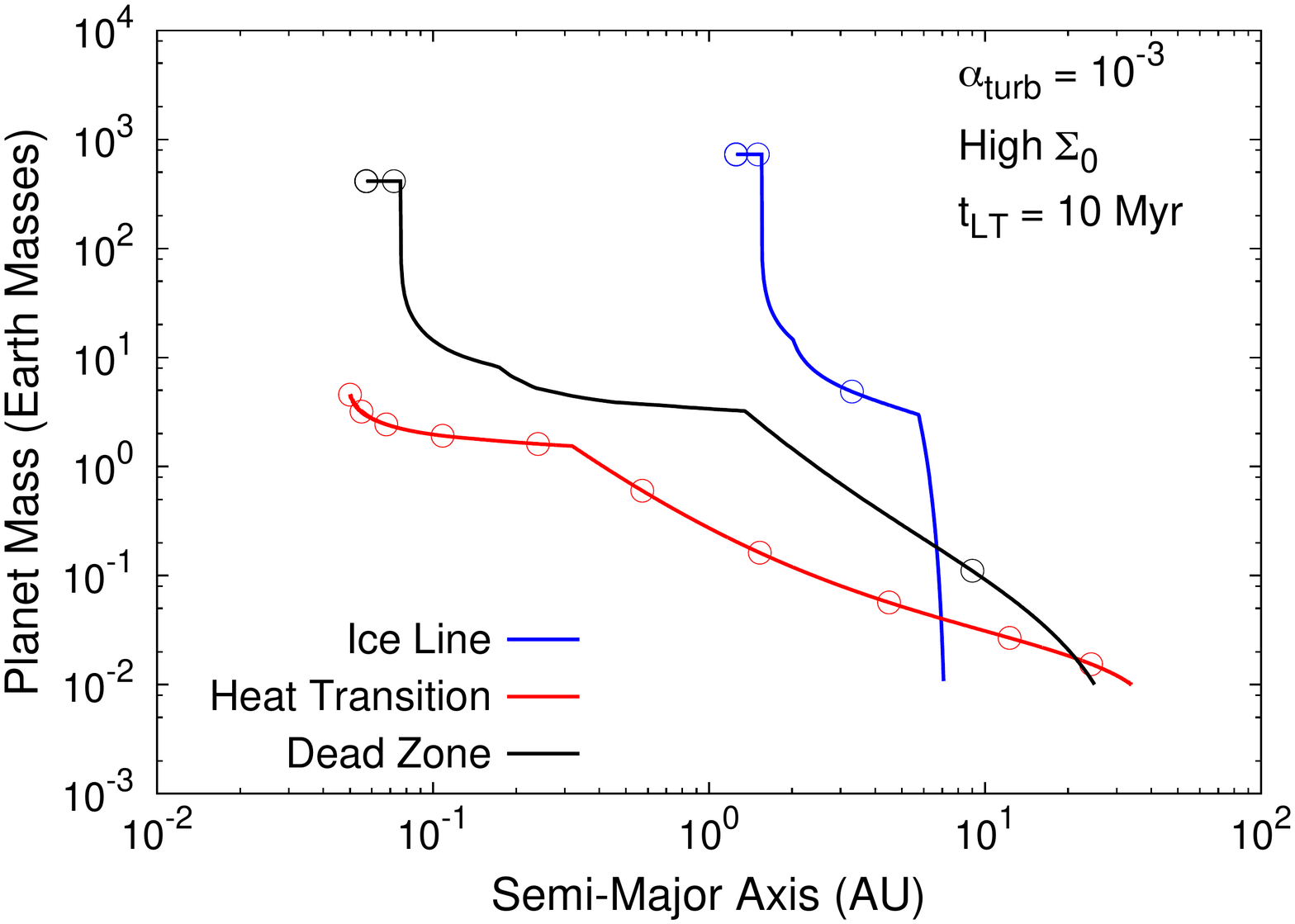} \includegraphics[width = 0.45 \textwidth]{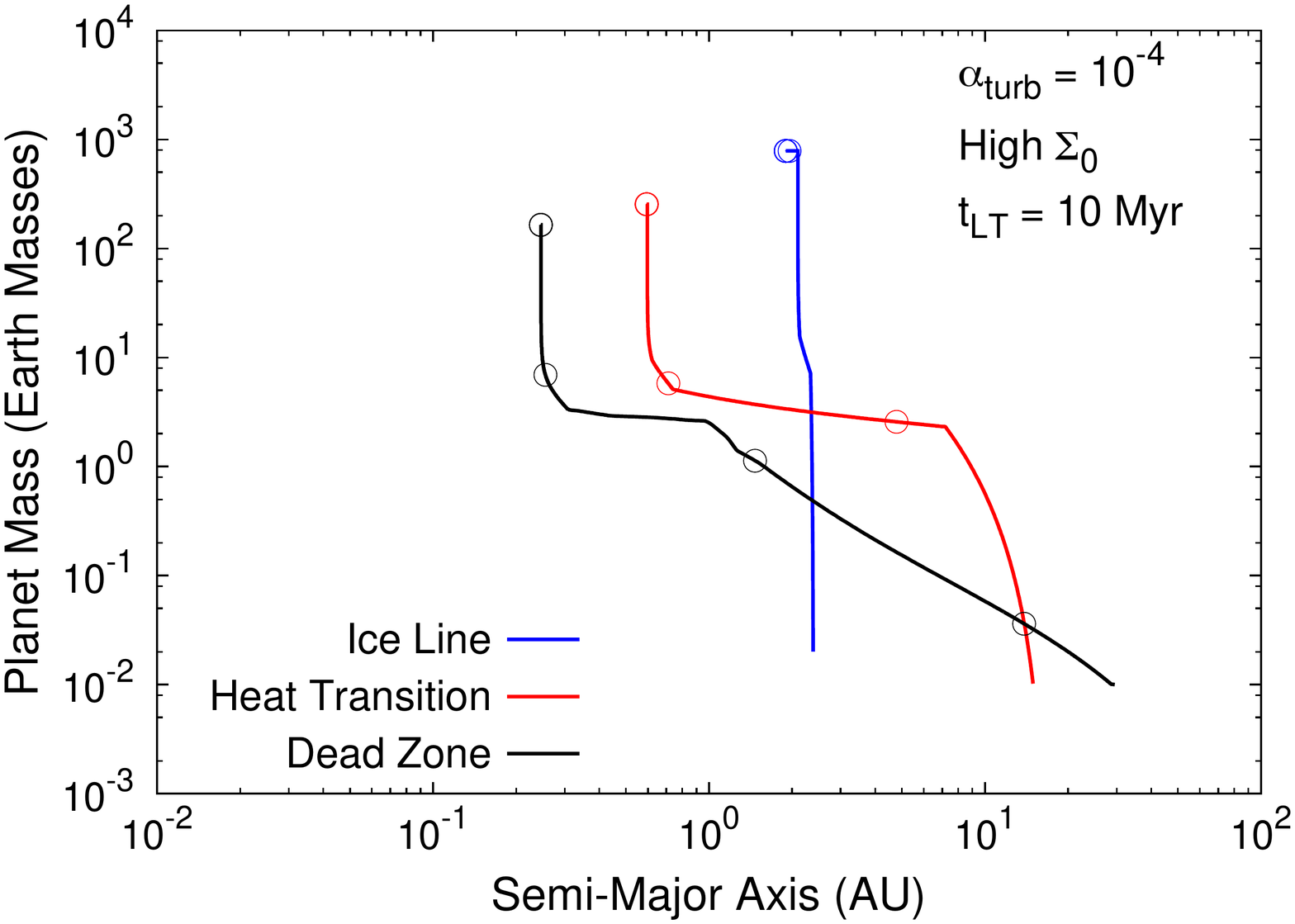} \\
\includegraphics[width = 0.45 \textwidth]{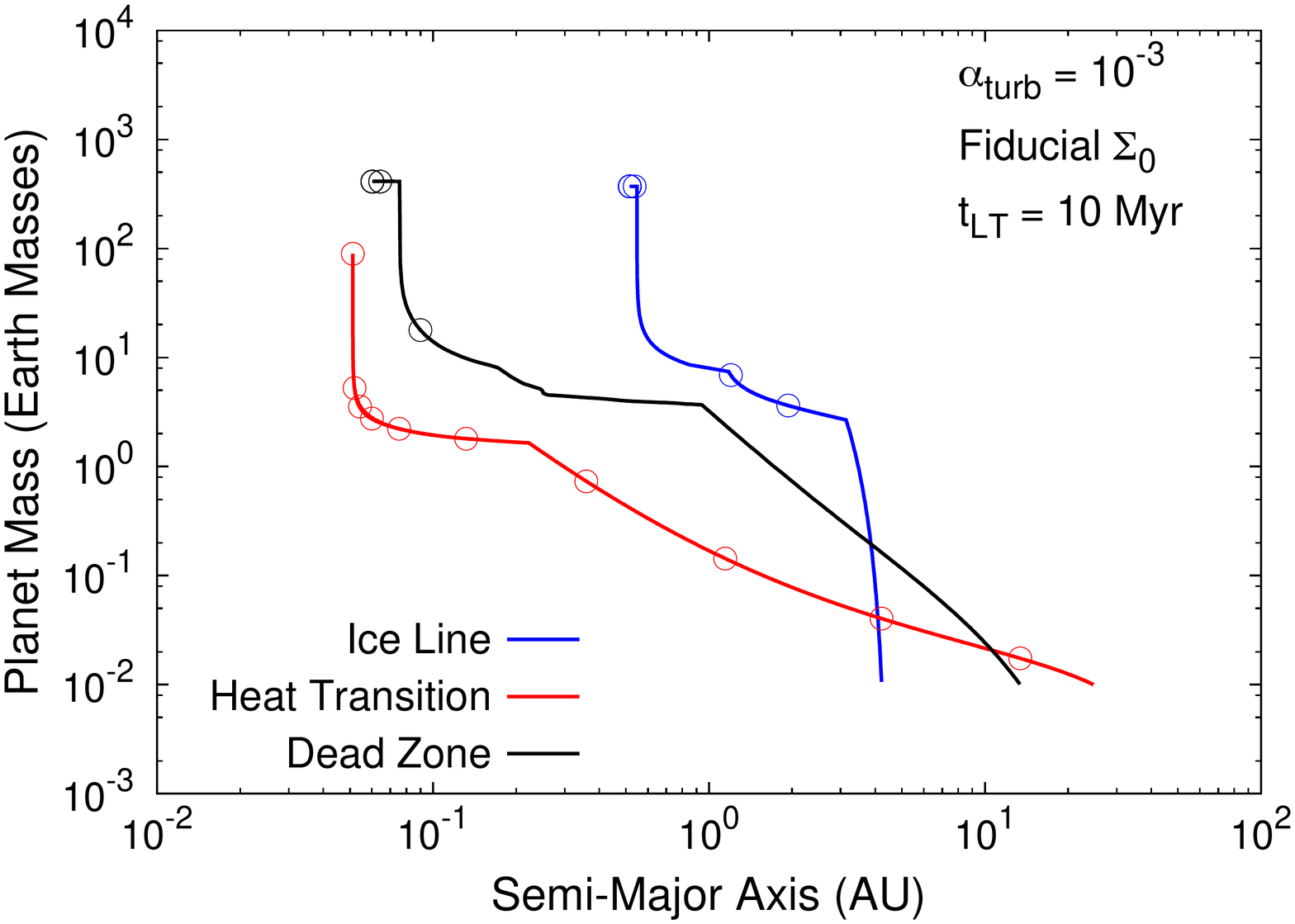} \includegraphics[width = 0.45 \textwidth]{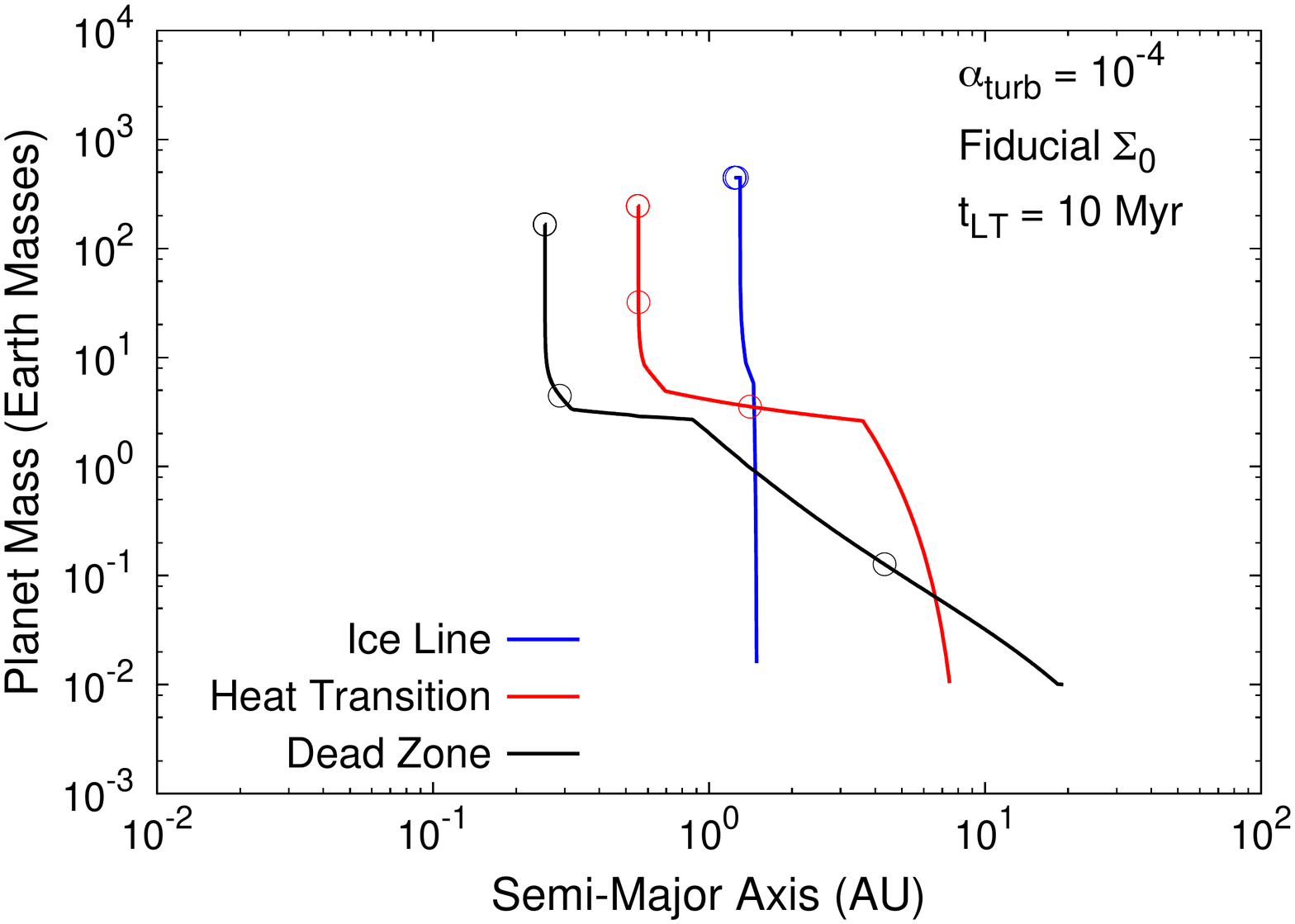} \\
\includegraphics[width = 0.45 \textwidth]{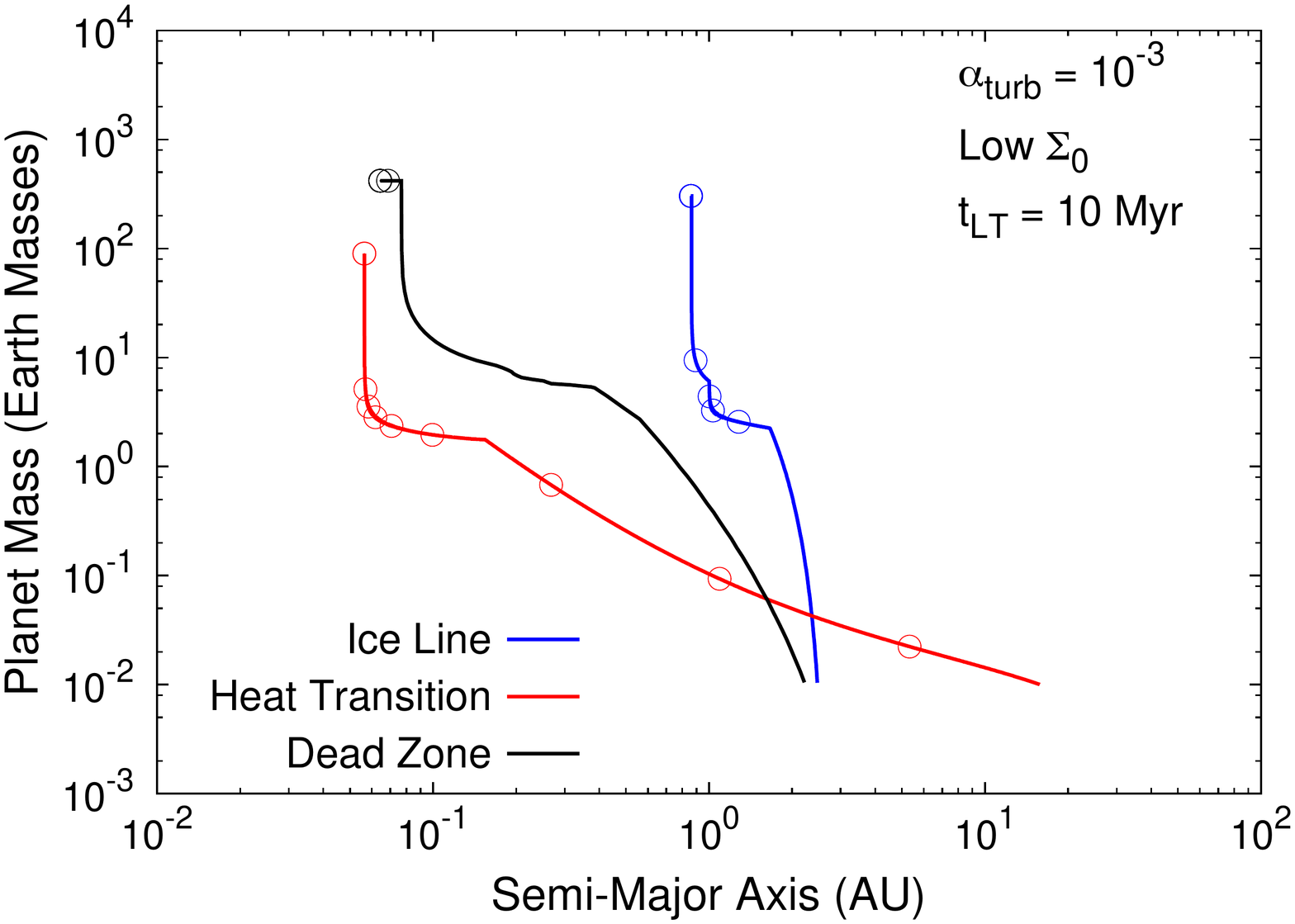} \includegraphics[width = 0.45 \textwidth]{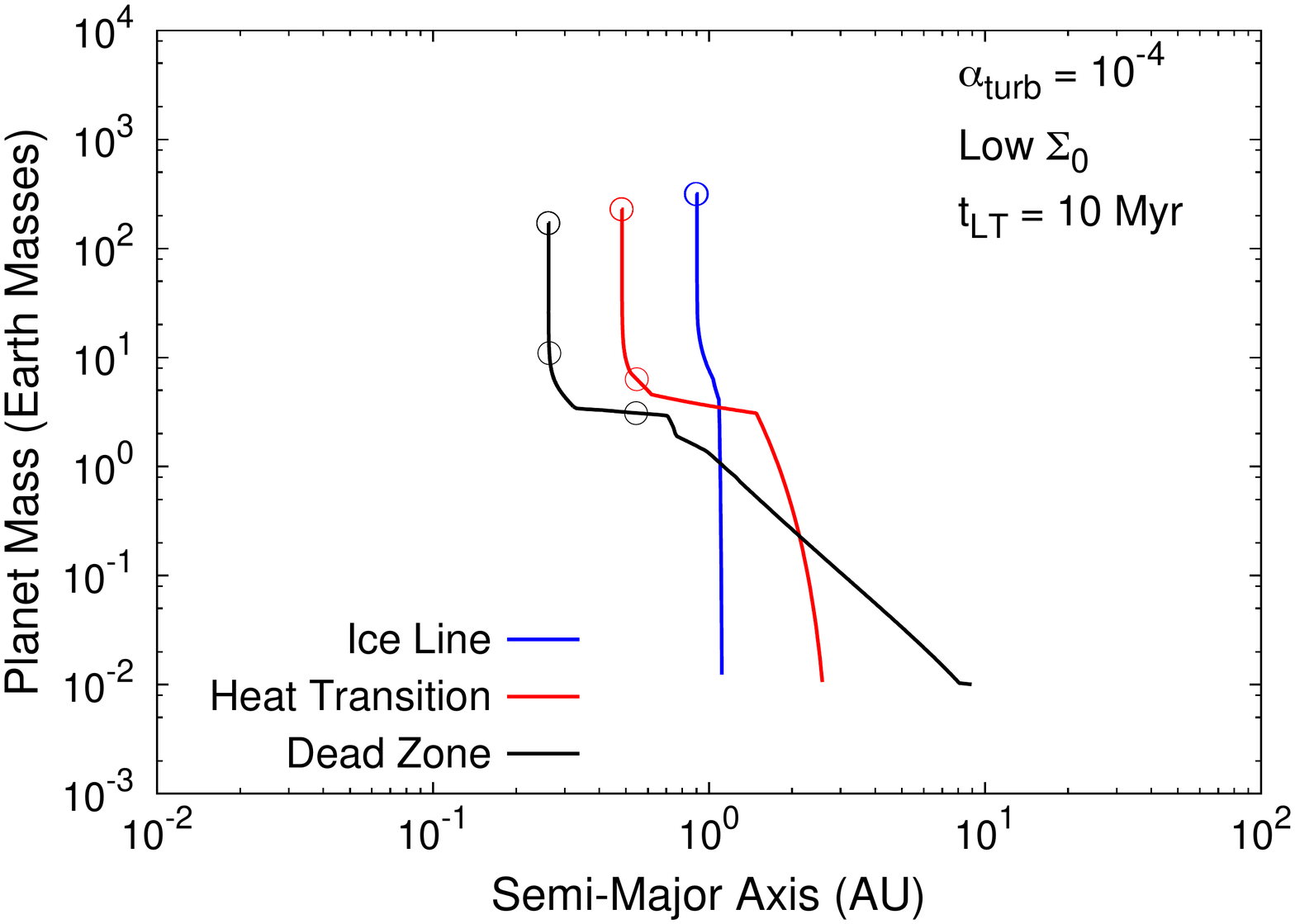}
\caption[Planet formation tracks: comparing effect of strength of disk winds and disk surface density]{We show a grid of planet formation tracks pertaining to the turbulent (left column) and combined (right column) disk models. We investigate the effect of disk surface density $\Sigma_0$ by considering three settings: high $\Sigma_0$ = 3 $\Sigma_{0,\rm{fid}}$ (top row), fiducial $\Sigma_{0,\rm{fid}}$ (middle row), and low $\Sigma_0$ = 0.33 $\Sigma_{0,\rm{fid}}$ (bottom row). All models assume a long disk lifetime of 10 Myr. Open circles along the formation tracks indicate planets' positions at 1 Myr intervals, indicating their positions if a shorter disk lifetime were to be considered.}
\label{Results1_Tracks}
\end{figure*}

Figure \ref{Results1_Tracks} shows a few representative planet evolution tracks in the M-a diagram. Planet formation models are calculated for each of the two models (turbulent and combined), and at 3 settings of $\Sigma_0$: fiducial, high, and low (as described in section \ref{Results4_DiskSettings}). While our planet formation models are calculated in long-lived 10 Myr disks, we include time marks along each planet formation track at 1 Myr intervals. This demonstrates the effect that disk lifetime will have on these models, as shorter lifetimes will simply truncate the formation tracks at positions in the M-a diagram as indicated by the time marks. This grid of planet formation models therefore contains a significant amount of information foreshadowing the outcomes of full population synthesis calculations, as it shows the effect of two of the most crucial varied disk parameters; namely the disk mass and lifetime.  

We first examine the fiducial $\Sigma_0$ case for the turbulent model. We see that formation within each of the 3 traps gives rise to gas giants across a wide span of orbital radii when a 10 Myr disk lifetime is used. The ice line produces a warm gas giant at $\sim$0.6 AU after 3 Myr. Its position in the disk within 5 AU for the entirety of the disk's evolution results in efficient planet formation in this trap. The dead zone forms a hot Jupiter at $\sim$0.07 AU after only 2 Myr, due to its rapid inward migration into the inner, high density regions of the disk. The heat transition also forms a hot Jupiter at $\sim$0.06 AU, but requires the full 10 Myr disk lifetime to do so. Disk lifetimes of 4-9 Myr would truncate the heat transition planet's formation within the super Earth region of the M-a diagram between 0.06-0.4 AU. This result confirms that the heat transition's large radius and slow inward migration results in the trap being relatively inefficient for planet formation in this model. 

Changing the disk surface density has two counteracting effects on planet formation rates. Increasing the disk surface density gives rise to higher accretion rates at a given orbital radius $r$. However, the second effect that opposes this higher accretion rate is that increasing the surface density also results in the positions of planet traps being shifted radially outwards to regions with lower $\Sigma$. This shift is as a result of the higher surface densities throughout the disk, and correspondingly larger viscous heating. Therefore, planets forming at a given trap in models with different $\Sigma_0$ values will accrete at different radii. When comparing the effect of disk surface density on planet formation, then, one should consider the net effect on the \emph{local} accretion rate at the traps' positions.

In the case of the high $\Sigma_0$ turbulent model, the ice line again produces a warm gas giant, but its final position is outside 1 AU and it forms in only 2 Myr. The dead zone planet's formation is similar to the fiducial model, resulting in a hot Jupiter at roughly 0.07 AU after 2 Myr. Even after 10 Myr of formation, the heat transition is only able to produce a super Earth at roughly 0.06 AU. In this case, the outward shift of the trap with the increase in $\Sigma_0$ was sufficient to reduce its accretion such that the trap failed to produce a gas giant. The high $\Sigma_0$ model is interesting in that it produces each of the three main classes of observed exoplanets: a super Earth, a hot Jupiter, and a warm Jupiter. 

\begin{table*}
\centering
\caption[Final planet masses and orbital radii: comparing planet formation results between the turbulence-dominated and combined turbulence \& winds models]{A summary of the final planet masses and orbital radii (at the end of each disk's 10 Myr-lifetime) in each model shown in figure \ref{Results1_Tracks}, where we investigated the effect of different $\Sigma_0$ values on planet formation results in both the turbulent-dominated and combined turbulence \& winds models.}
\begin{tabular}{|c| c| c| c|}
\hline
\hline
& Ice Line & Heat Transition & Dead Zone \\
Model & $a_p$, $M_p$& $a_p$, $M_p$ & $a_p$, $M_p$\\
\hline
\hline
\multicolumn{4}{|c|}{Turbulence-Dominated Model} \\
\hline
High $\Sigma_0$ & 1.3 AU, 729 M$_\oplus$& 0.05 AU, 4.6 M$_\oplus$& 0.06 AU, 414 M$_\oplus$\\
Fiducial $\Sigma_0$ & 0.5 AU, 372 M$_\oplus$& 0.05 AU, 89 M$_\oplus$ & 0.06 AU, 411 M$_\oplus$\\
Low $\Sigma_0$ & 0.86 AU, 303 M$_\oplus$ & 0.06 AU, 90 M$_\oplus$& 0.06 AU, 418 M$_\oplus$ \\
\hline
\multicolumn{4}{|c|}{Combined Turbulence \& Winds Model} \\
\hline
High $\Sigma_0$ & 1.9 AU, 786 M$_\oplus$ & 0.6 AU, 256 M$_\oplus$& 0.25 AU, 165 M$_\oplus$ \\
Fiducial $\Sigma_0$ & 1.2 AU 446 M$_\oplus$& 0.55 AU, 246 M$_\oplus$& 0.25 AU, 166 M$_\oplus$\\
Low $\Sigma_0$ &0.9 AU, 316 M$_\oplus$ & 0.5 AU, 229 M$_\oplus$ & 0.26 AU, 170 M$_\oplus$ \\
\hline
\end{tabular}
\label{Results4_1_Table}
\end{table*}

When examining the low $\Sigma_0$ turbulent scenario, we find that this model also results in the ice line producing a warm gas giant at roughly 1 AU. However, its formation time increases to 5 Myr. Since the ice line's location in the disk is relatively insensitive to $\Sigma_0$, the formation timescale at this trap intuitively scales with disk surface density. This formation track shows that an appreciable range of disk lifetimes, 1-4 Myr, will truncate this planet's formation and result in a super Earth or Neptune between 1-2 AU. The decrease in $\Sigma_0$ from the fiducial setting has a significant effect on the locations of the dead zone and heat transition traps, and ultimately results in their formation timescales decreasing. The dead zone again forms a hot Jupiter at 0.07 AU, but in this model does so within 1 Myr. The heat transition forms a hot Jupiter at 0.06 AU in 9 Myr. This formation track also shows that a large range of disk lifetimes, in this case 3-8 Myr, will result in the formation of a super Earth with orbital radius between 0.06-0.3 AU.

Tracks for the turbulent-dominated model  reproduce a main result from \citet{Alessi2018} for which this model is most directly comparable; i.e. $\alpha_{\rm{turb}}=10^{-3}$, constant $f_{\rm{dtg}}=0.01$ -  even though we are using a different disk model.   The point is that when using an X-ray ionized disk and its related dead zone, there is a clear separation between warm gas giants formed near 1 AU in the ice line, and hot Jupiters formed within 0.01 AU from the dead zone. Based on the results of figure \ref{Results1_Tracks}, we expect that this result will remain even when a full population is considered with distributions of disk lifetimes and $\Sigma_0$. With that said, it should be recalled that this is indeed a different model, with disk winds carrying a significant amount ($\sim$ 80\%) of angular momentum.

Figure \ref{Results1_Tracks} shows the planet formation tracks corresponding to the combined turbulence and winds models. We find a much more compact configuration among gas giants that are formed in each of the three traps at all settings of $\Sigma_0$ than we found in the turbulent model. In this circumstance, the formed gas giants have the most compact orbital configuration at low $\Sigma_0$, and the largest span of orbital radii is encountered in the high $\Sigma_0$ case. The ice line forms a gas giant within 1 Myr at each investigated $\Sigma_0$. The ice line forms gas giants at the largest separation from the host star among each of the traps. Their radii range from 1-2 AU, shift outwards slightly as $\Sigma_0$ is increased. The ice line is thus extremely efficient in gas giant formation in the combined model, which is a result of its location within 3 AU for the entire disks' evolution.

In contrast to the turbulent model, the heat transition produces gas giants at $\sim$ 0.6 AU in 2-3 Myr of disk evolution in the combined model, with shortest formation times corresponding to the low $\Sigma_0$ case. The final orbital radius of the resulting gas giant is insensitive to the setting of $\Sigma_0$. In the combined model, the smaller amount of viscous heating results in the heat transition existing at a smaller orbital radius where disk surface densities are higher. This shift to smaller orbital radii increases the efficiency of planet formation in this trap when compared to the turbulent model.

In the combined model, formation in the dead zone trap is also quite different. The trap produces gas giants between 0.2-0.3 AU whose orbital radii are also quite insensitive to $\Sigma_0$. These planets' formation timescales are between 3-4 Myr, with shortest formation timescales pertaining to the lower $\Sigma_0$ cases. Comparing to the turbulent model, the dead zone's location in the combined scenario is at a larger radius. The trap rapidly evolves inwards in both cases. While in the turbulent model, this trap exclusively produced hot Jupiters within 1-2 Myr, the larger initial radius of the dead zone in the combined scenario results in gas giants forming with longer formation times and larger final orbital radii. 

In table \ref{Results4_1_Table}, we summarize the results of our planet formation models shown in figure \ref{Results1_Tracks}.  We find that a larger range of planet formation outcomes are achieved in the turbulent case when varying the disk lifetime and surface density. Not only are the final configurations of planets more compact in the combined scenario, but their orbital radii and formation times are quite insensitive to variation in $\Sigma_0$. For example, across the range of $\Sigma_0$ settings investigated, the combined model does not produce any hot Jupiters or short-period super Earths. Additionally, we generally found shorter gas giant formation timescales at all settings of $\Sigma_0$ in the combined disk models, indicating that the super Earth population will be limited at the lower setting of $\alpha_{\rm{turb}}=10^{-4}$. As a result, gas giants will form too efficiently to achieve a reasonable comparison with the data. 

Varying $\Sigma_0$ has a larger effect on planet formation models in the turbulent scenario, which affects planets' formation times and/or orbital radii. We also find that the turbulent model readily shows that the formation of each of the observed classes of planets can be achieved. The ice line produces warm gas giants and super Earths near 1 AU; the dead zone forms hot Jupiters, and the heat transition results in super Earths forming over a range of orbital radii, as well as hot Jupiters in the longest-lived disks. 

Our planet formation results thereby suggest that a full population model in the turbulent scenario with $\alpha_{\rm{turb}} =10^{-3}$ may achieve a better comparison with the observed M-a distribution than the ``combined'' turbulence and disk winds model with $\alpha_{\rm{turb}} = 10^{-4}$. To fully explore this idea, we continue our investigation using full planet populations in the following section \ref{Results3}. 

Finally, we refer the reader to Appendix \ref{Results4_2}\footnote{The appendix for this work is included as online supplementary material}, where we show disk evolution and planet formation results corresponding to disk models with low levels of turbulence ($\alpha_{\rm{turb}} = 10^{-6}$), where the effects of the disk wind outflow strength are analyzed. In this investigation, we found that low settings of the $\alpha_{\rm{turb}} = 10^{-6}$ give rise to planet formation that is entirely insensitive to changes in disk parameters such as lifetime or mass. This indicates that, in a full population run as shown in section \ref{Results3}, minimal variability in final planet masses or orbital radii would be found even if disk parameters were stochastically sampled. Additionally, we found that high settings of the wind outflow strength ($K$ = 1) result in interesting effects on disk evolution, namely a local maximum in $\Sigma$ at $\sim$ 10 AU (which was also found in \citealt{Chambers2019}) giving rise to an additional planet trap. We emphasize, however, that the high setting of the wind outflow parameter produce large, unconstrained outflow rates $\dot{M}_{\rm{wind}} \gtrsim \dot{M}_{\rm{acc}}$ that are in contention with the theoretical and observational constraint provided by equation \ref{Outflow_Constraint}.

\section{Results II: Planet Populations \& the Effect of Relative Turbulence and Disk-Winds Strength} \label{Results3}

\begin{figure*}
\centering
\includegraphics[width = 0.32 \textwidth]{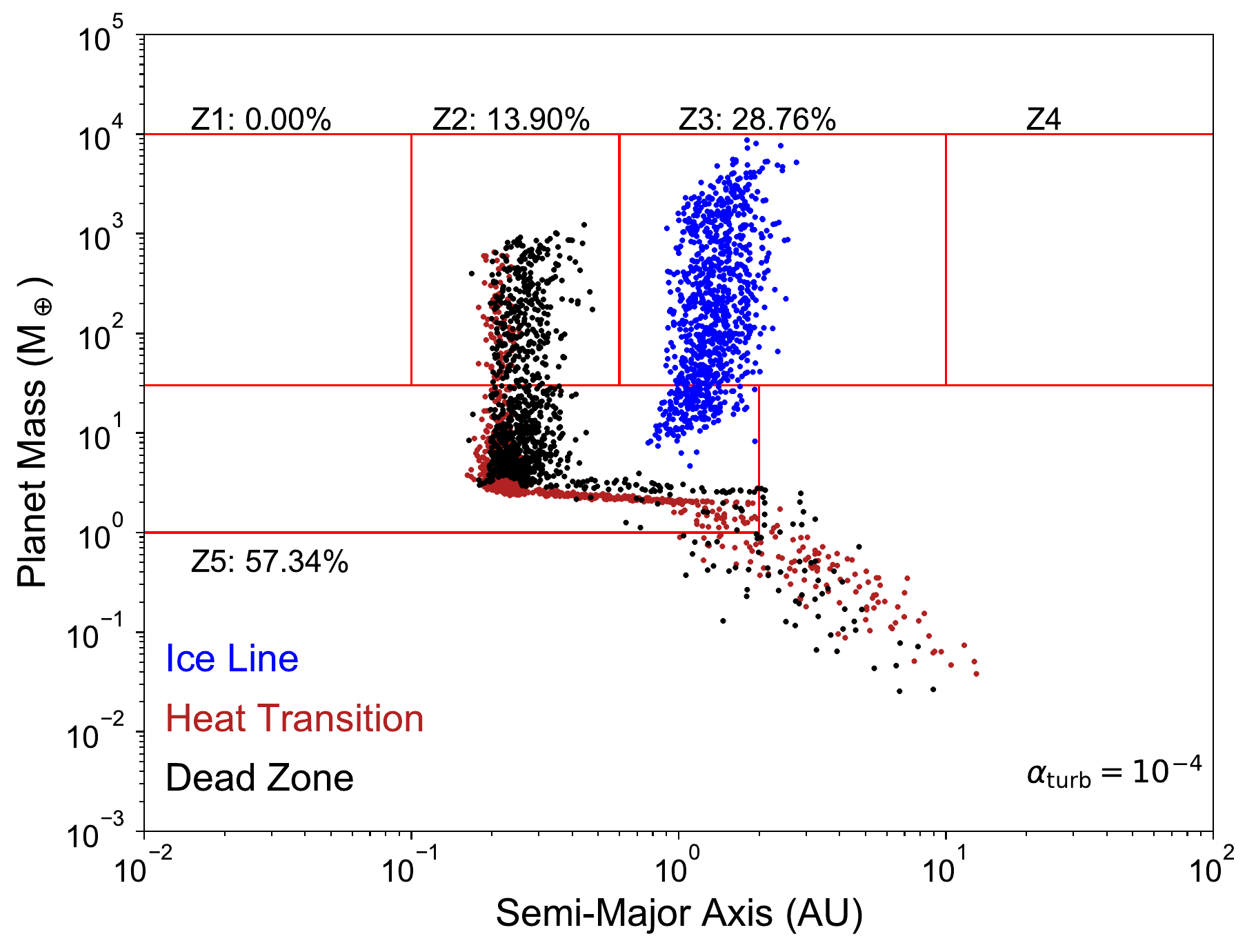} \includegraphics[width = 0.32 \textwidth]{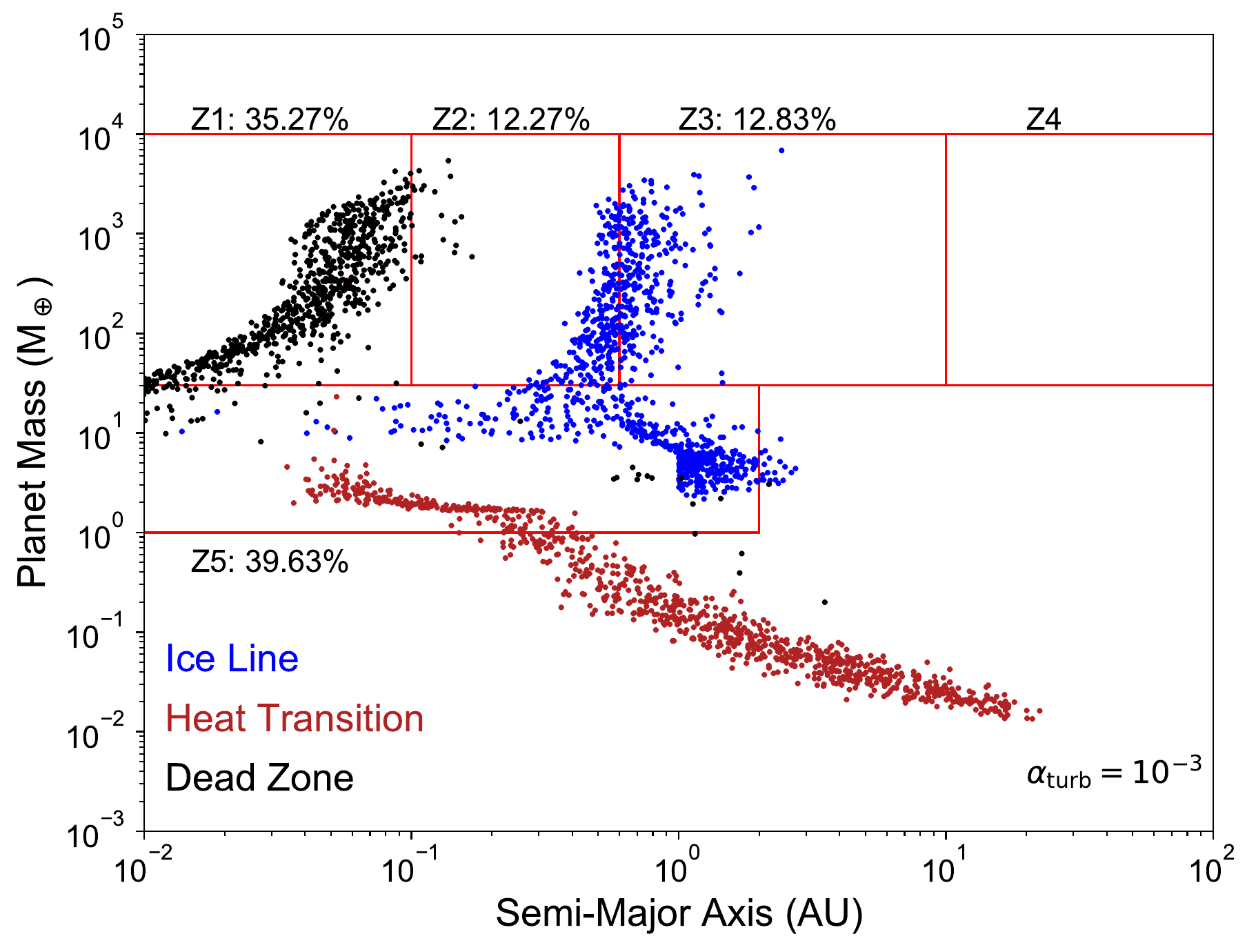} 
\includegraphics[width = 0.32 \textwidth]{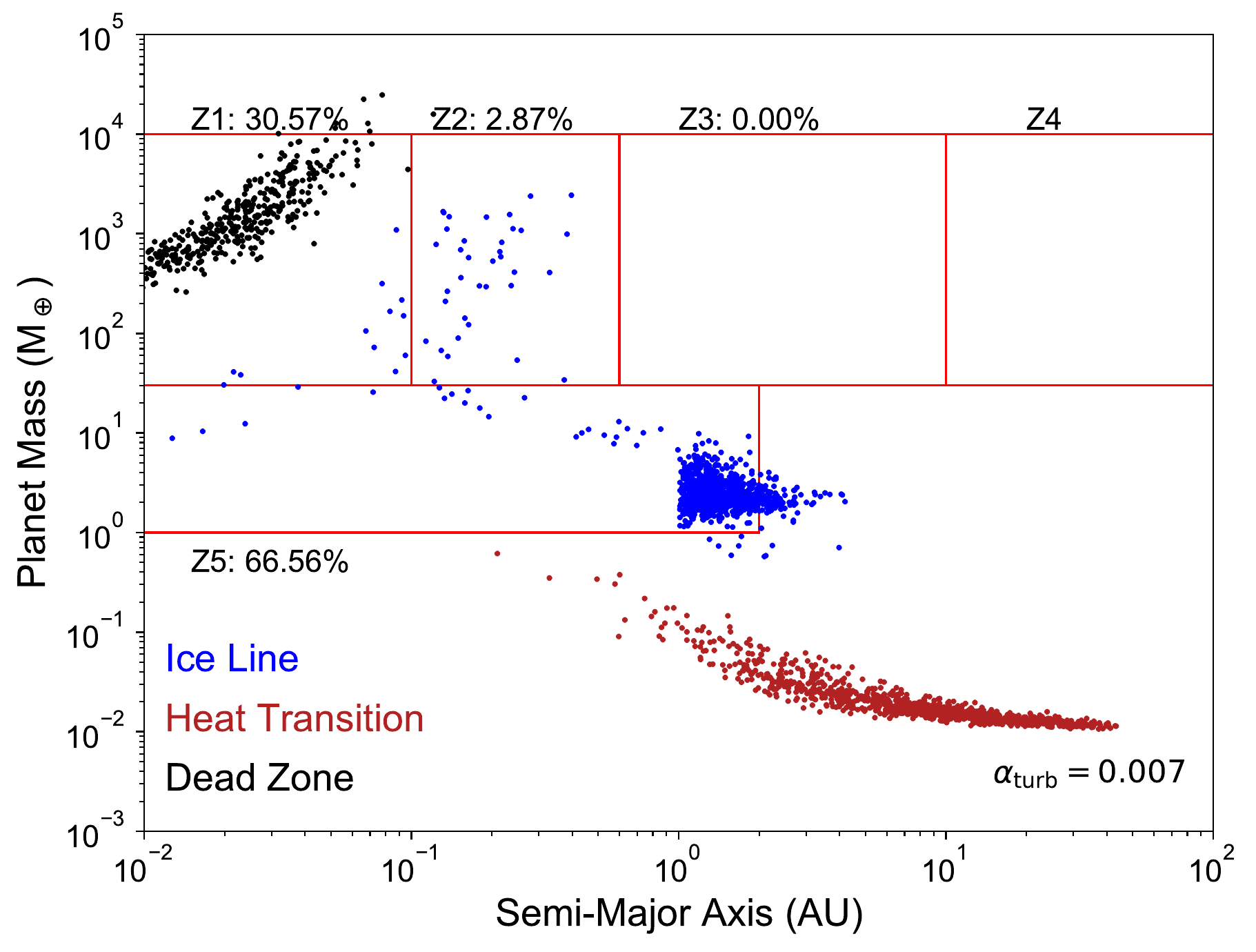}
\caption{A sequence of populations' M-a distributions is shown to compare the effect of the $\alpha_{\rm{turb}}$ value. Within each population, a constant $\alpha_{\rm{turb}}$ values is used, for which we consider 10$^{-4}$ in the left panel, $10^{-3}$ in the middle, and 0.007 in the right.}
\label{Alpha_Compare}
\end{figure*}

We now turn to examining data for whole synthetic populations of planets. Within these populations, we account for variability in planet formation environments by stochastically varying disk lifetime, initial surface density, and metallicity.   In all of the migration tracks that we compute for planets in the M-a diagram, we truncate our computations at an inner radius of  0.01 AU.    In so doing, we neglect the effects of the inner disk radius cutoff or the planet trap at the inner edge of the dead zone that are known to halt further radial inward migration at such radii  \citep{Masset2006}.  

In figure \ref{Alpha_Compare}, we begin by considering populations at three different $\alpha_{\rm{turb}}$ values, held constant within each population. We show populations corresponding to $\alpha_{\rm{turb}} = 10^{-4}$ (corresponding to $\alpha_{\rm{wind}} = 3\times 10^{-4}$) and 10$^{-3}$ ($\alpha_{\rm{wind}} = 3\times 10^{-4}$)\footnote{the same values used in the previous section \ref{Results4_1} to analyze individual disks and planet formation models} as well as a high setting of $\alpha_{\rm{turb}} = 0.007$ ($\alpha_{\rm{wind}} = 1\times 10^{-4}$) . The high setting was chosen to correspond with the upper limit found from observations of the TW-Hya disk in \citet{Flaherty2018}. 

The M-a distributions corresponding to the $\alpha_{\rm{turb}} = 10^{-4}$ and $10^{-3}$ populations represent the collection of individual tracks shown at these settings of $\alpha_{\rm{turb}}$ from the previous section \ref{Results4_1}, whose scatter comes from sampling a distribution of disk lifetimes, surface densities, and metallicities. At the low $\alpha_{\rm{turb}} = 10^{-4}$ value, each trap populates a somewhat restricted region of the M-a space, producing no planets within 0.1 AU. The dead zone and heat transition traps have comparable planet distributions, forming most planets at $\sim$ 0.1-0.4 AU with a range of masses between super Earths and gas giants. Additionally, both traps' distributions have a tail of super Earths extending 2 AU, with a small amount of failed cores at yet larger $a_p$ and lower masses. The ice line forms mostly warm gas giants, with a smaller fraction of Neptune-mass planets with $a_p$ mostly in the 1-2 AU range. These results are very much in line with what we found in section \ref{Results4_1} with individual formation tracks. The M-a distribution in the $\alpha_{\rm{turb}} = 10^{-4}$ case covers a limited region of the M-a space due to planet formation being insensitive to disk parameters; most importantly the disk surface density.

In the $\alpha_{\rm{turb}} = 10^{-3}$ population, there is a significant increase in the diversity of the planet formation outcomes, but it remains the case that the synthetic population covers a limited region of the M-a space and does not have sufficient scatter to compare reasonably with the observed M-a distribution. Again, the resulting population is a direct extension of the results found for individual formation tracks, covered in section \ref{Results4_1}, and each traps' output in the full population shows similar results. The ice line mostly produces super Earths and Neptunes between 1-2 AU, as well as gas giants with $a_p$ between roughly 0.5-2 AU. The dead zone shows rapid inward migration and planet formation timescales, resulting in almost exclusively hot Jupiter formation. Lastly, the heat transition forms super Earths at fairly small $a_p$ between roughly 0.05-0.5 AU, with a tail of failed cores extending to lower masses and large orbital radii.

The final panel of figure \ref{Alpha_Compare} shows the population resulting from a high setting of $\alpha_{\rm{turb}} = 0.007$, for which individual tracks were not shown in the previous section \ref{Results4_1}. The ice line predominantly forms super Earths between 1-2 AU with a small number of gas giants formed at 0.1-0.5 AU. The dead zone only forms hot Jupiters, which in this population are even more massive than in the $\alpha_{\rm{turb}}= 10^{-3}$ case. Lastly, we find that the heat transition exclusively forms failed cores at low $M_p$ and large $a_p$, failing to even form Earth-mass planets during planet formation. 

To summarize, we find that as we transition from low to high $\alpha_{\rm{turb}}$, planet formation in the ice line and heat transition become less efficient, while the distribution of planets from the dead zone shifts to smaller orbital radii and larger masses. To explain this sequence, we recall that viscous heating is reduced at lower values of $\alpha_{\rm{turb}}$, reducing the midplane temperature. Since all traps' positions are dependent on disk temperature, their position in the disk changes (i.e. recall figure \ref{Results1_Traps}). Both the ice line and heat transition's positions change to smaller $a_p$ at lower settings of $\alpha_{\rm{turb}}$, leading to planets accreting from regions of the disk with higher solid surface densities, thus resulting in more efficient growth. The position of the dead zone depends sensitively on both the disk surface density and temperature, and was shown in figure \ref{Results1_Traps} to shift to larger radii at lower settings of $\alpha_{\rm{turb}}$, although in both cases the trap quickly evolves to $\sim$ 0.1 AU within 1-2 Myr. This rapid inward evolution leads to the dead zone producing massive planets at all settings of $\alpha_{\rm{turb}}$. 

\begin{figure*}
\centering
\includegraphics[width = 0.45 \textwidth]{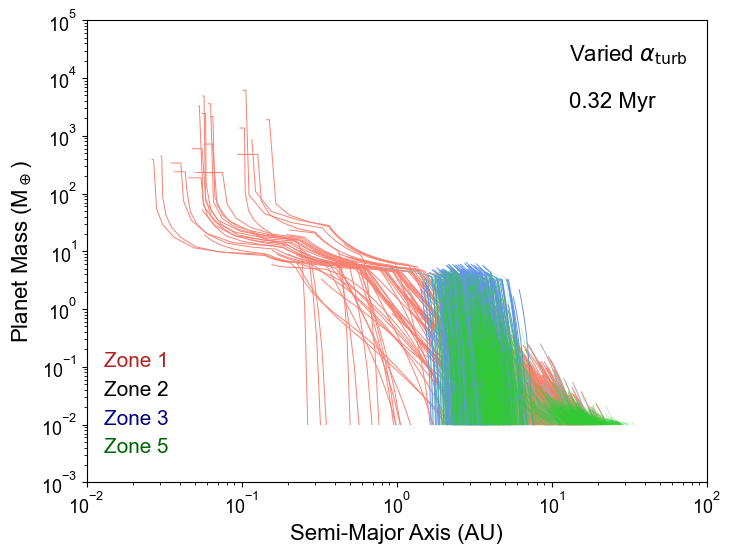} \includegraphics[width = 0.45 \textwidth]{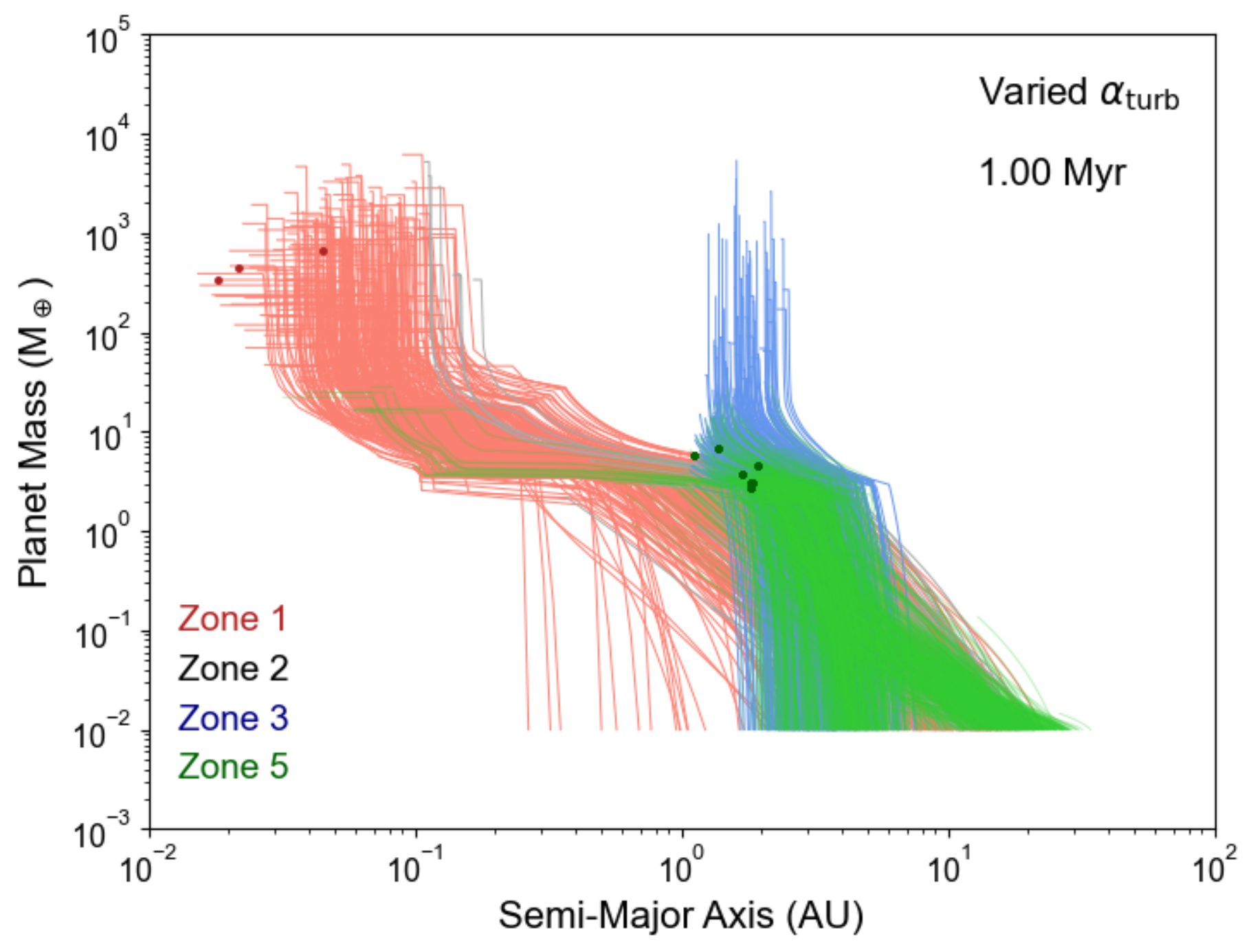} \\
\includegraphics[width = 0.45 \textwidth]{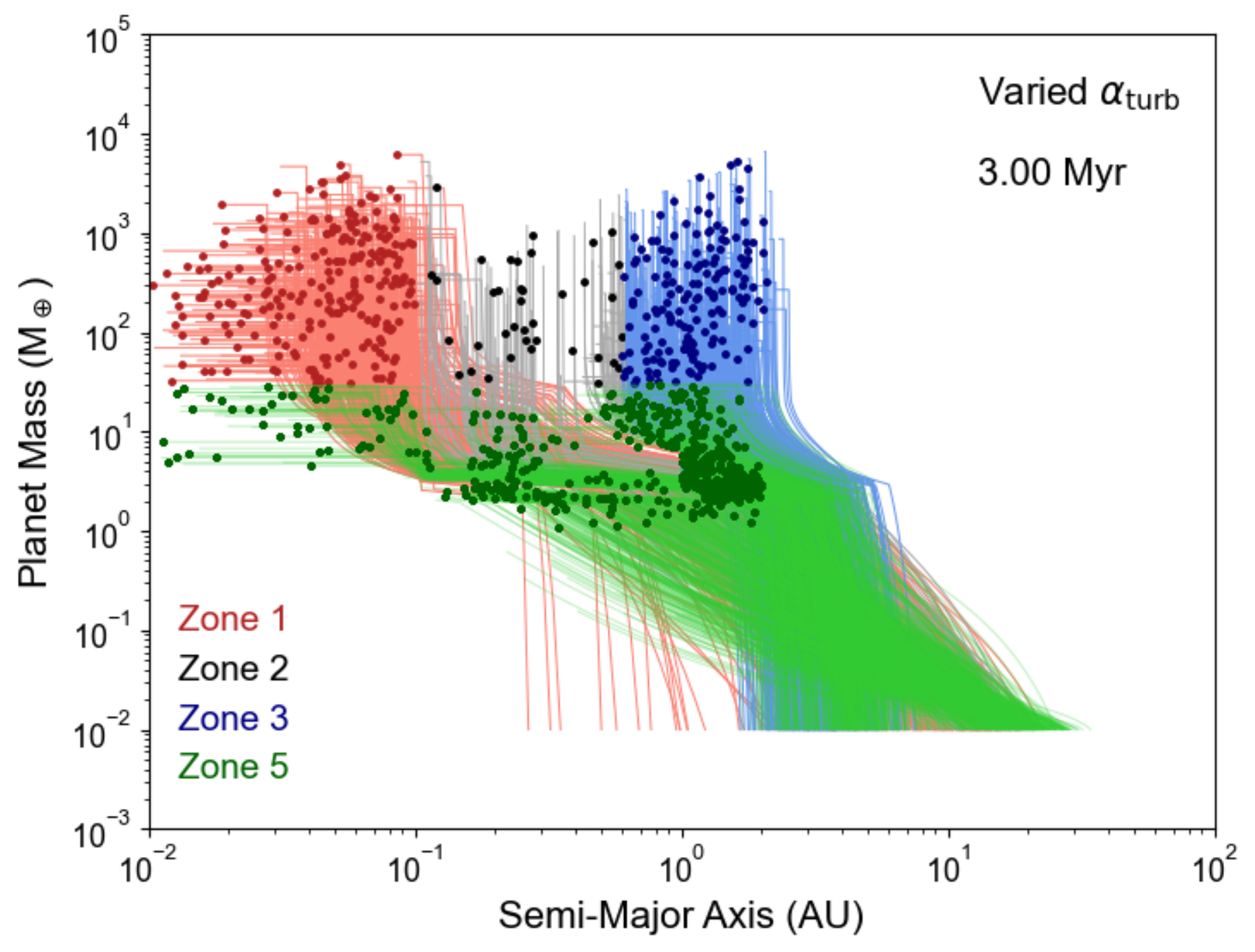} \includegraphics[width = 0.45 \textwidth]{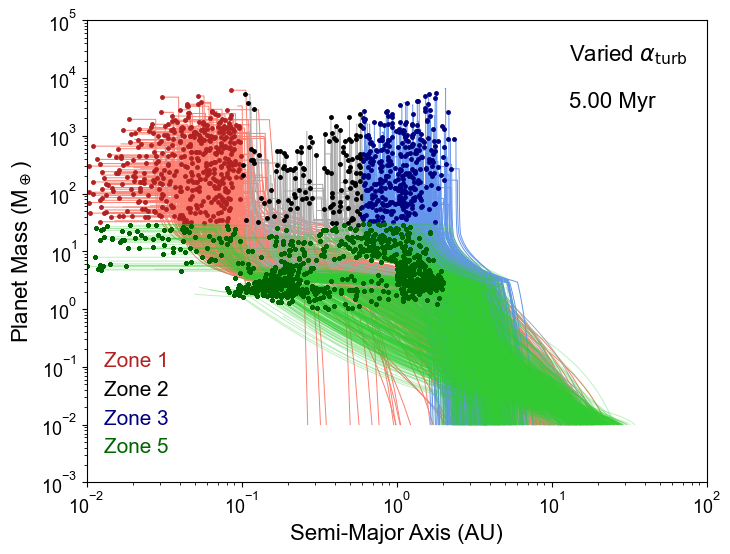} \\
\includegraphics[width = 0.45 \textwidth]{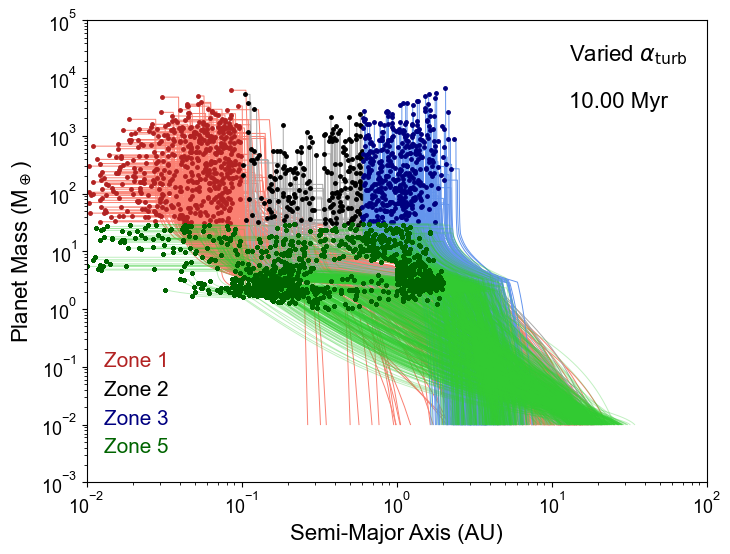} \includegraphics[width = 0.45 \textwidth]{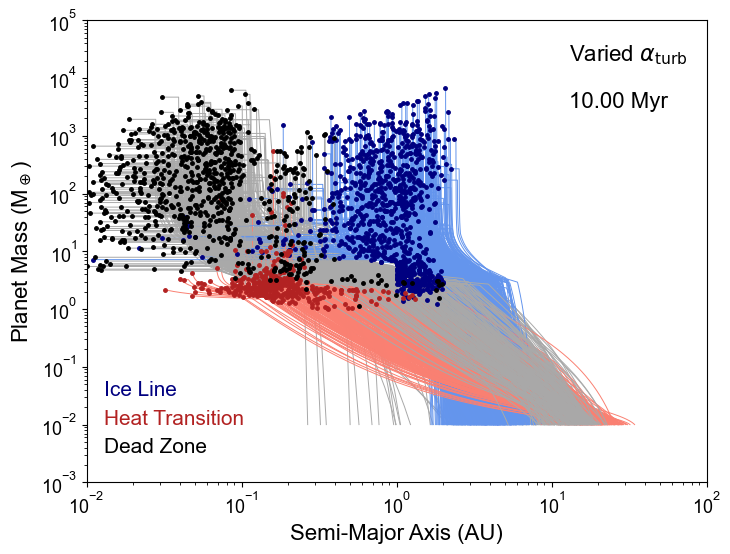}
\caption{We show a time series of the varied $\alpha_{\rm{turb}}$ population's formation tracks at 4 intermediate times leading to their final M-a distribution at 10 Myr. Tracks are coloured based on the zone the resulting planet populates. Data points denote the termination of a planet formation track at the end of its disk lifetime. In the bottom right panel, we show the final population's M-a distribution and tracks with colour now indicating the trap each planet formed in during the trapped type-I migration regime. A complete animation of this time-series is included as part of this paper's online supplementary material: 
\protect\href{https://youtu.be/0zGDZ6i14w8 }{[click here]}}
\label{Track_Population}
\end{figure*}

The trend of the dead zone planets to shift inwards at higher $\alpha_{\rm{turb}}$ is natural and that may explain the tendency to form more massive hot Jupiters at smaller $a_p$ with increased $\alpha_{\rm{turb}}$: planets will accrete from regions of the disk with higher surface densities.  On the other hand, there are reasons to think that building hot Jupiters at close in distances are difficult.   We have already noted that depending on the detailed physics of the opacities in these accreting atmospheres, it may prove too difficult to cool the gas effectively.   Another consideration is that the forming planet has a smaller Hills radius at small disk radii limiting the core growth there substantially \citep{Coleman2017} .   We see however that in our planet tracks  the cores of planets destined to form Hot Jupiters in our models have already reached the SuperEarth mass scale well outside of the inner disk regions, at around 1 AU.  This can be clearly seen by examining the tracks in figure \ref{Track_Population}, which we discuss next. 

Overall, we find that any individual setting of $\alpha_{\rm{turb}}$ produces a population whose M-a distribution does not compare reasonably with the observed distribution. In all cases, there is insufficient scatter and the synthetic populations cover a limited region of the M-a space. It is certainly intriguing that the populations at different settings of $\alpha_{\rm{turb}}$ cover different regions of the M-a diagram. 

\subsection{Populations from a Distribution of Relative Disk Turbulence and Wind Strength}
We now compute populations whose $\alpha_{\rm{turb}}$ parameter is sampled from a distribution. As covered in section \ref{Model_Population}, we construct an $\alpha_{\rm{turb}}$ log-normal distribution ranging from 10$^{-4}$ to 0.007, with a mean of 10$^{-3}$. We recall that the distribution's upper limit coincides with the observationally-inferred upper limit of the turbulence strength in TW-Hya \citep{Flaherty2018}. Our model results motivate a lower-limit of $10^{-4}$ to be reasonable, as planet formation results become increasingly less sensitive to disk parameters as $\alpha_{\rm{turb}}$ is decreased, and linking scatter in the M-a and M-R relations to variability in disk parameters remains the underlying motivation of using a population synthesis approach. We have continued our investigation of planet formation in disks with low levels of turbulence in Appendix A by considering disks with $\alpha_{\rm{turb}} = 10^{-6}$. We found that this trend persisted, with planet formation results becoming \emph{entirely} insensitive to changes in disk lifetimes and surface densities. Our approach of varying $\alpha_{\rm{turb}}$ in our populations is also motivated by the probable variability of disk $\alpha$ parameters between different systems, and we are reflecting that variability in our models by stochastically varying this key input parameter.

In figure \ref{Track_Population}, we show a time-sequence of formation tracks resulting from including $\alpha_{\rm{turb}}$ in our list of stochastically-varied parameters\footnote{The time-sequence shown includes several snapshots from a full animation of the formation of this population, included as part of this paper's online supplementary material}.   Tracks are coloured based upon the zone of the M-a diagram the resulting planet populates. Data points indicate the termination of a given planet formation track at a particular planet's disk lifetime. Most formation tracks terminate at times near 3 Myr which is the average of the disk lifetime distribution. At the final time of 10 Myr, all formation tracks have terminated revealing the final planet distribution. For clarity, we have only shown planet tracks that populate the various zones of the M-a space (omitting, for example, formation tracks of sub-Earth-mass cores). 

\begin{figure*}
\centering
\includegraphics[width = 0.45 \textwidth]{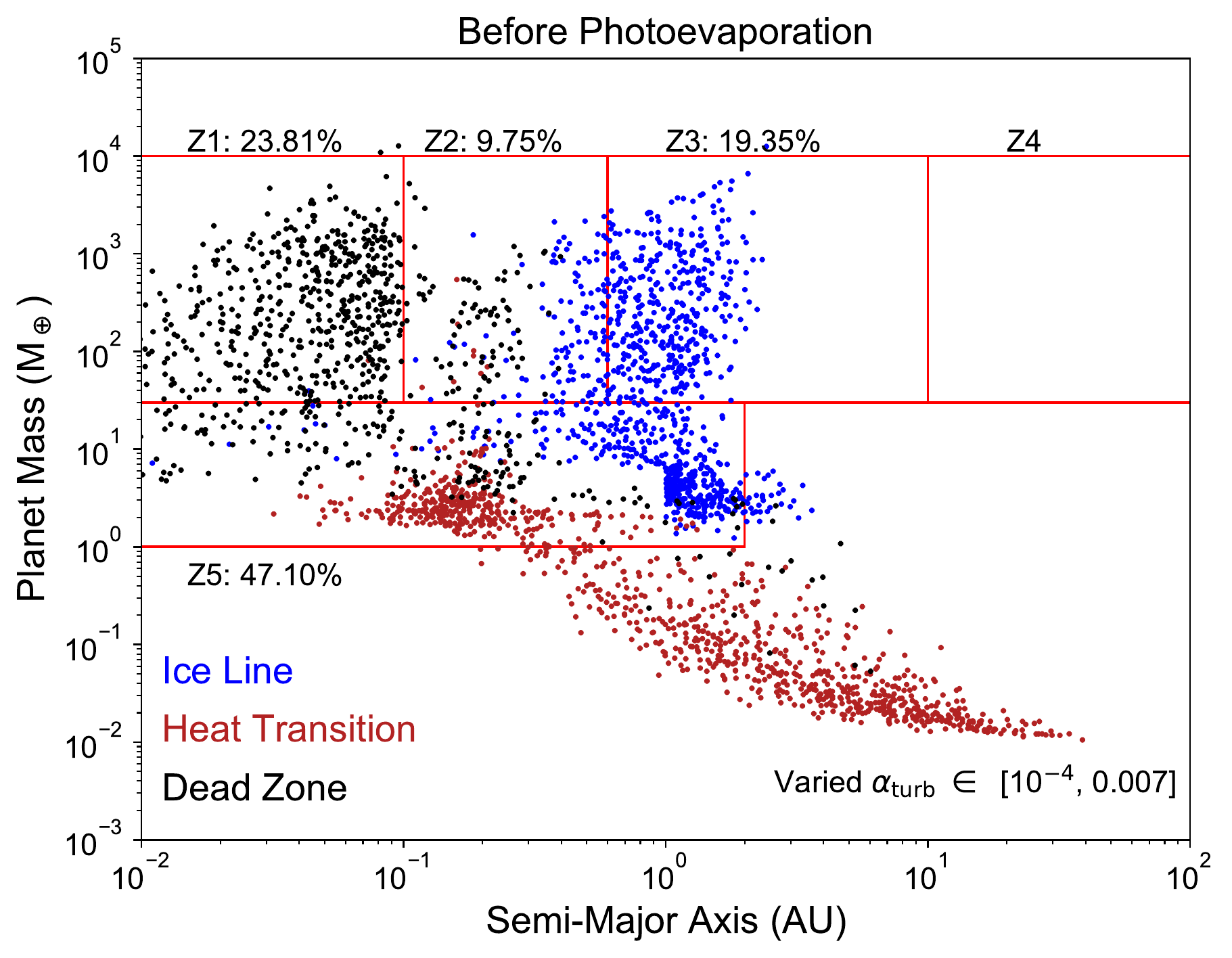} \includegraphics[width = 0.45 \textwidth]{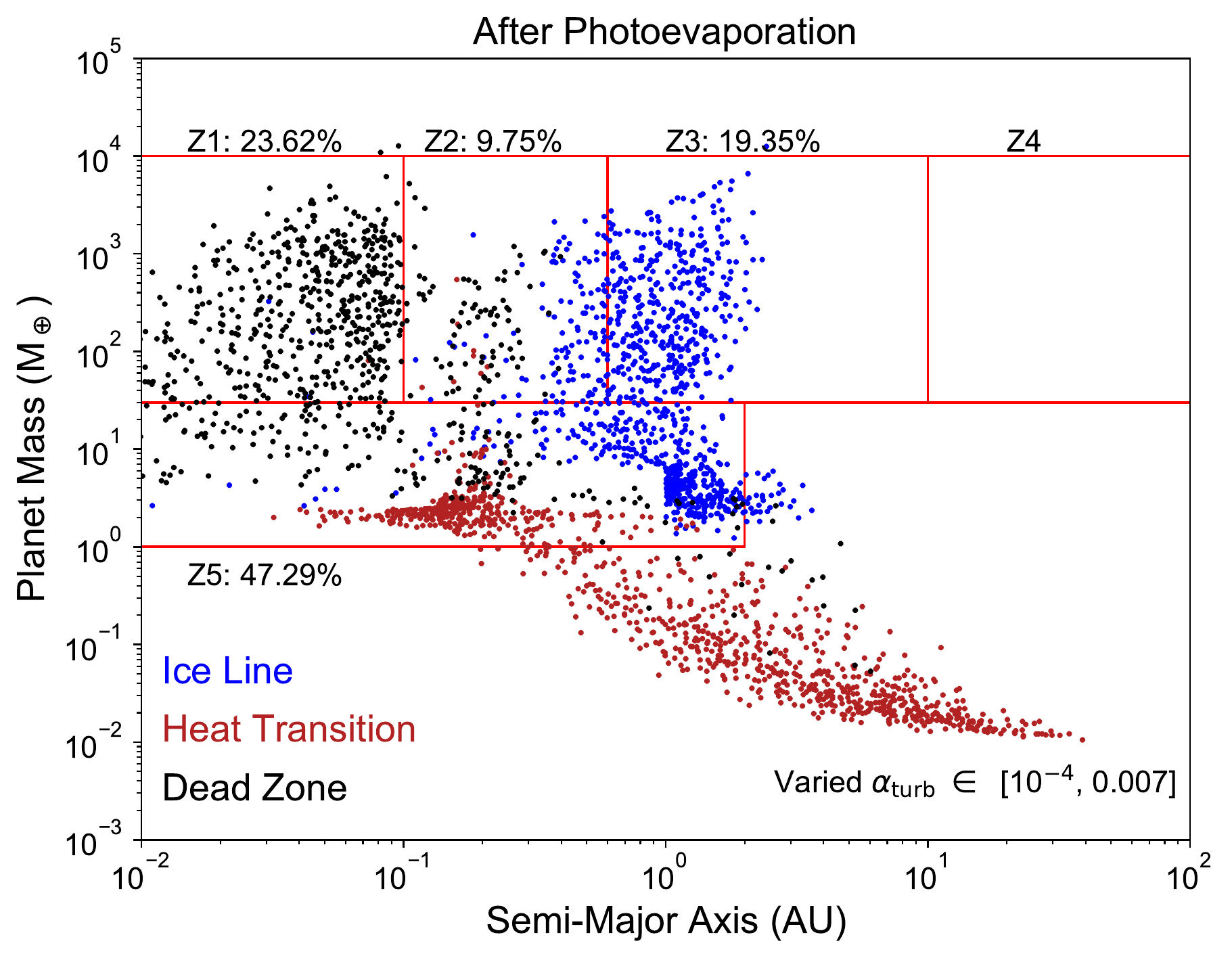} 
\caption{The final M-a distribution of the varied $\alpha_{\rm{turb}}$ population is shown directly after the disk phase (i.e. before photoevaporation in considered) in the left panel, and after up to 2 Gyr of photoevaporation in the right panel. We include the frequencies by which planets populate various zones of the M-a space.}
\label{Population_Ma}
\end{figure*}

Lastly, at the final time in the sequence presented in figure \ref{Track_Population}, we also show formation tracks with colour indicating the trap each planet formed in during their trapped type-I migration phase. This reveals that hot Jupiters arise from the dead zone, warm Jupiters from the ice line, and super Earths from each of the three traps. However, the small $a_p \lesssim$ 1 AU portion of the super Earth distribution is predominantly formed in the heat transition, while super Earths at larger orbital radii have formed in the ice line.

An important result here is that many of the hot Jupiters form quite rapidly, reaching the zone 1 region of the M-a space prior to 1 Myr. These tracks result from planet formation in the dead zone, whose evolution brings planets into the inner regions of disk quickly, allowing them to accrete from high-density regions leading to rapid formation. Many warm gas giants near 1 AU also form quickly, accreting substantial material within 1 Myr. These tracks correspond to formation in the ice line, showing that the trap efficiently forms planets due to its location in the inner, high surface density region of the disk  at early times in disk evolution ($\lesssim$ 3 AU). Super Earth formation tracks terminate at a range of times from intermediate ($\sim$ 3 Myr) to long disk lifetimes near 10 Myr. 

We have argued in earlier works that Super Earths are likely to be failed cores in the sense that their formation timescales exceed their disk lifetime \citep{ Alessi2017}.   Figure \ref{Track_Population} reveals that this can arise by either short disk lifetimes (in the case of terminating formation by 3 Myr), or notably long formation times (for which the disk lifetime is long, over several Myr).  Many tracks that populate the super Earth region of the M-a space with long formation timescales correspond to planet formation in the heat transition trap. Since the heat transition resides outside of 5-10 AU for the first 1-2 Myr of disk evolution, solid accretion rates onto cores remain small at these times leading to inefficient planet formation (not producing many gas giants).

\subsection{General Properties of Planet Populations in the M-a Diagram}

In figure \ref{Population_Ma}, we show the varied $\alpha_{\rm{turb}}$ population's M-a distribution both before (immediately following formation) and after photoevaporation. We see that the resulting populations range over a substantial portion of the M-a space. When compared with figure \ref{Alpha_Compare} with its constant settings of $\alpha_{\rm{turb}}$, we see significantly more scatter and an M-a distribution with more correspondence to observations.   It is also clear that photoevaporation does not remove much mass from planets which is why the two panels in figure \ref{Population_Ma} look so similar.   It is well known that photoevaporation contributes mostly to changing the radius of close in SuperEarths, not their masses. 

A striking feature in this M-a diagram is that the super Earth and Neptune planets are of decidedly mixed origins, with contributions from  each of the three traps. We also see direct formation of short period super Earths within $a_p \lesssim 0.03$ AU from the protoplanetary disk phase.  Our previous population synthesis models in the pure turbulent disk scenario was limited in its ability to directly form short-period super Earths \citep{Alessi2020b}.  However, in the combined turbulence and disk winds scenario, we find super Earths with quite small $a_p$ forming directly from the protoplanetary disk.

\begin{figure*}
\centering
\includegraphics[width = 0.45 \textwidth]{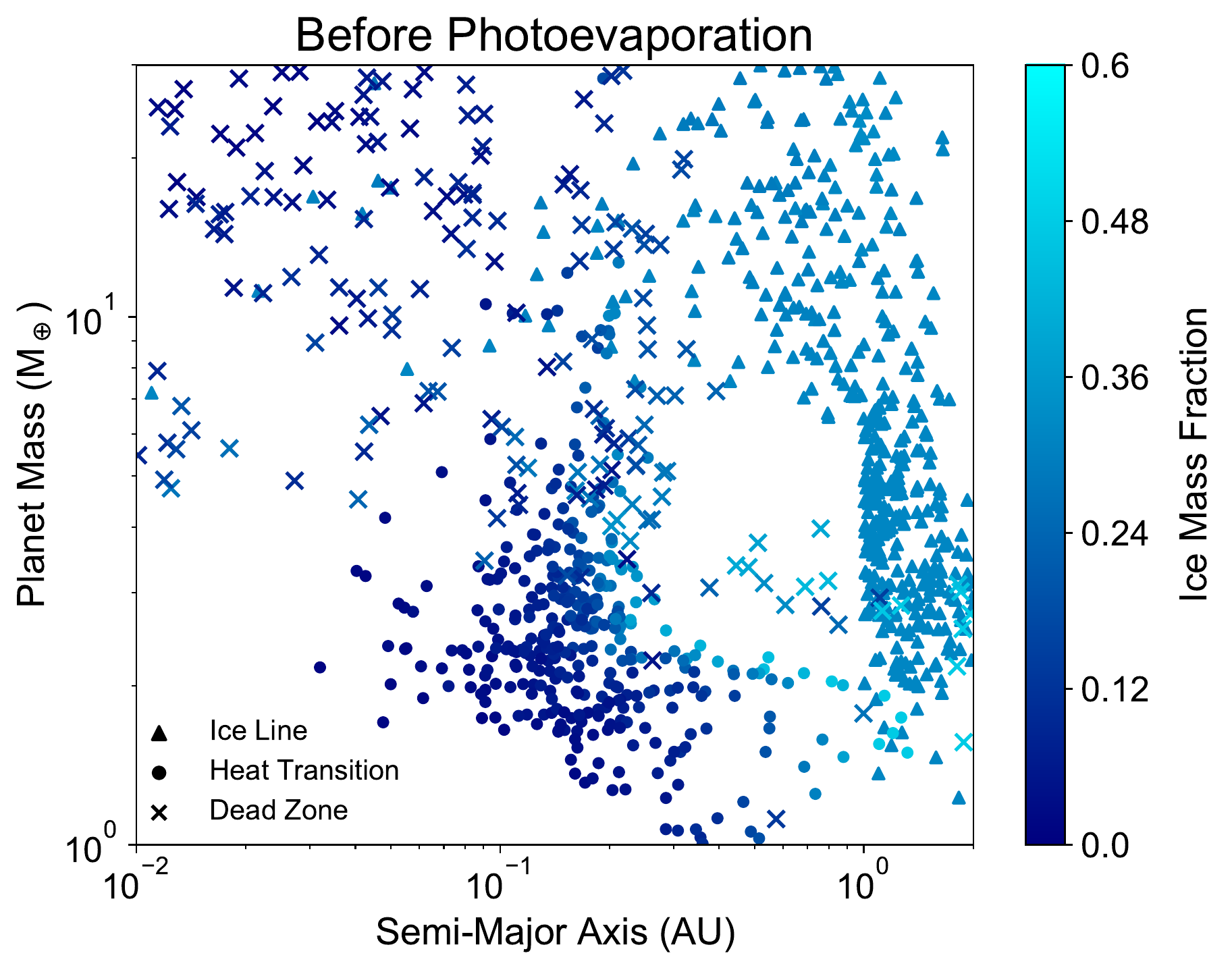} \includegraphics[width = 0.45 \textwidth]{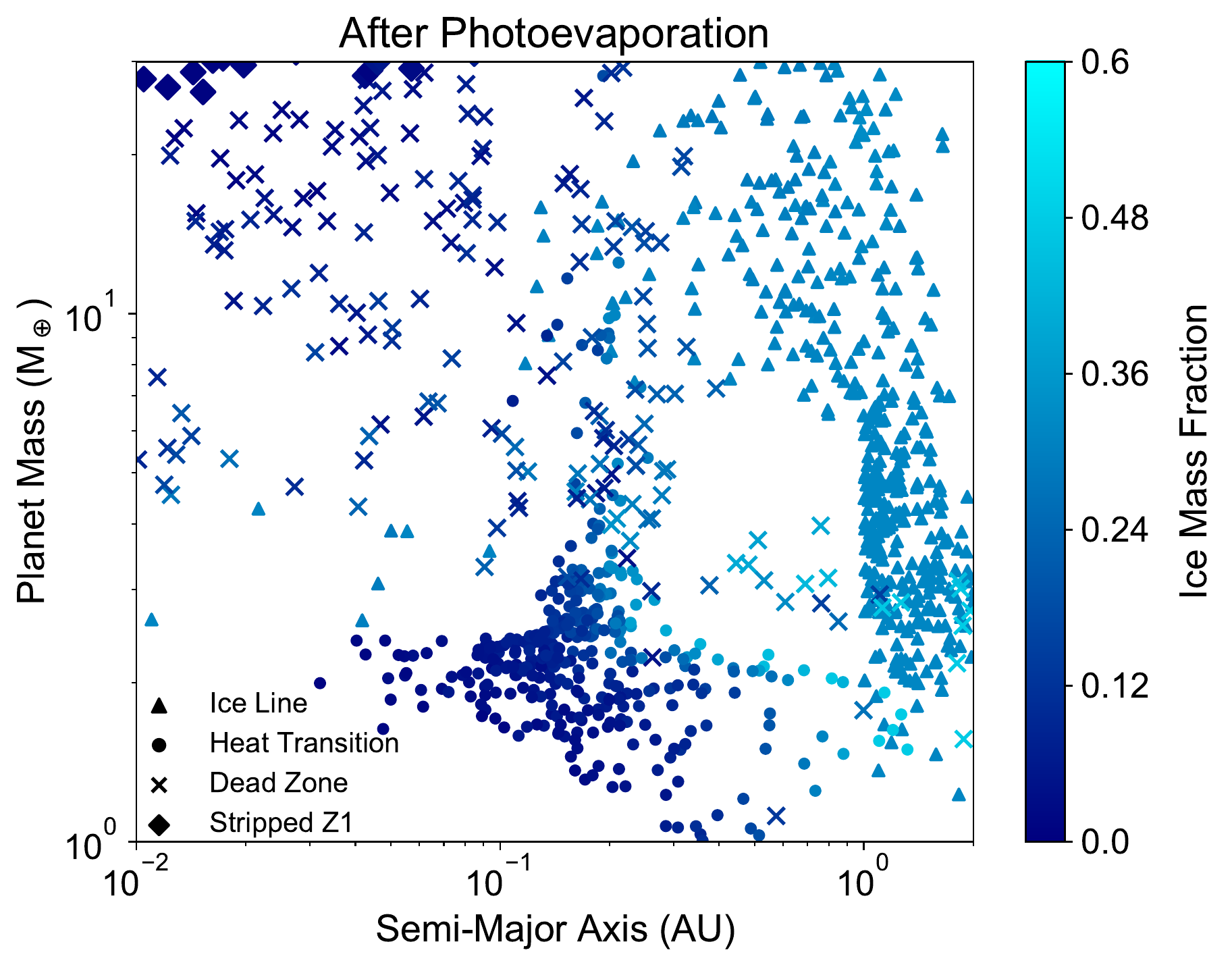} 
\caption{The super Earth and Neptune region (zone 5) of the M-a space is shown for our varied $\alpha_{\rm{turb}}$ population before (left) and after (right) photoevaporation. Planets' colours denote their ice compositions, the main indicator of their overall solid composition. Datapoint shapes indicate the trap each planet formed in, or in the after photoevaporation case, the small number of planets that formed as zone 1 sub-Saturns and lost atmospheric mass to produce a short-period Neptune.}
\label{Z5_Composition}
\end{figure*}

The varied $\alpha_{\rm{turb}}$ population's gas giant distribution consists of a large number of hot Jupiters formed within the dead zone, and a large number of warm gas giants formed in the ice line between 0.5-2 AU, with a reduction in gas giant frequency between roughly 0.1-0.5 AU. This reduction in gas giant frequency at these orbital radii corresponds with the reduced gas giant frequency found near orbital periods of 10 days in occurrence rate studies \citep{Santerne2016, Petigura2018}. This also reproduces the result we found in \citet{Alessi2018} where our populations showed a clear separation between the hot and warm Jupiter populations. However, occurrence rate studies also suggest that the gas giant frequency should have an overall increasing trend with increasing orbital radii, reaching a maximum between 3-10 AU \citep{Cumming2008}.  There is tension between our varied $\alpha_{\rm{turb}}$ population's M-a distribution  and the observed trend in gas giant frequency in two aspects: 1) the population forms too many hot Jupiters due to efficient formation in the dead zone trap; and 2) the warm Jupiter distribution formed in the ice line sharply falls off at $a_p >$ 2 AU, instead of extending to larger radii in the 3-10 AU range. 

Each trap leaves different and somewhat distinct imprints on the M-a diagram. The dead zone efficiently forms hot Jupiters with $a_p \lesssim$ 0.1 AU as well as some Neptunes with $a_p \lesssim$ 0.5 AU. The trap quickly evolves into the high density, inner regions of the disk leading to rapid formation of massive planets. The ice line forms many warm gas giants, as well as super Earths and Neptunes mostly in the 0.5-2 AU range. This trap is situated within 5 AU for the entirety of the disk's evolution, leading to rapid growth of cores as they accrete from a high surface density region of the disk. The heat transition has the most inefficient planet formation as it resides in the outer disk for the majority of evolution. It contributes a substantial amount of super Earths between 0.1-0.6 AU, with a tail of sub-Earth mass cores extending to larger orbital radii that fall below observational limits. 

Photoevaporation has the greatest effect on planets at smaller orbital radii (as it is a flux-driven effect) and also for planets with lower surface gravity as they are more susceptible to mass loss. Comparing the zone 5 M-a distribution before and after photoevaporation, we indeed see that planet masses reduce within $a_p \lesssim$ 0.2 AU. In particular, planets that formed in the heat transition show the most noticeable mass reduction. This is because the heat transition planets form with lower core masses due to their delayed growth compared to planets in the ice line and dead zone traps. This leaves more atmospheric mass available for stripping, leading to a greater overall mass difference after photoevaporation. Many of the short-period super Earths that form in the dead zone are mostly/entirely stripped, but their mass difference is not as noticeable due to their higher core masses (and lower atmospheric mass fractions). Our analysis of this population's M-R distribution in what follows will provide more insight into this argument.

\begin{figure*}
\centering
\includegraphics[width = 0.47 \textwidth]{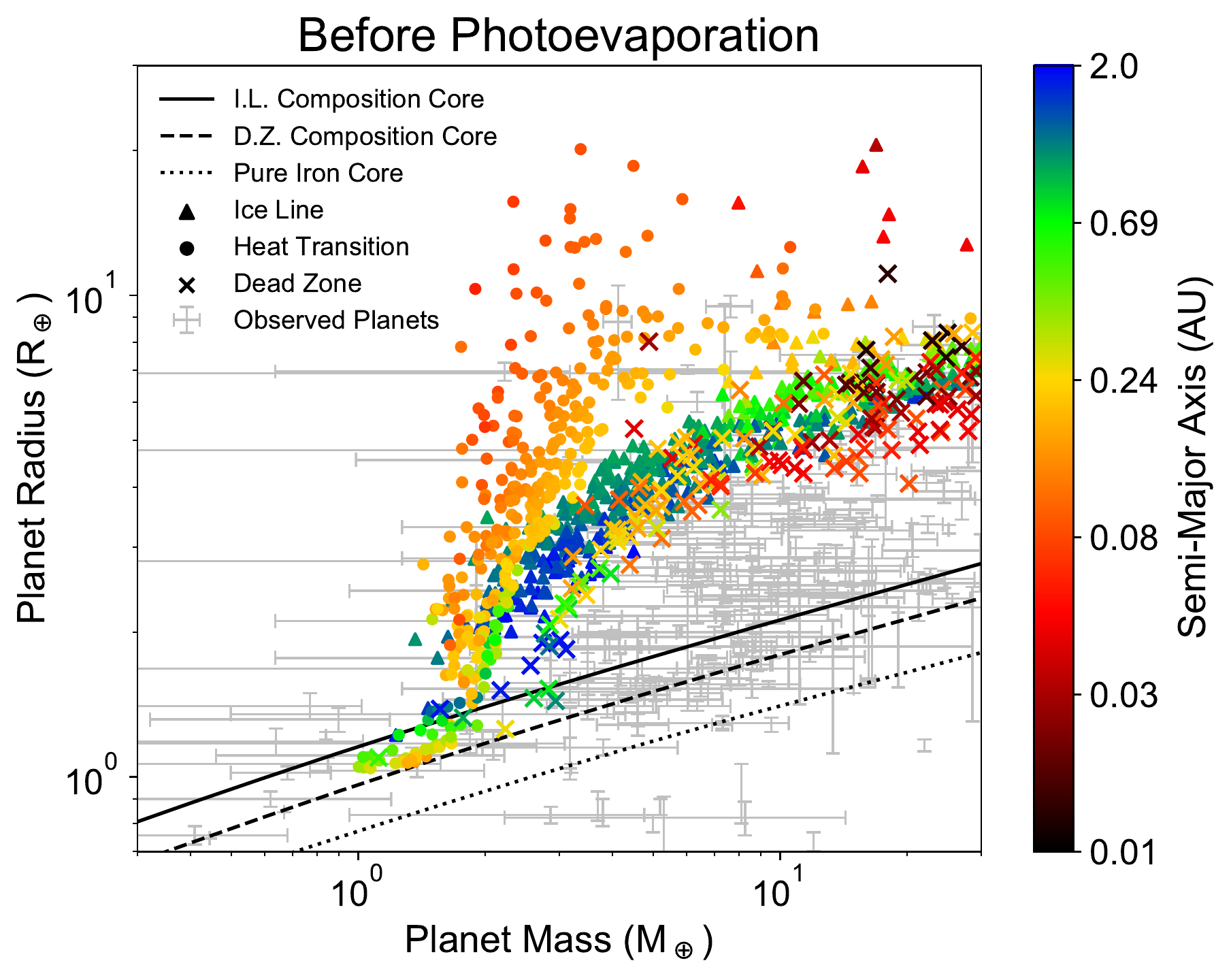} \includegraphics[width = 0.47 \textwidth]{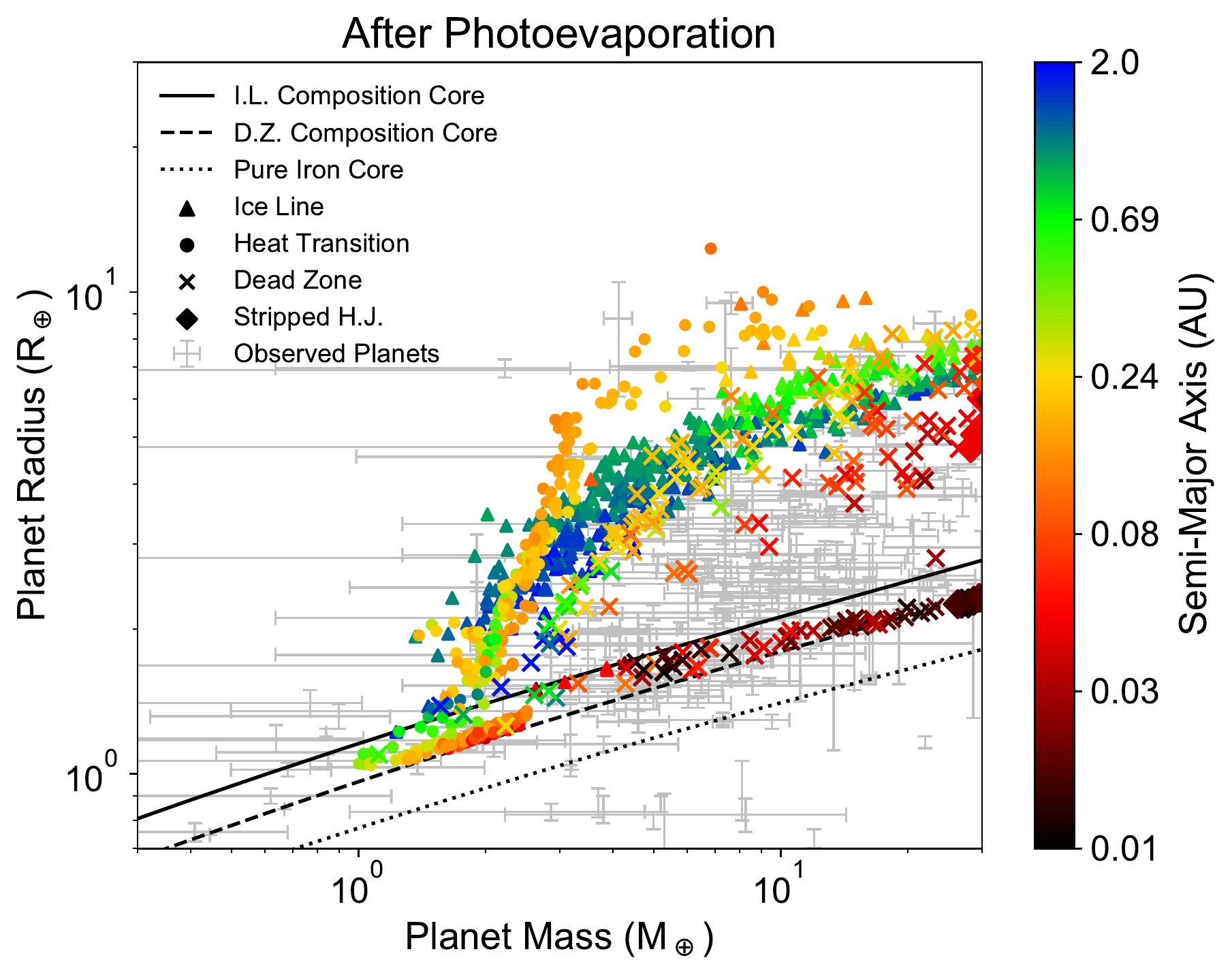} 
\caption{We show the varied $\alpha_{\rm{turb}}$ population's M-R distribution before (left) and after photoevaporation (right). Data point colours indicate planets' orbital radii, and the shapes denote the trap they formed in. We also indicate, in the right panel, planets that populate zone 1 immediately following the disk phase, but experience mass loss leading to Neptune-mass planets. Photoevaporation improves our comparison with the observed data (plotted in grey with error bars), particularly for planets whose radii are heavily increased due to accreted gas ($M_p \gtrsim $ 2 M$_\oplus$). Finally, we show M-R contours corresponding to three different uniform solid compositions: ice rich planets (taking ice line core compositions from \citealt{Alessi2020b}), dry, Earth-like compositions (using dead zone core compositions), and pure iron cores.}
\label{Population_MR}
\end{figure*}

\subsection{Planetary Compositions and the M-R diagram for Populations}

Figure \ref{Z5_Composition} shows the ice fractions of planets throughout zone 5 as computed via disk equilibrium chemistry models. As we found in \citet{Alessi2020b}, a planet's ice composition is the main indicator of its overall solid composition, with only small variations in the iron-to-silicate ratio existing throughout the disk.  We find a large variety in ice abundances on super Earths and Neptunes as is expected given their varied origins. Planets that form in the ice line have roughly one third of their solid mass comprised of ice. Both the dead zone and heat transition traps form a mix of dry and rocky super Earths (Earth-like compositions), and ice-rich planets depending on where the planet accretes with respect to the ice line. Many of the planets forming in these two traps accrete solids inside the ice line leading to dry planets, however a portion accrete outside and can achieve up to $\sim$ 50\% of their solid mass in ice. The position of the traps relative to the ice line at the timing of solid accretion is dependent on the exact disk parameters in the population.

Our composition results found in zone 5 planets differ from those found in our previous work that considered the pure turbulent disk scenario, \citet{Alessi2020b}, in three ways. First, the dead zone planets here show a range of solid compositions from dry and rocky to a smaller portion with substantial ice contents, whereas in our previous work the trap exclusively formed planets with nearly zero ice content. We attribute the difference in results to a difference in the trap's evolution in the two disk models. Second, we find a larger variety of super Earth solid compositions in this work's population since zone 5 has planets formed in each of the three traps. In our previous work, depending on the initial disk radius considered, at most two of the three traps contributed a substantial portion of super Earths. Lastly, while there is an overall trend for dry planets to exist at small $a_p \lesssim$ 0.5 AU, and ice-rich planets at $a_p \sim$ 0.5-2AU, there is not a clear division in planet composition at any $a_p$ since the heat transition and dead zone form planets with a range of compositions, and the traps overlapping regions of zone 5. In our previous work, we found clear divisions between super Earths that formed in different traps, and therefore having markedly different compositions.

The M-R distribution for the varied $\alpha_{\rm{turb}}$ population in figure \ref{Population_MR}, compares the distributions before and after atmospheric photoevaporation. Planets are coloured to show their orbital radii, a key parameter to consider when examining photoevaporation's effect.  We overlay our computed populations with the observed exoplanet M-R distribution.  The data set we have used is the same as that  used in  \citet{Alessi2020b} which is for planets around solar mass stars and extracted from the NASA Exoplanet Data Base in March 2020.  To guide our analysis, we also plot three M-R relations corresponding to three different uniform solid compositions: an average solid composition found in the ice line, which is an ice-rich planet having roughly a third of its solid mass in ice, a dry dead zone planet composition which corresponds to a rocky Earth-like planet, and lastly a pure iron core. These contours show the range of planet radii a planet can achieve from variations in solid compositions alone. They are useful in our analysis of our M-R distributions as they indicate planets with little/no atmospheres which will lie near or within the contours. Planets significantly above the M-R relations indicate planets with some atmospheric mass. For this point, we note that since gas is the lightest constituent material out of which planets are composed, only a small amount of gas can significantly increase a planet's radius \citep{Lopez2013}. 

Our computed populations' M-R distribution before photoevaporation shows a division between core-dominated planets and planets whose radii are significantly influenced by gas at roughly 2 M$_\oplus$. Planets at higher masses have accreted sufficient gas to lie significantly above the solid M-R contours. At $M_p \gtrsim$ 2 M$_\oplus$, a large portion of our computed M-R distribution compares reasonably with the upper portion of the observed M-R distribution. However, a fraction of these planets, many of which formed in the heat transition, have radii that lie significantly above the observed distribution. These planets have particularly high atmospheric mass fractions, and lower core masses than planets formed in the other two traps at comparable total planet masses. Our M-R distribution before photoevaporation has many planets that are at large radii for a given mass in comparison to the observed distribution, which is a reproduced result from \citet{Alessi2020b}. We note that all of these planets form as planets whose formation was terminated during their slow gas accretion phase, so it follows that their radii are somewhat influenced by atmospheres. Since we are considering our M-R distribution as formed from the disk, we see that fairly massive atmospheres can be accreted in the absence of photoevaporatiive losses. As gas accretion is extremely sensitive to the setting of atmospheric opacity during formation \citep{Mordasini2014, HP14}, our M-R distribution may differ were we to explore envelope opacity values other than those we found in \citet{Alessi2017}. 

Photoevaporation will  reduce the radii determined by transits for close in planets.  We quantify this by plotting the final M-R distribution in the right panel of figure \ref{Population_MR}. Photoevaporation indeed improves our comparison to the observed distribution in several ways. First, the planets that formed in the heat transition whose radii lied well above the observed distribution have their radii reduced to compare reasonably with the upper portion of the observed M-R distribution. These planets have rather low core masses and are more susceptible to atmospheric mass-loss due to their lower surface gravities. Planets with orbital radii $\lesssim$ 0.05-0.1 AU are significantly or entirely stripped due to their proximities to their host stars and large XUV-flux received, while planets at larger $a_p$ have unaffected radii. We see that for all masses up to 30 M$_\oplus$, there are planets at small $a_p$ that are entirely stripped as shown by their radii coinciding with the solid-core M-R contours. At the higher $M_p$ end of the M-R distribution, there are a small number of stripped cores that originally formed as zone 1 planets (sub-Saturns). We note that many of these planets whose atmospheres are entirely stripped have quite large core masses, explaining why they show significant radius differences due to photoevaporation while only showing modest differences in planet masses (i.e. figure \ref{Z5_Composition}).

Photoevaporation is therefore important in understanding the fate of gas accreted by planets that migrate too close to their host stars. However, its results are largely bimodal in the sense that planets for the most part either retain their atmospheres or are fully stripped. This is seen in our population's distribution as planets distribute either along the solid M-R contours (stripped planets) or along the high radius distribution that coincides with the upper portion of the observed distribution. However, our models are limited in their ability to compare with the portion of the M-R distribution that exists at intermediate radii for a given mass. We find very few planets that, after photoevaporation, achieve radii in this intermediate state (between fully stripped or retained atmospheres). Finally, the effect of planets' solid compositions on the M-R diagram can be seen among fully stripped planets, whose radii slightly differ depending upon their compositions. However, these radius differences are small and are within typical radius observational uncertainties.

\section{Discussion \& Conclusions} \label{Results4_Conclusion}

We have investigated the effects of MRI-turbulence and MHD disk winds as well as a range of initial disk conditions  on  disk evolution,  and the properties and compositions of the resulting planetary populations. The relative strength of turbulence and disk winds is a critical factor in these models. We first considered various constant settings of turbulence and wind strengths initially, focusing simulations on just three characteristic values for  $\alpha_{\rm{turb}}$,  ($10^{-4}$, $10^{-3}$, and 0.007) that cover the range of values quoted in the observational literature. Disk winds in our model are essential to carry off the angular momentum from the inner dead zone region of the disk, as recent simulations have clearly shown \citep{BaiStone2013, Gressel2015}.  Even at a setting of $\alpha_{\rm{turb}} = 10^{-3}$, over 80\% of the disk accretion is being driven by winds as opposed to turbulence in our models (since typical settings of $f_w \gtrsim 0.8$).   We found that each of these constant $\alpha_{\rm{turb}}$ models produced only a rather limited variety planets populating the M-a diagram even though we included the full observationally-constrained distributions of initial disk surface densities and lifetimes in accordance with previous population synthesis efforts (i.e. \citealt{IdaLin2004b, Mordasini2009, HP12}).    While all of the constant $\alpha_{\rm{turb}}$ populations had insufficient scatter to compare reasonably to the observed M-a distribution, the $\alpha_{\rm{turb}} = 10^{-3}$ population showed the most promising results  as it produced the widest variety of outcomes in terms of planet classes formed. At this setting, super Earths were formed over a range of orbital radii between $\sim$0.05-2 AU, and there was a separation between warm Jupiters at $a_p \sim$ 0.6-2 AU formed in the ice line, and hot Jupiters formed in the dead zone. Both higher and lower constant settings of $\alpha_{\rm{turb}}$ resulted in populations that were much more confined in terms of their M-a distributions. 

This result reinforces our argument that if the scatter in populations is significantly  linked to initial conditions for the disk population, then a single constant values of turbulence strengths do not well represent the physics of turbulence across a whole population of initial disks. We therefore carried out simulations by adding a  turbulence-strength parameter, $\alpha_{\rm{turb}}$, to our set of stochastically-varied disk parameters in our population synthesis calculations.   We chose an average value of $\alpha_{\rm{turb}} = 10^{-3} $ for the distribution that was sampled in the varied $\alpha_{\rm{turb}}$ population. It is interesting that this optimal value of $\alpha_{\rm{turb}}$ is in agreement with that used in our previous population synthesis works approached within a pure viscous disk framework. Perhaps this is a consequence of our normalization of the fiducial disk surface density $\Sigma_0$ and initial accretion rate $\dot{M}_0$, recalling that we set these initial disk parameters in accordance with our fiducial viscous disk from \citet{Alessi2020} in order to directly compare the disk models' results.

We found  a drastic improvement in the final planetary M-a distribution when a distribution of.$\alpha_{\rm{turb}}$ was used.  As different individual settings of $\alpha_{\rm{turb}}$ were shown to populate nearly distinct regions of the M-a diagram (i.e. figure \ref{Alpha_Compare}), sampling this parameter from a distribution introduced a significant source of scatter in the final planetary distribution. The ``varied $\alpha_{\rm{turb}}$'' population formed a substantial amount of each of the observed planetary classes: warm Jupiters, hot Jupiters, and super Earths. The super Earth population now extends to small orbital radii $\lesssim 0.03$ AU, a region of the M-a diagram that our previous models in the pure viscous framework did not populate. Additionally, the varied $\alpha_{\rm{turb}}$ population readily produces a separation between the hot and warm Jupiter populations, in line with occurrence rate studies \citep{Petigura2018}. 

 \subsection{Comparison with observed populations in M-a diagram} 
 
These results demonstrate that the relative strength of turbulence and disk winds could be one of the key features that shape the outcomes of planet formation and planetary populations. This is especially noteworthy as disk models frequently prescribe a single value for wind or viscous strengths.  Here we address the successes and caveats regarding comparison of our models with the data in the M-a diagram.

Although the current data is still somewhat limited, it is to be expected that disks  will exhibit a \emph{range} of turbulence and disk winds strengths.  One reason for this is that more massive disks with their higher column densities, are better able screen ionizing  X-rays over greater portions of the disk, which in turn will reduce MRI turbulence.  Thus there should be a correlation between MRI driven turbulence amplitudes and disk mass, as has been argued by \citet{Speedie2022}. Our results indicate that the distribution of their relative strengths will have crucial effects on planet formation.   Even within an individual disk, there is no theoretical basis for  having $\alpha$ parameters that are constant with disk radius or time. In fact, numerical MHD results show that the disk $\alpha$ parameters can vary within disks' radial and vertical extents in a non-trivial manner (i.e. \citealt{BaiStone2011, Lesur2014, Gressel2020}).  

Since all of our previous population synthesis investigations within the viscous regime adopted a constant value of $\alpha_{\rm{turb}} = 10^{-3}$ \citep{Alessi2018, Alessi2020}, this raises the important question of why the pure viscous models resulted in scatter in the final M-a distribution reasonably in line with the data, while constant $\alpha_{\rm{turb}}$ in this combined winds and turbulence model do not. We attribute this to the traps in this combined model being less sensitive to the initial disk parameters, most notably the initial disk surface density $\Sigma_0$. Since the location of the early stages of planet formation is set by the position of traps in the type-I migration regime, the traps' locations in large part shapes the radius distribution of the final planet population in our models. 

The main tension between our results and the M-a diagram distributions is the overpopulation of the Hot Jupiter region (by a factor of 2) and an underreprsentation of SuperEarths.  Specifically comparing populations shown in Figure 8b in this paper with the observational statistics quoted in  \citet{Alessi2020b} we find:  Hot Jupiters 23.6 \% vs 11.5 \%; Warm Jupiters 19.4 \% vs. 14.9 \%; and SuperEarths and Mini Neptunes 47.3 \% vs. 62.2 \%.    Clearly the overproduction of Hot Jupiters in our simulations is at the cost of the SuperEarth sector, as we have already indicated.   We  attribute this to efficient planet formation in the dead zone which is a result of the trapid migration of the trap to the inner disk where surface densities are high.  What are the caveats to this picture?  

One technical point is our neglect of dust evolution via radial drift, as we did in our pure viscous simulations in \citet{Alessi2020}.  In the absence of dust traps this process rapidly depletes solids outside the ice line with the consequence that solid accretion outside the ice line occurred at a very slow rate. 
 Formation in the dead zone trap is particularly interesting as planets experience delayed growth:  solid accretion only begins at an appreciable rate after the trap migrates inside the ice line. We have shown that this occurs in the new disk model after $\sim$ 1 Myr (recall figure \ref{Results1_Traps}). Radial drift would therefore have the effect of reducing the frequency of the dead zone trap's formation of hot Jupiters, with more planets ending up as super Earths and Neptunes due to this delayed growth effect. Otherwise, we do not expect radial drift to greatly affect our results of the varied $\alpha_{\rm{turb}}$ population.   As an example,  when dust evolution was included, the ice line was a prime location of efficient planet formation, resulting in a mix of super Earths and warm Jupiters. These are precisely the planet classes that are formed in this trap when radial drift is not included. 
 
A plausible reason for the noted differences in the relative size of our computed hot Jupiter and SuperEarth populations the observations is related to our neglect of possible  planet transfers during trap crossings - most notably where the dead zone trap crosses the ice line trap on its way to the inner part of the disk as its column density falls.    As previously noted, the ice line could intercept and capture the lower mass planets moving inwards with the dead zone trap.  The reason is that the corotation mass scale (which is the characteristic mass at which the outward co-rotation torque peaks) is higher in the active region beyond the dead zone - which the ice line enters for the first time.  While it is beyond this already overly long paper to compute this effect, we suggest that since the co-rotation mass scale at which the co-rotation torque peaks scales as
  $M_{corot} \propto \alpha^{2/3} h^{7/3}  $ \citep{Speedie2022}, lower mass planets moving with the dead zone would be trapped by the ice line as this crossing occurs.  Only the most massive planets on the dead zone trap would move inside the ice line.  This is likely to reduce the Hot Jupiter population substantially, because we have seen that the dead zone is the primary source for the hot Jupiter population.  At the same time, this would effectively increase the SuperEarth population particularly in the region out towards 1 AU where the ice line trap lingers.  We will need to compute this effect in another paper but suggest that this may be the most likely correction that would restore the correct balance between the SuperEarth population statistics.

 Planet formation near the heat transition trap is very inefficient due to its location outside of the ice line for the majority of disks' lifetimes.  Out there  radial drift has depleted solids. This causes the trap to form only failed cores, with the most massive planets only becoming super Earths. This is also quite similar to the heat transition's output in the varied $\alpha_{\rm{turb}}$ population, although the heat transition does produce an appreciable number of super Earths between 0.1-1 AU. We expect the formation frequency of super Earths from the heat transition would be reduced if dust evolution was included in our models.
 
 The relative amount of radiative versus viscous heating is important in understanding disk and planet evolution.   We recall that the position of all three of the traps depends on the midplane temperature structure - either directly in the case of the ice line and heat transition, or indirectly in the case of the dead zone outer edge.  Viscous heating of the disk is only contributed by MRI-turbulence, while the radiative heating profile is constant and independent of disk $\Sigma$ where radiative heating dominates.   As a result, only the fraction of disk evolution driven by turbulence translates to changes in viscous heating with $\Sigma$.   Ultimately this results in the trap positions being less sensitive to changes in $\Sigma$ in a population than they otherwise would be in a pure viscous model and hence to less scatter in the final M-a distribution.  This is is another physical reason why the variation of the $\alpha_{\rm{turb}}$ parameter is important and therefore plays an significant role in the properties of populations in the combined winds and turbulence disk picture.

The use of just single planet evolution in any disk leads to potential limitations in understanding observed populations.   Multiple planet can affect migration and accretion while still in the disk phase and planet-planet scattering effects will also play a role in the post disk phase of the evolution of a planetary system.  Multiple planet population forming in gas disks may enter into stable near mean-motion resonances with closer in planets which effectively reduces their rapid inward migration \citep{KleyNelson2012}.  Recent work has shown that such effects take place, but that even with multiple planet effects included, differences are not typically greater than factors of two in final orbital radii  \citep{Emsenhuber2021}.  
 After the gas is dispersed,  we first note that a single planet-planet scattering of equal mass planets in which one is removed from the system, would reduce the orbital radius by a factor of two or less \citep{Nagasawa2008}.  Outward scattering resulting from planet-planet interactions in the interior of the planetary system (within 10 AU) to distances of the order 100 AU would be difficult to achieve.   These and other dynamical effects arising from the inward scattering of planets by the Kozai effect need to be considered  \citep{Bryan2016} need to be considered, but our point remains that variable viscosities are important.  
 
In extending our current model into multiplanet simulations as performed by \citet{Emsenhuber2021}, we note that planets on traps could anchor resonant chains  with other forming planets.  The fact that the dead zone trap ultimately becomes becomes the innermost of the three traps that arise in disk interiors and that it would only trap the lowest mass planets may have important implications for the origin of densely packed planetary systems.

\subsection{Comparison with oberved populations in M- R diagram}

The varied $\alpha_{\rm{turb}}$ population also gave rise to a M-R distribution that, after accounting for post-disk atmospheric mass loss driven by photoevaporation, corresponds reasonably to the observed distribution. Atmospheric photoevaporation leads to a clear improvement in our population's M-R distribution over that which arises directly after the disk phase. In particular, a fraction of super Earths form in our models with quite large radii due to accreted gas, and lie above the observed data. Atmospheric photoevaporation reduces all of these planets' atmospheres to better correspond with the data. 

 We find that planets at masses $\lesssim$ 3 M$_\oplus$ have radii that compare well with the data, while super Earths and Neptunes at higher masses compare with the high-radius end of the observed data at a given mass.  This is similar to our previous work, \citet{Alessi2020b}, that investigated populations' M-R distributions in the pure viscous framework.  However, we do achieve a somewhat better M-R distribution than what we found in the pure viscous scenario due to the varied $\alpha_{\rm{turb}}$ population forming more super Earths at small orbital radii $\lesssim 0.1$ AU, which in turn leads to more atmospheric mass loss from photoevaporation. We therefore find a larger fraction of planets with masses $\gtrsim$ 3 M$_\oplus$ being partially or completely stripped, evolving their positions on the M-R diagram that better compare with average observed exoplanet radii for a given mass. 

However, a large fraction of super Earths do form at yet larger orbital radii where photoevaporation has a less-severe effect, and these planets remain at somewhat large radii and at the upper envelope of the observational data in the M-R diagram. As we found in \citet{Alessi2020b}, photoevaporation only has significant effects over a somewhat limited range of orbital radii $\lesssim$ 0.1 AU. This limits the mechanism's ability to improve our populations' M-R distribution, particularly at larger super Earth masses where planets in our model tend to accrete substantial atmospheres, leading to larger planetary radii than the bulk of the data. An additional means of atmospheric mass loss that has significant effects over a larger range of orbital radii would be advantageous for this purpose. This perhaps could be the case for core-powered atmospheric mass loss, however both photoevaporation and core-powered mass loss have only been investigated for planets at short orbital periods of less than 1 year \citep{Owen2013, Gupta2019}. 

\subsection{Comparison with other disk wind results}

One of the important aspects of disk winds for planet formation is that their efficient transport of disk angular momentum  drives the bulk of disk accretion flow so that high levels of turbulence are not required in contrast with the standard viscous disk picture.    Low levels of turbulence that provide minimal resistance to efficient dust settling to the disk midplane \citep{Hasegawa2017}.  However, as we have noted, our population results suggest that some levels of turbulence are likely present and contribute to the diversity of planetary populations.   A number of models of disk winds and planet formation have explored various regimes including a mix of turbulence and disk winds,  heavily mass loaded vs light disk winds, much lower levels of turbulence leading,, or no turbulence at all.   It is interesting to understand the differences that these regimes might have in the kind of populations that would arise.   

 First we explore in the Appendix A our own experiments with very low levels of $\alpha_{\rm{turb}}$ as well as cases in which very inefficient, heavy  disk winds dominate; $K \simeq 1$.   Using a value of $\alpha_{\rm{turb}} = 10^{-6}$ (see Appendix) as typical of very low levels of turbulence, we find that the disparity between the synthetic populations at constant $\alpha_{\rm{turb}}$ and the observed M-a distribution was exacerbated.  In this regime of extremely low turbulence, the planet formation tracks were entirely insensitive to disk parameters. This result  persisted regardless of whether we used the constrained disk outflow ($\dot{M}_{\rm{wind}}/\dot{M}_{\rm{acc}} \simeq$ 0.1; the fiducial setting), or relaxed this criterion to allow for a strong wind-driven outflow as is often prescribed for winds-driven disk models. High winds mass-loss rates resulted in a rapidly-decreasing disk surface density on timescales shorter than typical core accretion timescales, producing failed cores irrespective of disk surface density or lifetime. Our results indicate that winds-dominated scenarios alone have difficulties in explaining the diversity of planetary masses and orbital radii observed on the M-a diagram - if indeed this is one of the major sources of scatter. However, this pertains only to a trapped type-I migration scenario, and it may be the case that laminar disks do not require trapping as we have prescribed. 

The idea that planets can be prevented from undergoing rapid Type I migration has been the focus of many investigations for more than a decade  \citep{ HP11, Dittkrist2014, Bitsch2015, Coleman2016,  Bitsch2019, Speedie2022}.
In this type of approach where type-I migration torques are applied to directly determine migration of forming cores, the presence of traps (null torque locations) throughout the disk can cause cores to converge to the traps' locations.   Type I migration can be significantly reduced if a planet opens a gap in the disk, leading to Type II migration.   Recent work shows that the gap opening mass is given by the traditional analysis  \citep{LinPap1986, Rafikov2002} wherein the angular momentum flux of spiral waves excited by the planet is balanced by the viscous torque.  This has led to a new gap opening criterion for the ratio of planet mass to stellar mass, q that opens a gap:   $q \ge 5 \alpha^{1/2}h^{5/2} $ \citep{Kanagawa2018}.  For a low viscosity disk with $\alpha_{\rm{turb}} = 10^{-4}$ and an aspect ratio of $ h=0.05$ this means that a Neptunian mass can open a gap. Moreover TypeI migration is slowed because it is driven by the gas surface density in the gap rather than by the unperturbed disk.   Population synthesis studies suggest that this can be accomplished by a two - alpha disk model in which the bulk of the accretion flow is driven by the wind, leaving a low value of the turbulence  that greatly limits Type I migration -  \citep{Ida2018, Matsumura2021} .  This leads to good agreement with the population of warm Jupiters. 
 
Our simulations assume that planets as they become more massive in their traps will enter conditions in which they begin to open a gap and undergo Type II migrations.   The co-rotation torque for sufficiently massive planets will saturate somewhat below the gap opening mass scale resulting in the release of the planets from such traps \citep{Coleman2016, Speedie2022}.  These earlier studies were not informed by the recent results of \citet{Kanagawa2018} in which the mass scale at which gaps are opened and slower type II migration occurs is much lower - in the Neptune mass range for low viscosity disks.  This is about the mass scale at which planets are released from planet traps.  We find therefore that while our higher viscosity planet tracks may lead to greater incursions of planets to smaller disk radii (perhaps by factors of 2) that the lower viscosity tracks should remain accurate given these new results.

 It is possible that planets may be able to form in a winds-dominated disk without rapid migration into the host star, making the trapped type-I migration regime unnecessary in some systems.  A number of studies \citep{Ogihara2018, Suzuki2016, Chambers2019}  have shown that for the case of strong disk winds, the inner regions of the disk are essentially evacuated.   Due to this hollowing out of the inner disk, the peak of the disk column density distribution can occur out at 10 AU.  This peak acts essentially as a planet trap, and it effectively reduces  rapid inwards Type I migration from occurring.   As noted in the Introduction, it is far from clear that disk wind mass loss rates are comparable to accretion rates in general.   Were this to be generally true, one must wonder how star formation would occur in such disks when at least half of the material moving through the disk would be ejected.   Star formation is inefficient since only about a third of the gas in a gaseous core actually accretes onto a star \citep{Andre2014}.  Models have shown that the protostellar outflow can drive off the bulk of the accreting envelope \citep{Matzner2000}.    

This brings us to the case of disks which evolve solely by disk wind torques.  In the limit of vanishing turbulence flows within disks are laminar.   Dynamical instabilities such as the formation of vortices by Rossby wave instabilities are no longer suppressed by viscous damping effects and can therefore play a role in co-rotation torques.   Detailed treatments of co-rotation torque  in wind-evolving disks  (\citealt{McNally2018, McNally2020, Kimmig2020})  have focused on the effects of winds in the immediate co-rotation region near the planet.    Disk winds with high enough mass loss rates can drive fast accretion flows which result in outward directed co-rotation torques \citep{Kimmig2020, Speedie2022}.  As one moves from 2D to 3D, simulations reveal the surprising importance of hydrodynamical buoyancy instabilities that do not occur in 2D disks.  These lead to a strong inward directed torque on the planet \citep{McNally2020}.  Detailed 3D MHD simulations are still needed to further explore this effect and what consequences it might have for planetary populations.    We showed that planet migration is inwards at all disk radii in the standard, constrained outflow case using the standard \citet{Paardekooper2010} type-I torque formula (figure \ref{Results2_Torque}), so it is unclear how cores could avoid encountering traps in our framework. The unconstrained outflow case does show that a null torque exists and the direction of migration is outwards in the inner disk.    This is counteracted by the rapidly decreasing surface density of the disk under the action of the assumed very strong wind, offering little time for core accretion to take place.

 \subsection{Conclusions} 
This investigation has focused on the relative importance of disk turbulence and winds as well as the initial distributions of disk properties and turbulence strengths, in defining planetary populations and their chemical compositions.   All of our thousands of evolution tracks are based on single planet formation within disks and we have discussed the likely effects of multi-planet interactions.  Our principle conclusions are as follows:
 
\begin{itemize}
\item Our best comparison with the observed M-a distribution arises from populations that incorporate a range of relative strengths of turbulence and disk winds. When varying the $\alpha_{\rm{turb}}$ parameter that controls the strength of turbulence in a population, more scatter is introduced into the final M-a distribution that significantly improves correspondence with the data compared to populations that consider constant settings of this disk parameter.
\item Most notably, this varied $\alpha_{\rm{turb}}$ population forms a large amount of super Earths across a large orbital radius extent: 0.01 - 2 AU. This result of the combined winds and turbulence disk model is therefore an improvement over populations arising from the pure viscous scenario that we investigated in previous works, as the combined model directly produces super Earths at small $a_p < 0.1$ AU from the disk phase, whereas the pure turbulent models rarely formed super Earths at these small orbital radii. Additionally, the super Earth region of the M-a distribution was populated by each of the three planet traps in our model, which leads to a diverse range of planetary compositions from dry-rocky super Earths to ice rich 
planets.
\item Using this varied $\alpha_{\rm{turb}}$ model within the combined winds and turbulent disk framework, we obtain a reasonable comparison to the observed M-R distribution within the super Earth-Neptune mass regime. Solid-dominated planets with masses $\lesssim$ 2-3 M$_\oplus$ compare well with the data. Planets at larger masses either retain their accreted atmospheres and correspond with observed planets with largest radii at a given mass, or are stripped via photoevaporation, reducing their radii to compare with average observed planet radii at the corresponding mass.
\item Post-disk atmospheric photoevaporation has a critical role in reducing super Earth radii and improving our comparison with the observed M-R distribution. Compared to the population arising from the pure turbulent disk of \citet{Alessi2020b}, a larger number of super Earths form at smaller orbital radii in this model, and therefore more planets are affected by photoevaporation. This increases the number of stripped planets at masses greater than 3 M$_\oplus$, thereby improving our comparison with the M-R data compared to that of the pure turbulent disk.
\item The varied $\alpha_{\rm{turb}}$ population also produces a separation between warm gas giants near 1 AU formed in the ice line and hot Jupiters formed in the dead zone. Each trap leaves different and somewhat distinct imprints on the M-a diagram. The dead zone efficiently forms hot Jupiters with $a_p \lesssim$ 0.1 AU as well as some Neptunes with $a_p \lesssim$ 0.5 AU.  The ice line forms many warm gas giants, as well as super Earths and Neptunes mostly in the 0.5-2 AU range.  The heat transition contributes a substantial amount of super Earths between 0.1-0.6 AU, with a tail of sub-Earth mass cores extending to larger orbital radii that fall below observational limits. 
\item When considering populations with constant settings of the disk turbulence parameter, a setting of $\alpha_{\rm{turb}} = 10^{-3}$ resulted in a population with the largest range of final planet properties and scatter on the M-a diagram compared to lower (10$^{-4}$) and higher (0.007) settings of this parameter, that resulted in populations that covered very limited regions of the M-a space. Planet formation in disks with $\alpha_{\rm{turb}} \simeq 10^{-3}$ is therefore most sensitive to disk parameters compared to lower and higher settings. However, all of the constant $\alpha_{\rm{turb}}$ models failed to produce populations with sufficient scatter in their M-a distribution to compare reasonably with the data, and populations the varied the turbulent strength were necessary to improve this comparison.
\item Winds-dominated models did not show variation in planet formation results.   They fail to reproduce the observed  variation in the M-a distribution.
\end{itemize}

Our results clearly indicate that disk evolution and planet formation are sensitive to the strengths of these two disk evolution mechanisms.  They strongly reinforce the idea that disk winds provide the dominant means of angular momentum loss for most of the initial disk populations.   Turbulence is still expected to be present and important and is likely produced by weaker dynamical instabilities that are left after MRI is suppressed.  A distribution of turbulence strengths naturally creates a greater variety of outcomes for planet populations.  We emphasize that planet traps continue to play a vital role in defining planet populations.   Disk winds should in general be highly efficient in carrying off disk angular momentum so that the hollowing out of the inner disk is not necessary to produce a population of warm Jupiters trapped at a resulting column density peak out at several AU.  The major caveat in our calculations is that we overproduce hot Jupiters at the expense of SuperEarths.   This can be remedied by including the stripping of lower mass planets migrating with the inward moving dead zone as it crosses the ice line at a few AU.    These results motivate further phyiscal modeling of disk evolution processes, particularly in 3D MHD simulation of the co-rotation region very near to accreting and migrating forming planets.   This is a major topic of current research to which we intend to make further contributions. 

\section*{Acknowledgements}
We thank the anonymous referee for an insightful and very useful report that improved the manuscript.  We also thank Jessica Speedie, Alex Cridland, Nienke van der Marel, and James Wadsley for many stimulating discussions.  M.A. acknowledges funding from the National Sciences and engineering Research Council (NSERC) through the Alexander Graham Bell CGS/PGS Doctoral Scholarship and from an Ontario graduate scholarship. R.E.P. is supported by an NSERC Discovery Grant. This work made use of Compute/Caclul Canada. 


\section*{Data availability}
The data underlying this article will be shared on reasonable request to the corresponding author.


\bibliographystyle{mnras}
\bibliography{research}

\appendix

\section{Wind-dominated models \& the effect of outflow strength} \label{Results4_2}

We now consider the case of a winds-dominated disk model with $\alpha_{\rm{turb}}=10^{-6}$, which is same value used within the combined model's dead zone. We will examine the effect of the strength of the wind-driven outflow by comparing two models: the constrained case where $K$ is solved for using equation \ref{Outflow_Constraint}, and the unconstrained case where we use a high setting of $K=1$. In this disk model we do not consider a dead zone as the turbulence strength is quite low throughout the entire disk's extent. The related trap at the Ohmic dead zone's outer edge is therefore not present in these models. 

\begin{figure*}
\centering
\includegraphics[width = 0.45 \textwidth]{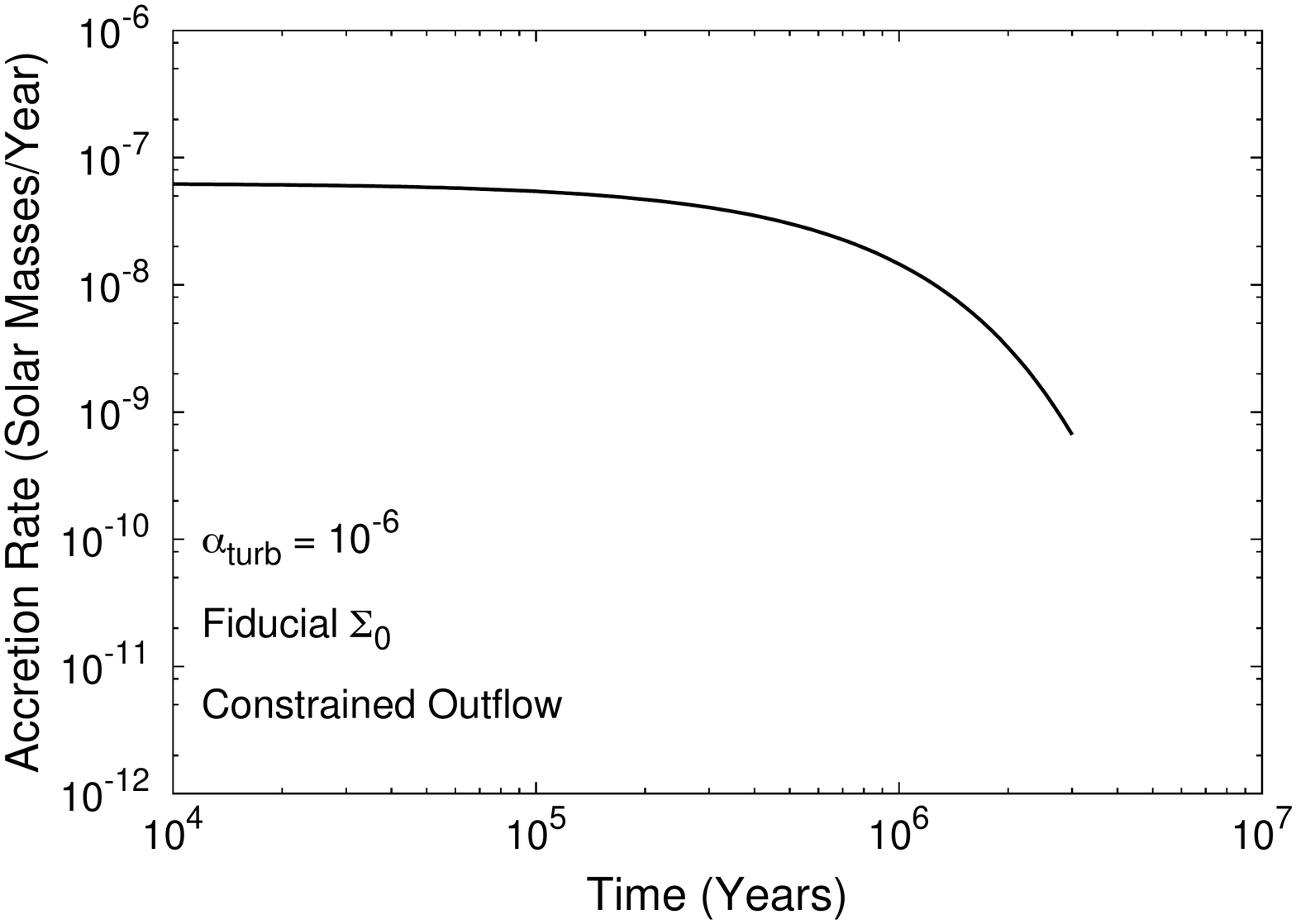} \includegraphics[width = 0.45 \textwidth]{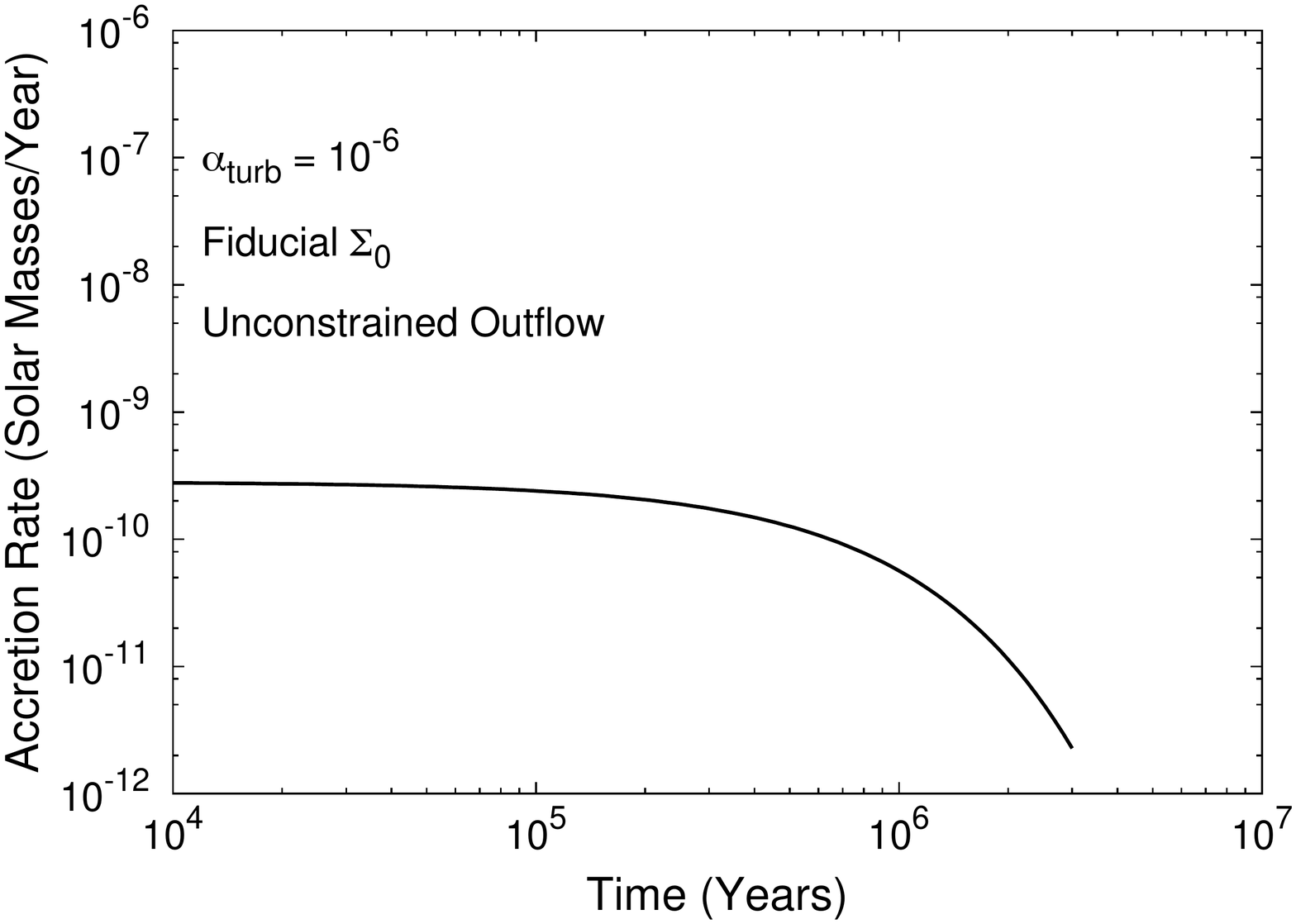} \\
\includegraphics[width = 0.45 \textwidth]{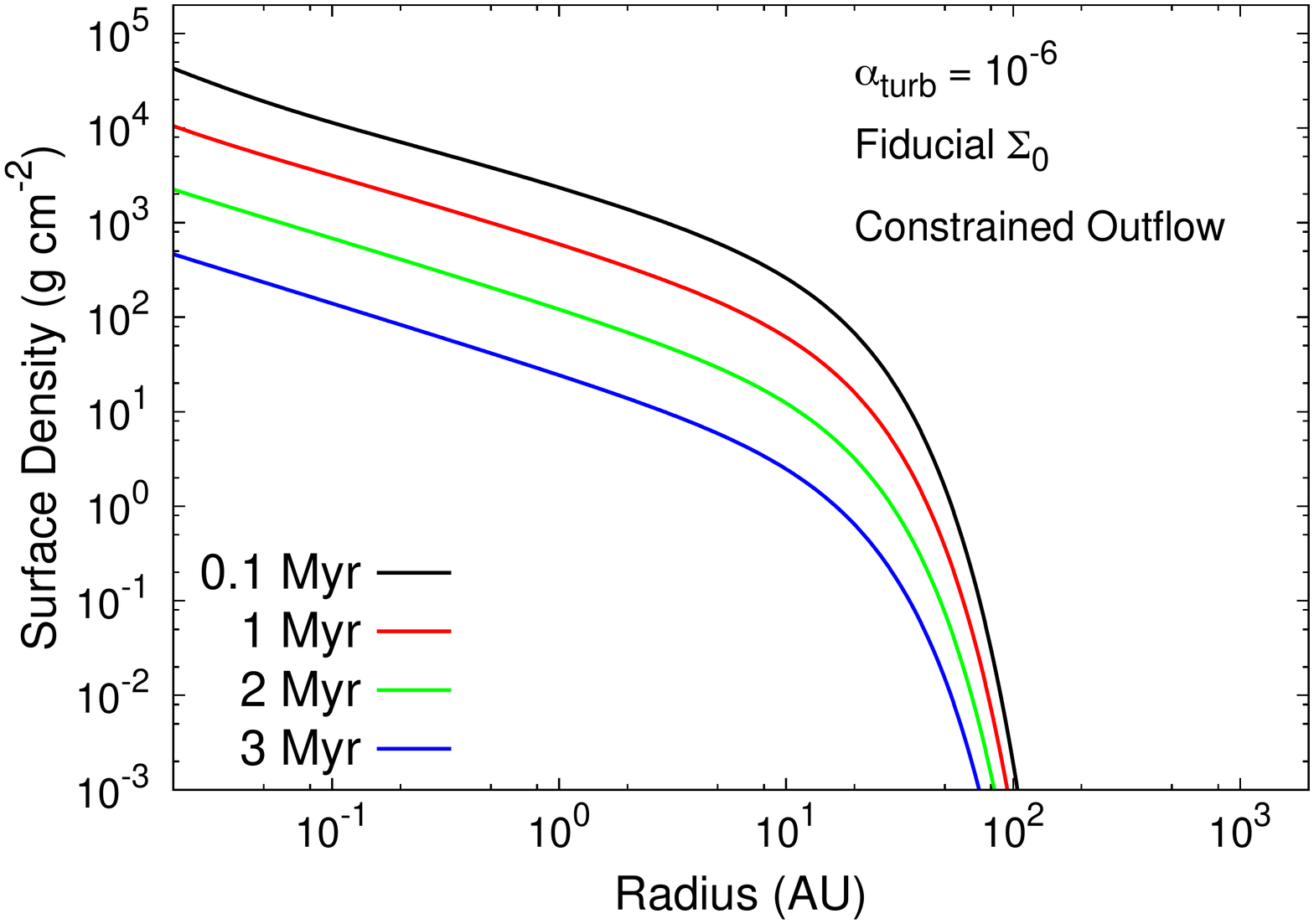} \includegraphics[width = 0.45 \textwidth]{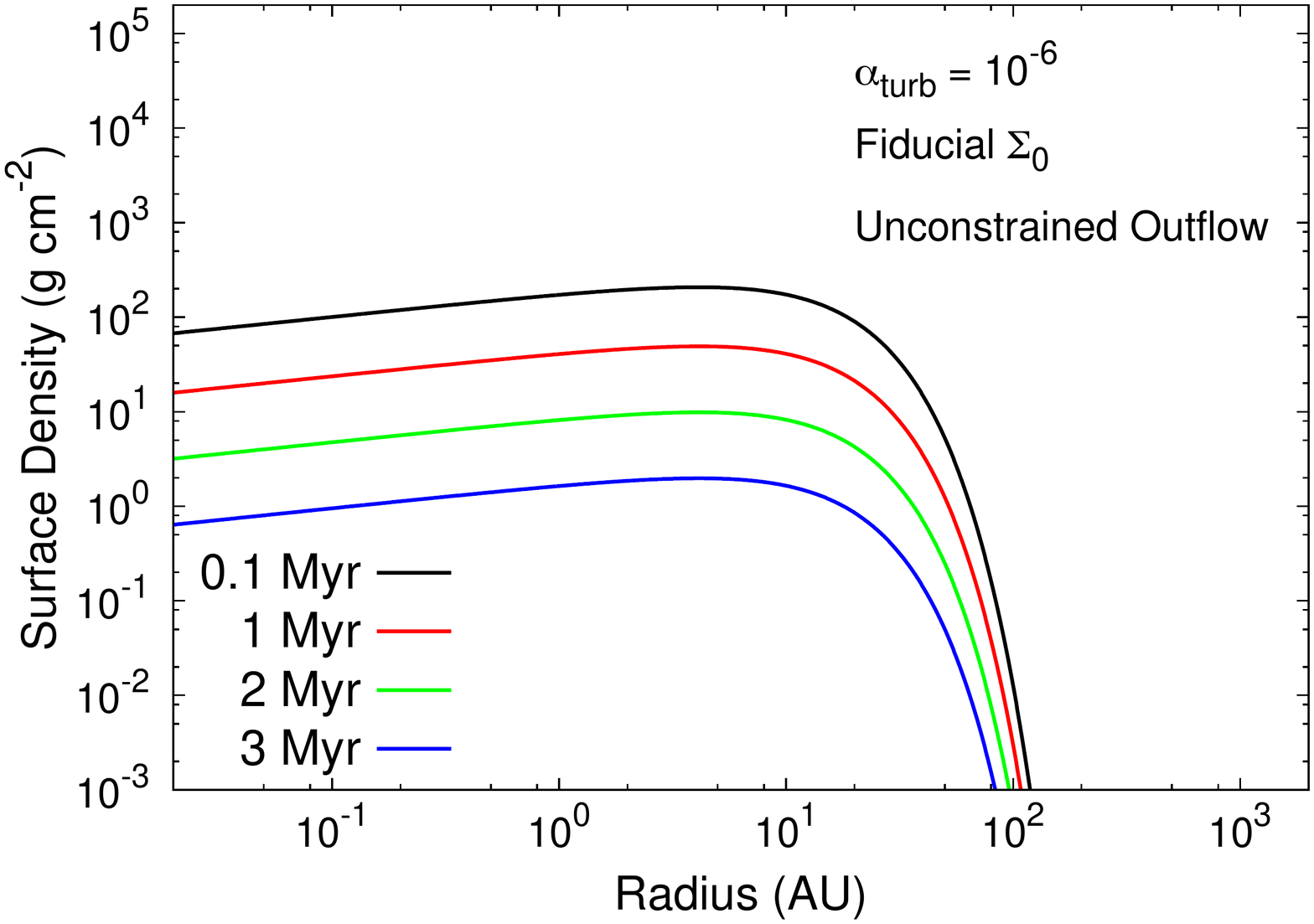} \\
\includegraphics[width = 0.45 \textwidth]{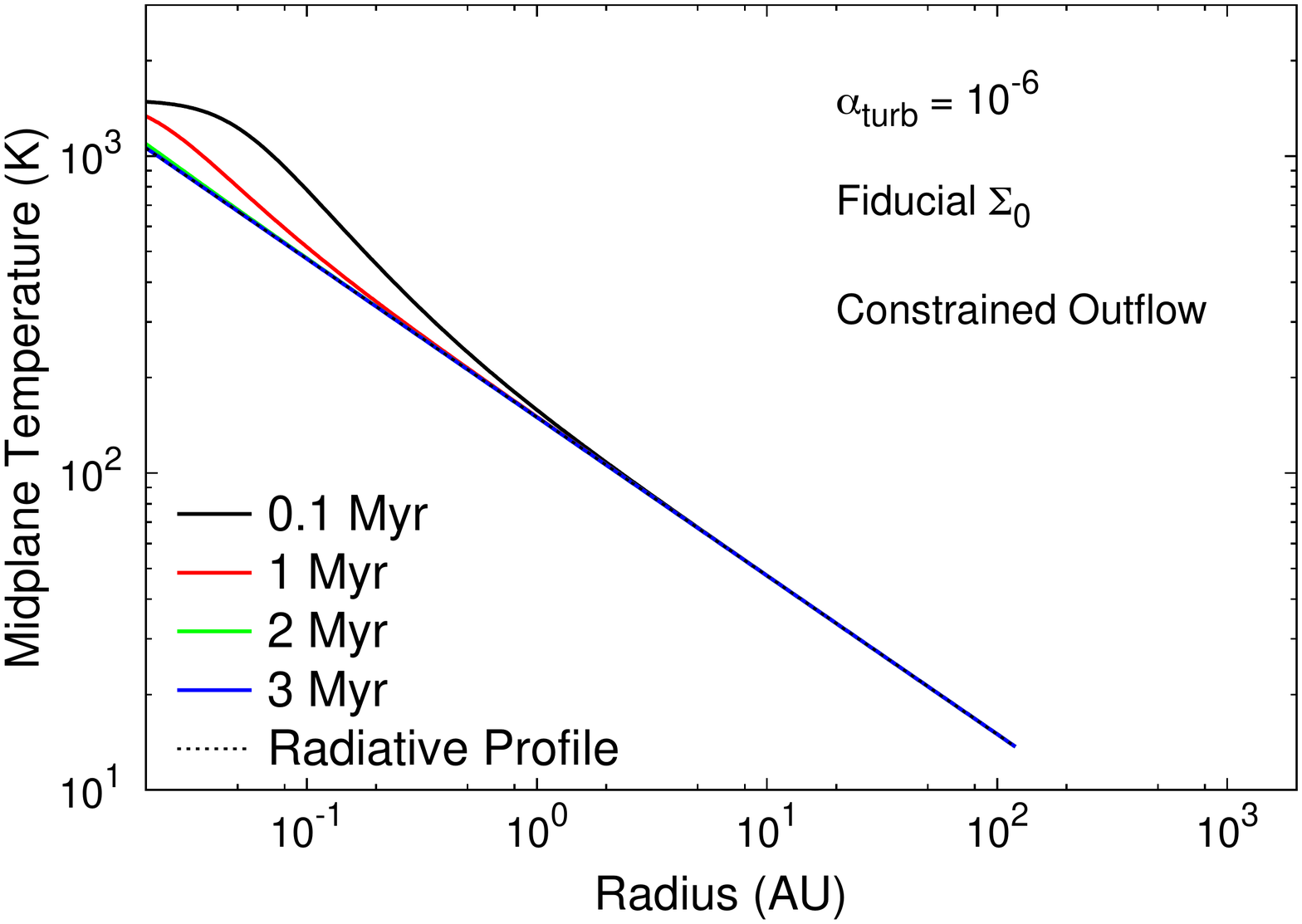} \includegraphics[width = 0.45 \textwidth]{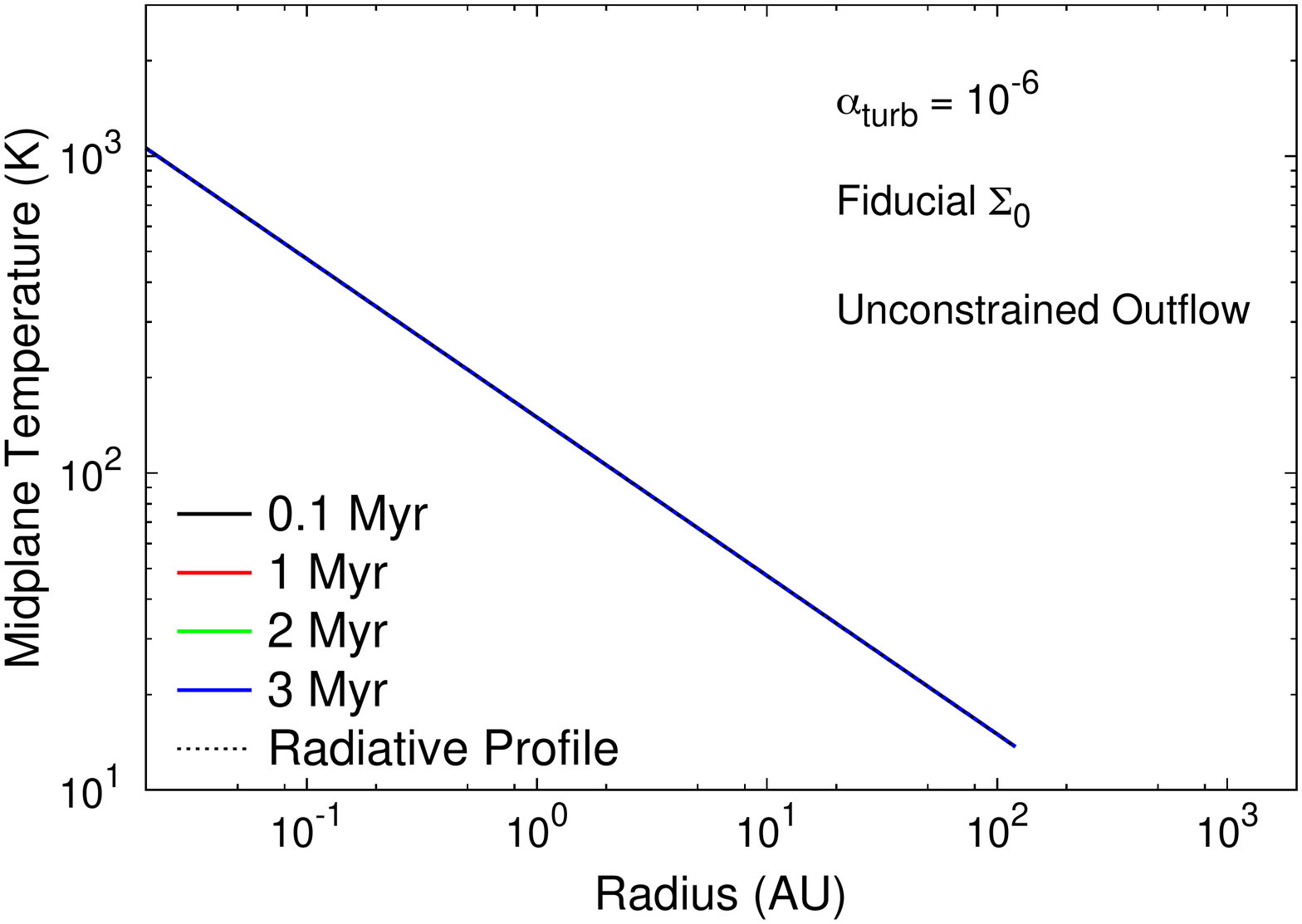}
\caption[Comparing disk evolution at two settings of outflow strength in winds-dominated disks]{We compare disk evolution in winds-dominated disks ($\alpha_{\rm{turb}}$ = 10$^{-6}$) at two settings of disk outflow strength. The left column pertains to the `constrained' outflow where $K$ is determined according to equation \ref{Outflow_Constraint}. The right column considers a high outflow strength with $K$=1. The top row shows evolution of disk accretion rate, and middle and bottom rows show radial profiles of surface density and midplane temperature, respectively.}
\label{Results2_Disk}
\end{figure*}

In figure \ref{Results2_Disk}, we plot the resulting disk evolution corresponding to each scenario. We find that, even though the strength of turbulence has been reduced by a factor of 100 compared to the combined model of the previous section, the ``constrained outflow'' model's accretion rate and surface density evolution is quite comparable to the $\alpha_{\rm{turb}}=10^{-4}$ case. The accretion rate decreases from its initial value of 6$\times10^{-8}$ M$_\odot$ yr$^{-1}$ to roughly $10^{-9}$ M$_\odot$ yr$^{-1}$ after 3 Myr of disk evolution. The surface density profiles also show the disk to contract with time in terms of its outer radius; a result of winds-dominated evolution. 

\begin{figure*}
\centering
\includegraphics[width = 0.45 \textwidth]{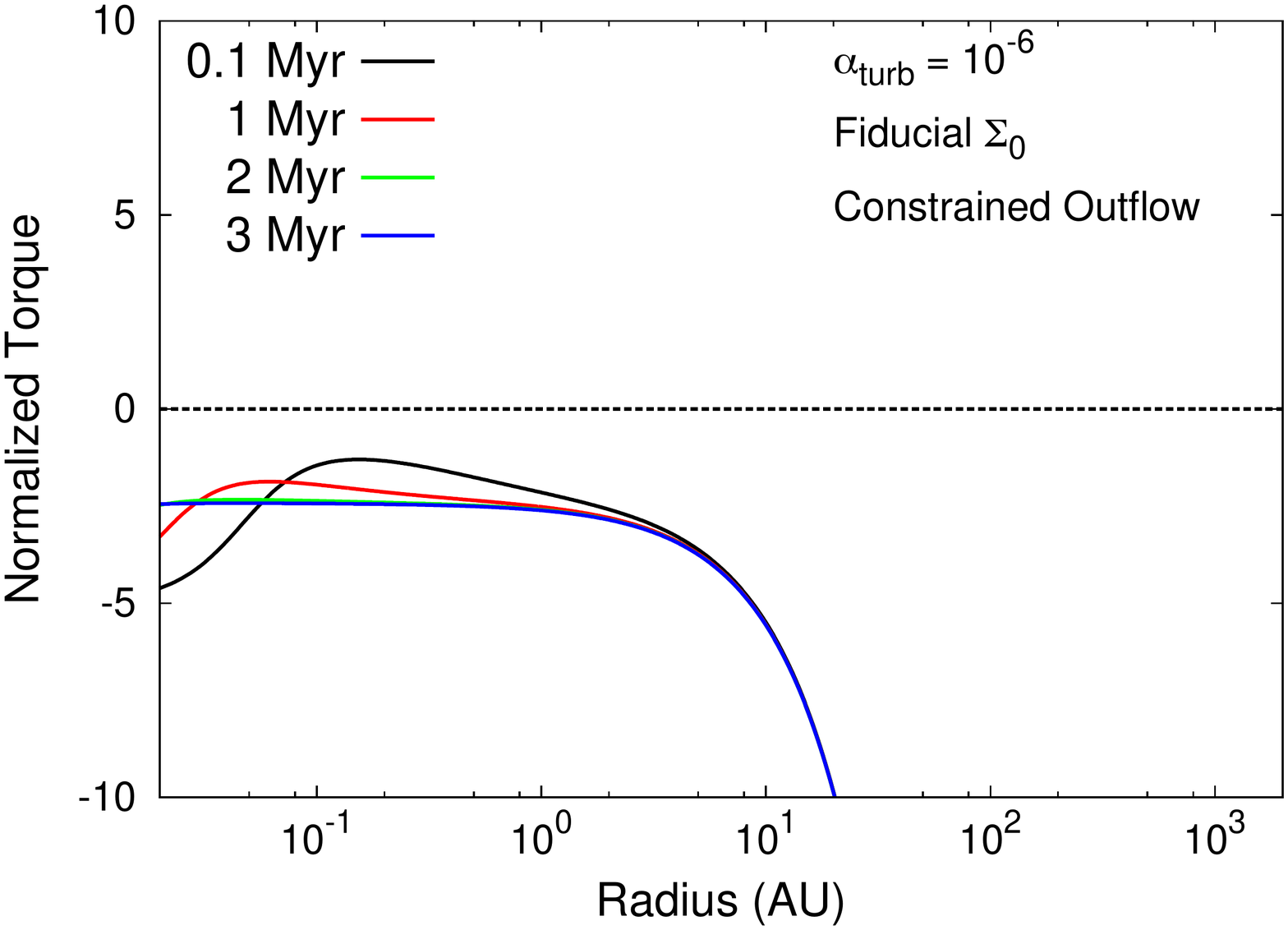} \includegraphics[width = 0.45 \textwidth]{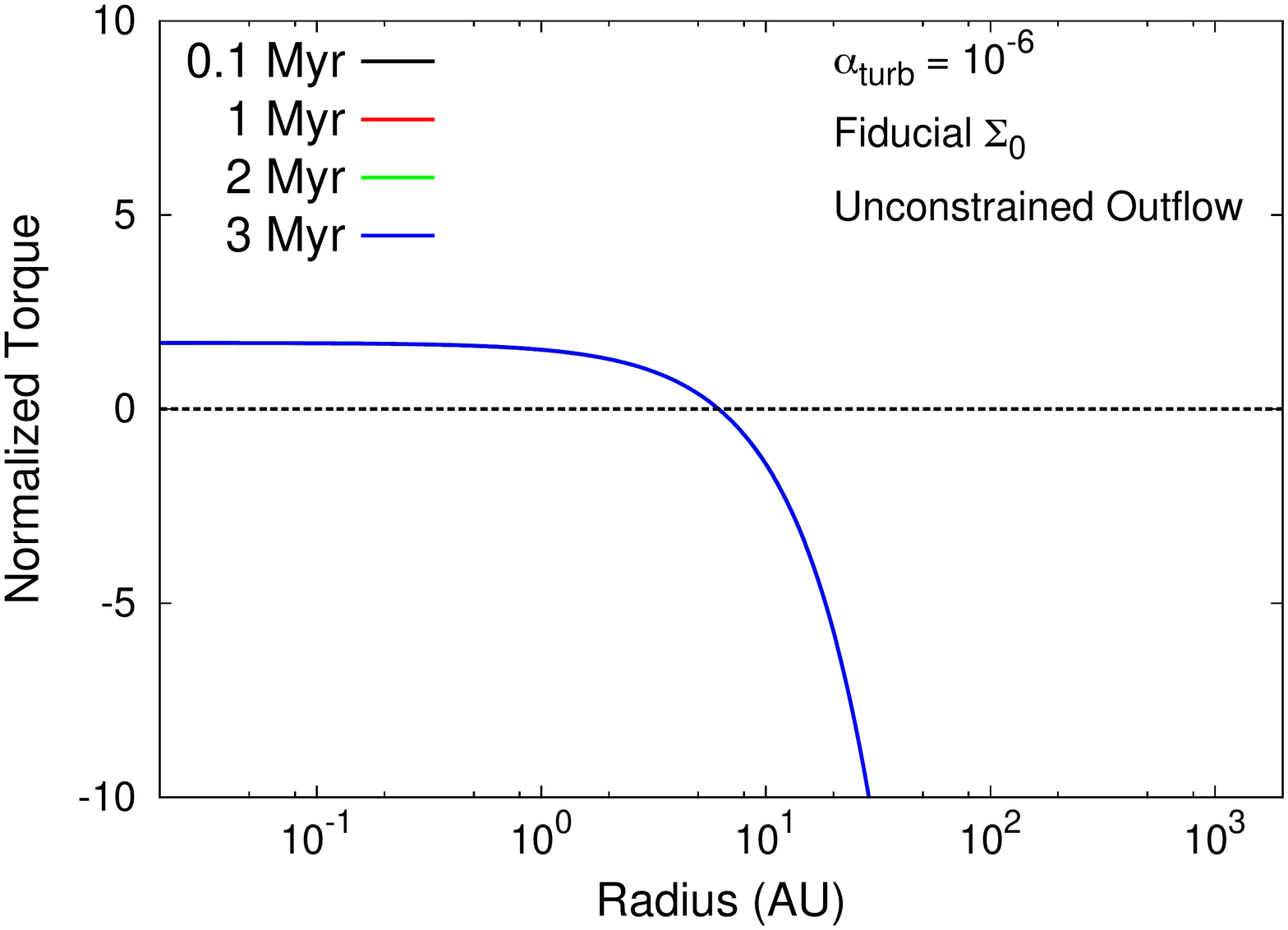} 
\caption[Type-I migration torques; comparing two settings of outflow strength in winds-dominated disks]{Profiles of the normalized type-I migration torque $\Gamma_{\rm{I}}/\Gamma_0$ are shown for both the constrained and unconstrained outflow models, calculated using equation \ref{Results4_Paardekooper}. Inward type-I migration will persist across the entire `constrained outflow' disk's extent. Conversely, we find a planet trap near 10 AU related to the surface density maximum in the `unconstrained outflow' model.}
\label{Results2_Torque}
\end{figure*}

The constrained outflow model's temperature profile is, however, significantly different from the $\alpha_{\rm{turb}}=10^{-4}$ case, as viscous heating has been substantially decreased due to the reduced strength of turbulence. This decrease results in the disk's viscous region being limited to the innermost region within 1 AU and early times in disk evolution $\lesssim$ 1 Myr. Between 1 and 2 Myr, viscous heating has decreased to the point where the entire disk's heating is dominated by host-stellar radiation. Once this has occurred, the midplane temperature profiles simply correspond to the radiative equilibrium profile (equation \ref{Results4_REQ}). As a consequence, the midplane temperature profile will become completely static, since the radiative temperature profile has no time-dependence. This will result in the ice line's position being stationary in the disk throughout its evolution. Furthermore, if disk chemistry were to be computed in this disk, the profiles would not evolve with time after the viscously-heated region has disappeared. We note that these implications of a static, radiative-equilibrium temperature profile are a result of the common assumption that we have used throughout this thesis that the host-star's luminosity is constant.

Now considering the unconstrained outflow scenario in the right column of figure \ref{Results2_Disk}, we can immediately see the interesting effect of a strong, winds-driven outflow on disk evolution. In this model, a larger fraction of the disk winds' stress is carried away in the outflow as opposed to contributing to disk accretion. This results in the disk accretion rate $\dot{M}$ being reduced by roughly two orders of magnitude compared to the constrained outflow scenario. 

Another compelling effect of a strong outflow can be readily seen in the disk surface density profiles. Rather than being a decreasing function of disk radius over the entire disk's extent, we see that $\Sigma$ increases until roughly 10 AU where it achieves a maximum value before decreasing in the outer disk. This is a result of of the outflow surface density rate being a decreasing function of radius $r$ (see equation \ref{Outflow_SD}), removing material more efficiently in the inner disk than the outer disk. This feature is also present in many of the surface profiles shown in the numerical treatment of \citet{Suzuki2016}. 

Since, in addition to having a low setting of turbulence, the unconstrained outflow model's accretion is substantially lower, there is no viscously-heated region in the disk. At all times, the midplane temperature is a result of heating through radiation alone, and the midplane temperature profiles are equal to $T_{\rm{req}}$ (equation \ref{Results4_REQ}) for the entirety of disk evolution. There are two consequences of this result. First, the heat transition and its related trap are not present in this model, as there is no transition into a viscously-heated region present in the disk. Second, as we have described above, the midplane temperature profile is completely static, which will result in an ice line radius that does not change as the disk evolves.

To summarize, since we are considering an extremely low setting of $\alpha_{\rm{turb}}$, there is no longer a dead zone or its related trap in either winds-dominated model. Additionally, in the unconstrained outflow case, there will be no heat transition, with only a static ice line remaining from our standard set of three planet traps. Both the heat transition and ice line will remain in the constrained outflow scenario. 

\begin{figure*}
\centering
\includegraphics[width = 0.45 \textwidth]{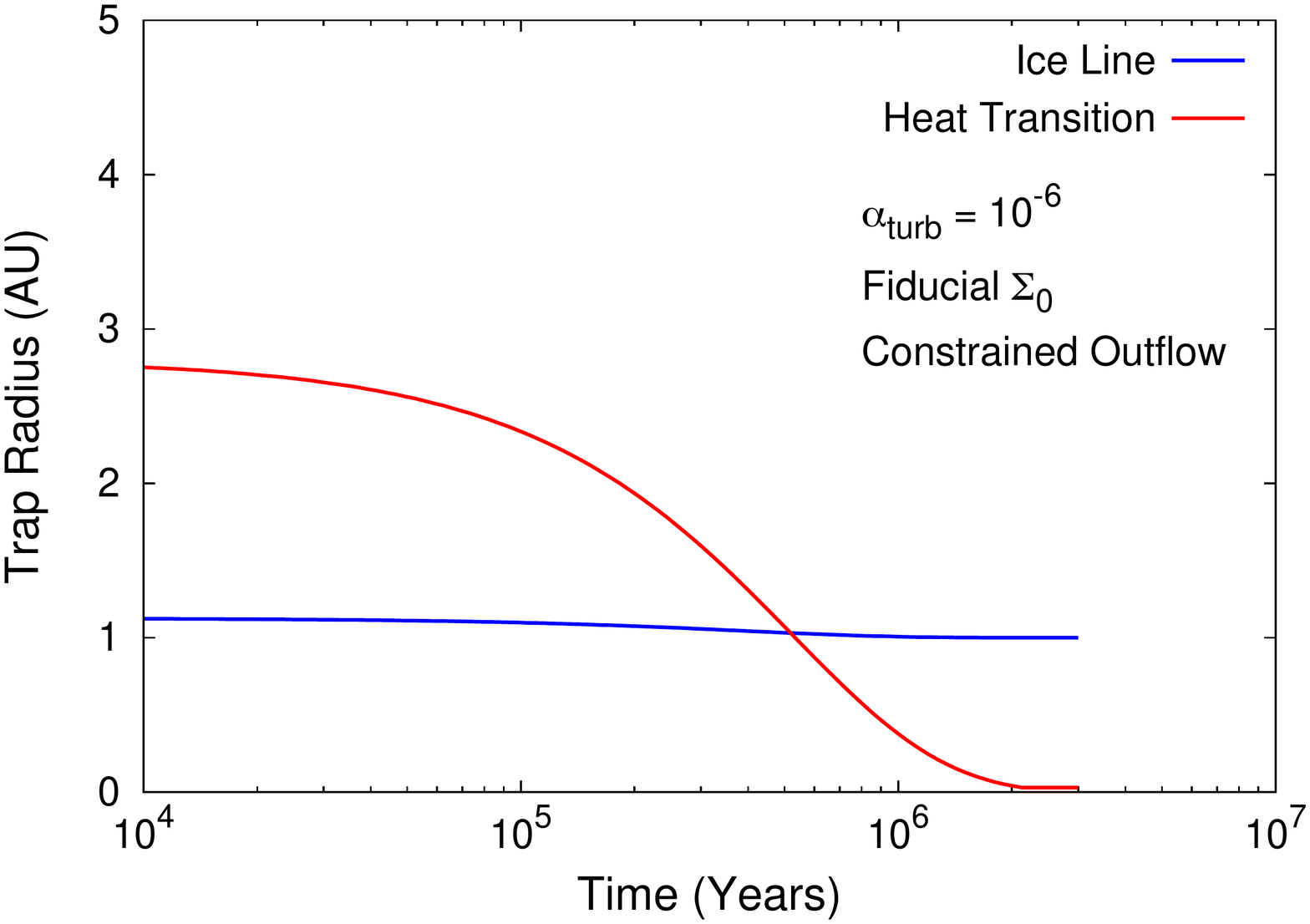} \includegraphics[width = 0.45 \textwidth]{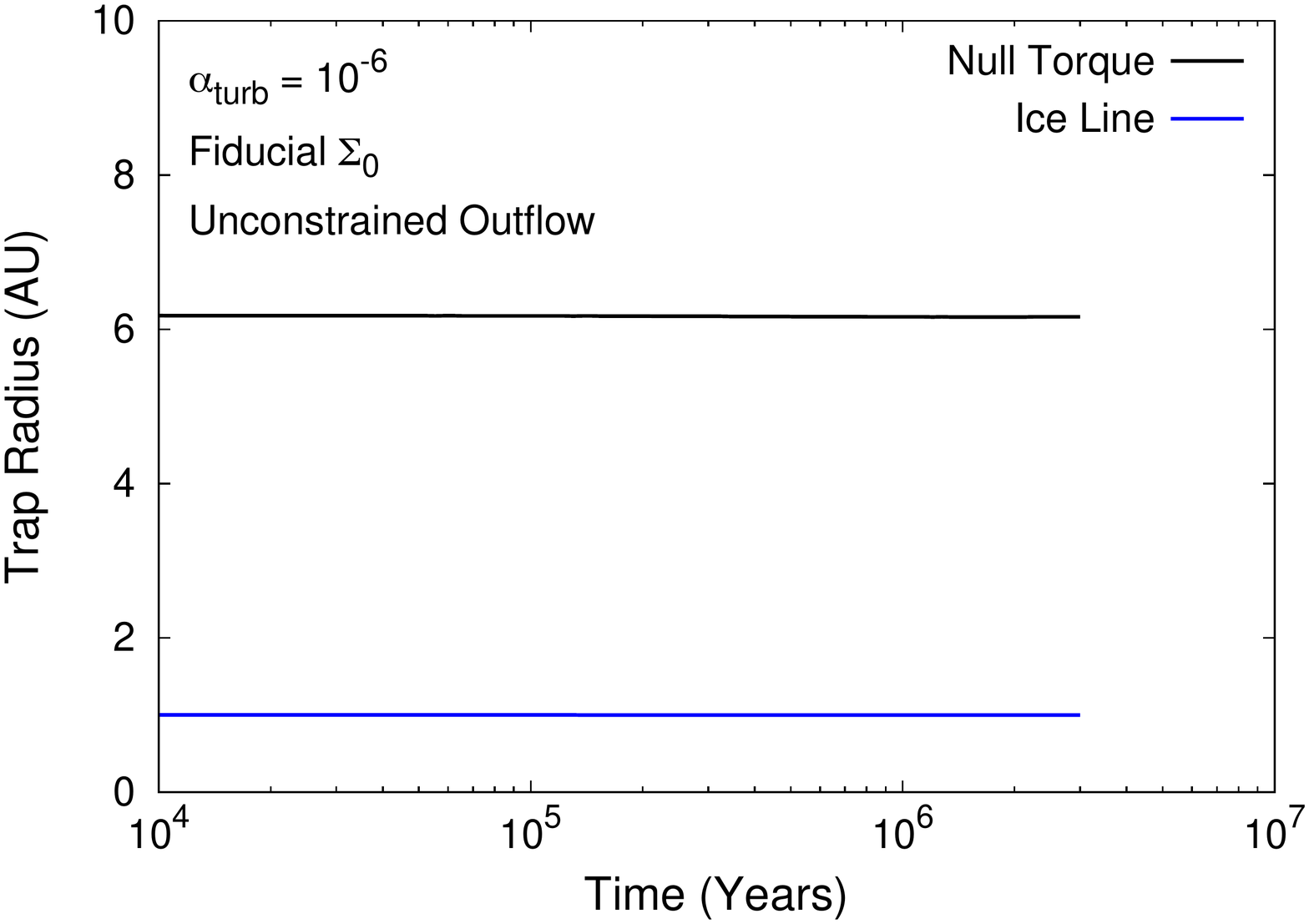} \\
\includegraphics[width = 0.45 \textwidth]{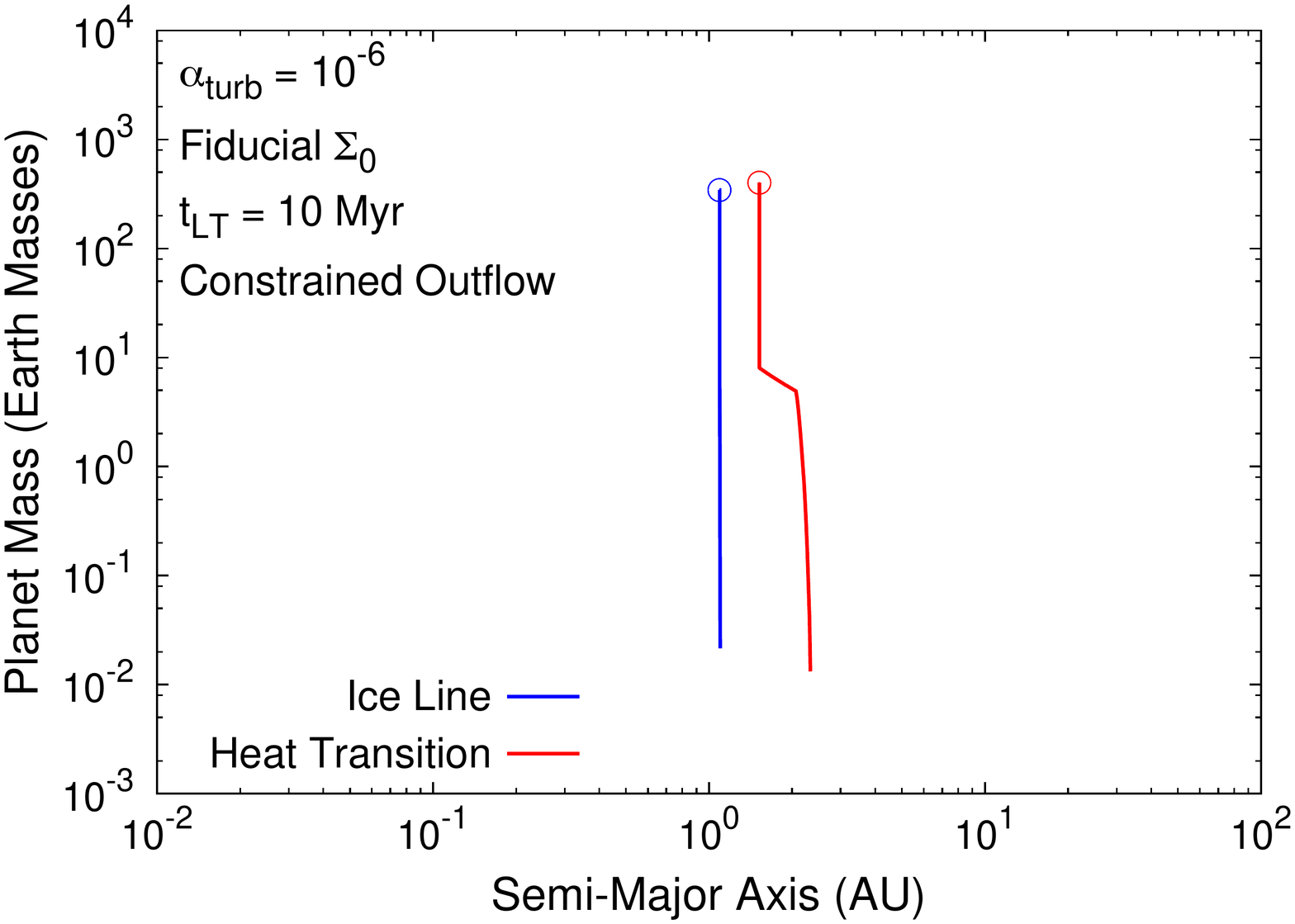} \includegraphics[width = 0.45 \textwidth]{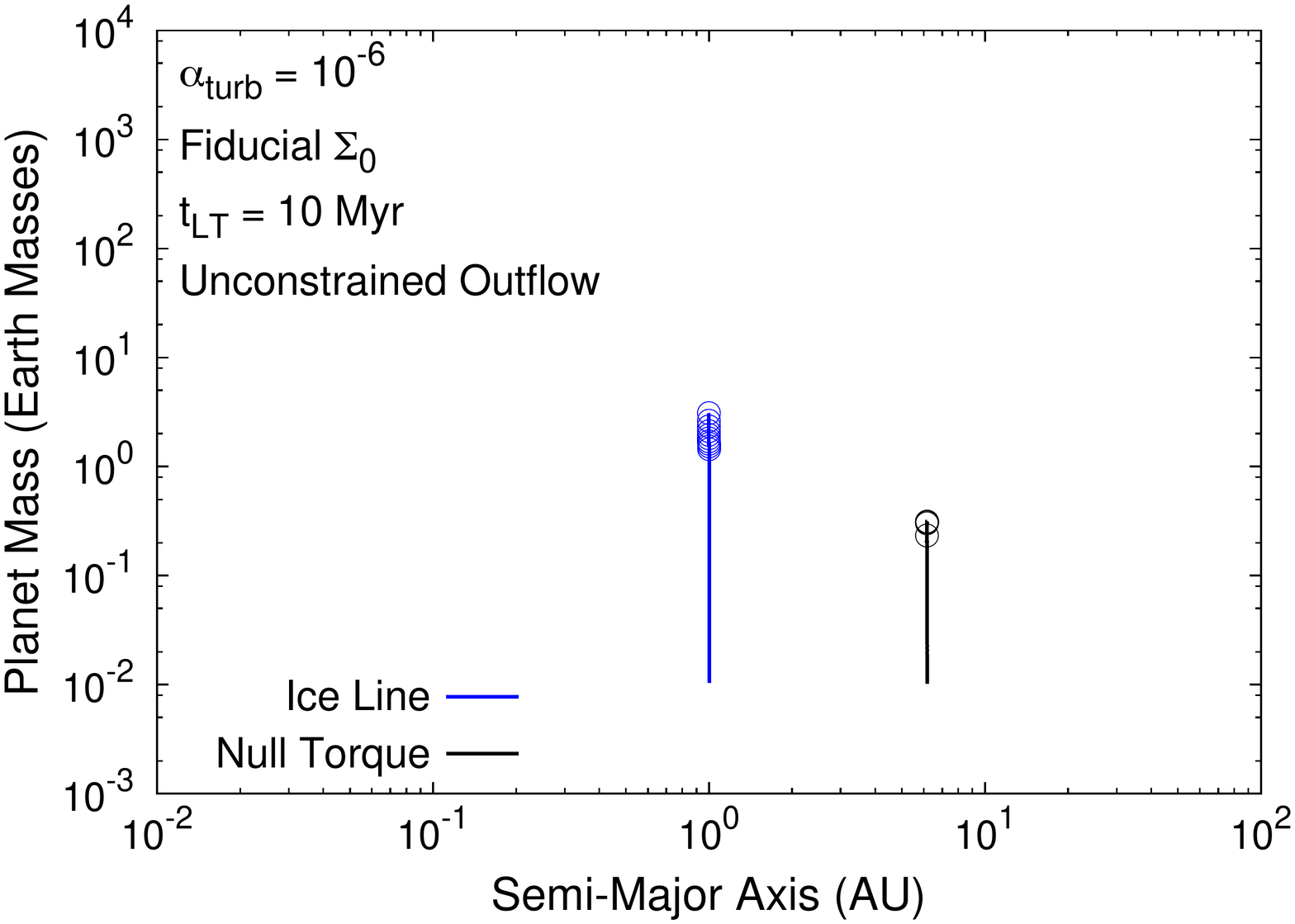} 
\caption[Planet traps' evolution and planet formation tracks; comparing two settings of outflow strength in winds-dominated disks]{We plot evolution of planet traps (top row) and planet formation tracks (bottom row) for both the constrained outflow (left column) and unconstrained outflow (right column) models. In the planet formation plots, open circles designate the planets' positions at 1 Myr intervals throughout their disks' 10 Myr-lifetimes.}
\label{Results2_Traps}
\end{figure*}

However, the surface density maximum that is encountered in the unconstrained outflow model has significant implications for planet migration. Type-I migration models have shown that forming cores will experience zero net torque at locations of surface density maxima (i.e. \citet{Sandor2011}). In fact, this is our main motivation for analyzing the unconstrained outflow model, despite the fact that it produces large outflow mass-loss rates that are in contention with disk winds theory and observations (i.e. equation \ref{Outflow_Constraint}).  

We adopt the standard type-I migration torque formula of \citet{Paardekooper2010} to determine if a null torque location (i.e. a planet trap) will exist in either model;
\begin{equation} \gamma \frac{\Gamma_{\rm{I}}}{\Gamma_0} = -2.5 -1.7 \beta + 0.1 \alpha +1.1(1.5-\alpha) +7.9 \frac{\xi}{\gamma} \;,\label{Results4_Paardekooper} \end{equation}
where $\Gamma_{\rm{I}}$ is the normalized type-I migration torque whose sign indicates the direction of planet migration ($\Gamma_{\rm{I}}<0$ indicating inward migration). $\alpha$ and $\beta$ correspond to the magnitudes of the local power-law indices of the surface density, and temperature profiles, respectively (i.e. $\Sigma \sim r^{-\alpha}$, $T\sim r^{-\beta}$). This torque formula illustrates that the type-I migration torque is sensitive to the local disk surface density and temperature profiles. We use an adiabatic index of $\gamma=5/3$ corresponding to a monatomic gas. $\xi$ is the disk's entropy gradient; $\xi = \beta - (\gamma-1)\alpha$. Lastly, the torque normalization factor $\Gamma_0$ is, 
\begin{equation} \Gamma_0 = \left(\frac{q}{h}\right)^2\Sigma_pr_p^4\Omega_p^2\;, \label{Results4_TorqueNorm} \end{equation}
where $q$ is the planet to star mass ratio, $h$ is the disk aspect ratio, and all quantities are computed at the planet's location. 

We will, however, only calculate the \emph{normalized} type-I migration torque $\Gamma_{\rm{I}}/\Gamma_0$ using equation \ref{Results4_Paardekooper}, to determine if any planet traps exist in either the constrained or unconstrained outflow models. These will correspond to locations where $\Gamma_{\rm{I}}$ = 0. We note that, since we are calculating the \emph{normalized} torque, we do not need to prescribe a core mass (which would be needed to determine the torque normalization, equation \ref{Results4_TorqueNorm}). 

The result of this calculation is shown in figure \ref{Results2_Torque}, where radial profiles of the normalized torque are shown throughout both disks' evolutions. We see that in the constrained outflow model, the type-I migration direction will be inward across the entire disk's extent, and for the entire disk's evolution. The constrained model's surface density profiles are similar to the `combined' turbulence and winds model of the previous section, so we expect this result to apply to that disk model as well. 

In the case of the unconstrained outflow model, we do indeed find a planet trap near 6 AU where $\Gamma_{\rm{I}}/\Gamma_0$ = 0, related to the maximum in the disk surface density profile. Interior to the trap, the direction of the type-I torque is outward, while exterior to the trap it is inward. The directions of the type-I migration torque are therefore appropriated to migrate planets towards the trap. We also find that the torque profiles in the unconstrained outflow scenario are time-independent. In addition to the static temperature profile, the radial profile of the surface density's power-law index $\alpha$ is also time-independent, which causes the static torque profiles. This can be seen in figure \ref{Results2_Disk}, where the unconstrained outflow model's $\Sigma$ profile decreases while maintaining its radial structure.

Based on this result of figure \ref{Results2_Torque}, we obtain a second planet trap in the unconstrained outflow case, in addition to the ice line. We refer to this trap simply as the ``null torque'' for this model. However, since the torque profiles and null torque radius are static, this trap will also be stationary in the disk throughout its evolution.

In figure \ref{Results2_Traps}, we plot the evolution of the traps' radii and planet formation tracks in both the constrained and unconstrained outflow models. Here, we only consider the fiducial setting of $\Sigma_0$. In the constrained outflow model, we see that the heat transition is located at a much smaller radius in the disk than either the turbulent or combined models of the previous section. This is as a result of the lower $\alpha_{\rm{turb}}=10^{-6}$ and reduced viscous heating. As previously outlined, the ice line's position is stationary in the disk near 1 AU, due to the static radiative equilibrium temperature profile. Since the traps exist at small disk radii, and the constrained outflow model maintains large surface densities within 10-20 AU, both traps efficiently produce warm Jupiters near 1 AU within only 1 Myr of formation. 

In the unconstrained outflow model, the radial locations of both the null torque and ice line traps are time-independent. In contrast to the constrained outflow model, planet formation in each of the two traps is extremely inefficient due to the rapid decrease in disk surface density caused by the strong outflow with $K=1$. Even considering a long disk lifetime of 10 Myr does not result in massive planets forming, since the disk surface density has decreased significantly by 1 Myr due to the outflow (see figure \ref{Results2_Disk}). The null torque only forms a sub-M$_\oplus$ core in this disk model, while the ice line forms a $\sim$ 3 M$_\oplus$ super Earth. In both cases, the planets form in-situ due to the time-independence of their traps' locations. 

We have also separately investigated how the positions of the traps change in both models with different settings of $\Sigma_0$. We find that the trap locations are extremely/entirely insensitive to the setting of $\Sigma_0$. This is mainly the case because the radiative equilibrium temperature profile, which dominates midplane heating, is independent of disk surface density (only depending on the radiative flux received by the disk). The resulting planet formation tracks are also very similar to those shown in figure \ref{Results2_Traps}. Based on these results that show limited planet formation and no variance with disk parameters, we do not expect a population synthesis model to be particularly interesting in either the constrained or unconstrained outflow models in terms of the planet traps framework. However, the low settings of $\alpha_{\rm{turb}}$ in a winds-dominated model may allow planets to form under a general type-I migration regime, as calculated with equation \ref{Results4_Paardekooper}, while not being restricted to the locations of traps. We have discussed this idea further in section \ref{Results4_Conclusion}. 


\bsp	
\label{lastpage}
\end{document}